\documentclass{article}

\usepackage{arxiv}

\usepackage{textcomp}
\usepackage[utf8]{inputenc} 
\usepackage[T1]{fontenc}    
\usepackage{url}            
\usepackage{booktabs}       
\usepackage{amsfonts}       
\usepackage{nicefrac}       
\usepackage{microtype}      
\usepackage{graphicx}
\usepackage[numbers]{natbib}
\usepackage{doi}
\usepackage{soul}
\usepackage{subcaption}
\usepackage{xcolor}
\usepackage{hyperref}
\usepackage{bm}
\usepackage{amsmath}
\usepackage{fancyvrb}
\usepackage{listings}
\usepackage{soul}
\usepackage{authblk}
\usepackage{mathtools}
\usepackage[toc,page]{appendix}

\newcommand{\figref}[1]{Fig.~\ref{#1}}

\newcommand{\eqiref}[1]{Eq.~\eqref{#1}}

\newcommand{\secref}[1]{Sec.~\ref{#1}}

\newcommand{\appref}[1]{App.~\ref{#1}}

\usepackage{pifont}

\usepackage{multirow}

\title{Towards scientific machine learning for granular material simulations - challenges and opportunities}

\author[1]{Marc Fransen\textsuperscript{*}}
\author[2]{Andreas Fürst\textsuperscript{*}}
\author[3]{Deepak Tunuguntla\textsuperscript{*}}
\author[4,15]{Daniel N. Wilke\textsuperscript{*}}
\author[2,16]{Benedikt Alkin}
\author[5]{Daniel Barreto}
\author[2,16]{Johannes Brandstetter}
\author[6]{Miguel Angel Cabrera}
\author[6]{Xinyan Fan}
\author[7]{Mengwu Guo}
\author[8]{Bram Kieskamp}
\author[9]{Krishna Kumar}
\author[10]{John Morrissey}
\author[1]{Jonathan Nuttall}
\author[10]{Jin Ooi}
\author[11]{Luisa Orozco}
\author[10]{Stefanos-Aldo Papanicolopulos}
\author[12]{Tongming Qu}
\author[6]{Dingena Schott}
\author[13]{Takayuki Shuku}
\author[14]{WaiChing Sun}
\author[8]{Thomas Weinhart}
\author[8]{Dongwei Ye}
\author[8]{Hongyang Cheng\textsuperscript{\dag}}

\affil[1]{Deltares, The Netherlands}
\affil[2]{Johannes Kepler University Linz, Austria}
\affil[3]{Saxion University of Applied Sciences, The Netherlands}
\affil[4]{University of Pretoria, South Africa}
\affil[5]{Edinburgh Napier University, UK}
\affil[6]{Delft University of Technology, The Netherlands}
\affil[7]{Lund University, Sweden}
\affil[8]{University of Twente, The Netherlands}
\affil[9]{University of Texas at Austin, USA}
\affil[10]{School of Engineering, The University of Edinburgh, UK}
\affil[11]{Netherlands eScience Center, The Netherlands}
\affil[12]{Hong Kong University of Science and Technology, Hong Kong}
\affil[13]{Tokyo City University}
\affil[14]{Columbia University, USA}
\affil[15]{University of the Witwatersrand, South Africa}
\affil[16]{Emmi AI GmbH, Austria}

\renewcommand{\shorttitle}{\textit{arXiv} GranML}

\begin{document}

\maketitle

\let\thefootnote\relax\footnotetext{\textsuperscript{*} Co-first authors. \textsuperscript{\dag} Corresponding author. All other co-authors are listed alphabetically.}

\begin{abstract}

Micro-scale mechanisms, such as inter-particle and particle-fluid interactions, govern the behaviour of granular systems. While particle-scale simulations provide detailed insights into these interactions, their computational cost is often prohibitive.
At a recent Lorentz Center Workshop on ``\href{https://www.lorentzcenter.nl/machine-learning-for-discrete-granular-media.html}{Machine Learning for Discrete Granular Media}'', researchers explored how machine learning approaches can aid the development of constitutive laws and efficient data-driven surrogates for granular materials while also addressing uncertainty quantification.
Attended by researchers from both the granular materials (GM) and machine learning (ML) communities, the workshop brought the ML community up to date with GM challenges.

This position paper emerged from the workshop discussions.
In this position paper, we define granular materials and identify seven key challenges that characterise their distinctive behaviour across various scales and regimes--ranging from gas-like to fluid-like and solid-like.
Addressing these challenges is essential for developing robust and efficient models for the digital twinning of granular systems in various industrial applications.

To showcase the potential of ML to the GM community, we present classical and emerging machine/deep learning techniques that have been, or could be, applied to granular materials. We reviewed sequence-based learning models for path-dependent constitutive behaviour, followed by encoder-decoder type models for representing high-dimensional data in reduced spaces.
We then explore graph neural networks and recent advances in neural operator learning. The latter captures the emerging field evolution of interacting particles via efficient latent space representation.
Lastly, we discuss model-order reduction and probabilistic learning techniques for high-dimensional parameterised systems, both of which are crucial for quantifying and incorporating uncertainties arising from physics-based and data-driven models. 

We present a typical workflow aimed at unifying data structures and modelling pipelines and guiding readers through the selection, training, and deployment of ML surrogates for granular material simulations.
Finally, we illustrate the workflow’s practical use with two representative examples, focusing on granular materials in solid-like and fluid-like regimes.

\end{abstract}


\section{Introduction}
\label{sec:intro}

Granular materials (GMs), ranging from beach sand to raw materials such as iron ore, are integral to various industrial processes.
They play an essential role across many engineering disciplines, including geotechnical \cite{OSullivan2011}, coastal \cite{Montella2023},  hydraulic engineering \cite{Lopez2018}, pharmaceutical \cite{Lopez2018}, additive manufacturing \cite{PARTELI201696,roy2024role,Alvarez2024}, agriculture \cite{HORABIK2016206,SHMULEVICH200737}, bulk handling \cite{SCHOTT202129,Fransen2022,LOMMEN2019273} and robotics \cite{Li2009a,Dieter2020}.

Although simple in appearance, GMs exhibit solid-, fluid-, and gas-like behaviour \cite{forterre2008flows}, making it one of the most complex materials to handle. A first-principle approach for simulating GM behaviour involves modelling individual grains' interactions. However, this is computationally expensive. Alternatively, continuum approaches can be adopted, where constitutive laws are required to describe the material's bulk behaviour that emerges from grain-scale interactions by relating stress and strain and/or their rates under specific conditions.

For certain solid-like behaviour, existing theories from computational solid mechanics, e.g., Mohr-Coulomb, Drucker-Prager and more advanced models incorporating micro-structural information like fabric tensors \cite{Li2002,Gao2014} can be utilised.
Similarly, for fluid-like GM processes, such as dense granular flows, theories from computational fluid mechanics \cite{tunuguntla2014mixture} provide useful frameworks. However, these models are typically crafted by ``domain experts'' within specialised fields and often limited to specific conditions.
As a result, they often fail to fully capture granular behaviour across the broad range of regimes and conditions encountered in practice.

\begin{figure}[b]
	\begin{minipage}[t]{0.5\textwidth}
		\centering		\includegraphics[height=0.8\textwidth]{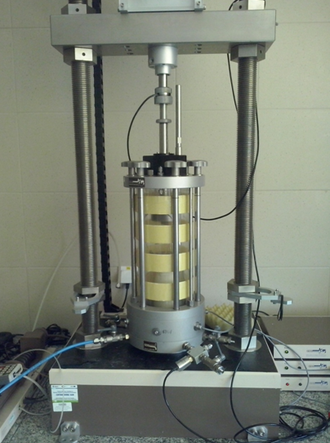}
	\end{minipage}
	\begin{minipage}[t]{0.5\textwidth}
		\centering		\includegraphics[height=0.5\textwidth]{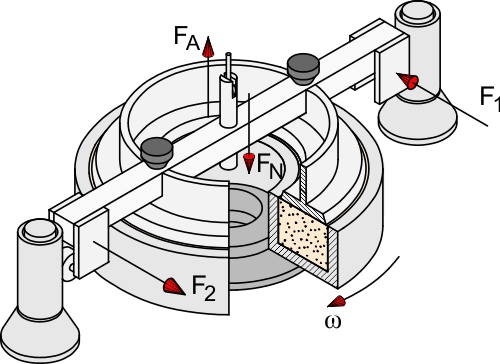}
	\end{minipage}
	\caption{Examples of laboratory devices for granular materials in (a) triaxial compression and (b) continuous ring-shear conditions. Copyright: \href{https://www.dietmar-schulze.de/ringschergeraete_e.html}{Dietmar Schulze}.}
	\label{fig:compress_and_shear}
\end{figure}

It is important to recognise that these conventional approaches were developed by combining theoretical frameworks with sparse experimental data, typically obtained under highly controlled laboratory settings (e.g. \figref{fig:compress_and_shear}).
While they perform adequately within these limited domains, extending them beyond their calibration range becomes increasingly challenging.
Unlike fluid and solid mechanics, granular mechanics lack a unified continuum theory, necessitating expensive, phenomenon-specific, multi-scale approaches to develop effective continuum models.

\paragraph{Current research interests towards modelling granular materials}

Current research focuses on (1) developing a unified theory capable of capturing GM phenomena across solid-like, fluid-like and gas-like regimes \cite{forterre2008flows,kamrin2024advances}; (2) upscaling particle-level methods to handle the size and complexity of industrial-scale problems; and (3) incorporating real-world variability, including boundary condition uncertainties, measurement noise and the inherent stochastic nature of granular systems \cite{Cheng2024,Fransen2025}.
From (1)--(3), a central question emerges:
can we derive or learn unified constitutive theories and governing equation solvers for granular materials directly from particle-scale information, resulting in models that are not only more efficient but also provide deeper insights than contemporary, phenomenological approaches? 

To address this, researchers increasingly rely on grain-scale numerical simulations, primarily employing the Discrete Element Method (DEM), in conjunction with micro-scale observations, e.g., using the micro-Computed Tomography \cite{Vlahinic2016}. 
DEM \cite{Cundall1979ADN} remains the most widely used technique for modelling granular systems and has demonstrated success across multiple fields, including process engineering \cite{Kieckhefen2020}, geotechnical engineering \cite{Khosravi2020}, bulk handling \cite{Schott2023, FRANSEN2023118526}, mining \cite{ABOUSLEIMAN2020123}, chemical engineering \cite{Golshan2020} and additive manufacturing \cite{Alvarez2024,shaheen2021influence}.

\paragraph{From physics-based solvers to data-driven alternatives}

DEM simulations are information-rich, capturing particle positions, velocities and orientations, and interaction forces with other particles and geometries.
This data can be subsequently mapped to macroscopic stress and strain fields, ultimately enabling the construction of closed continuum-scale models.
While effective, this multi-step approach remains highly phenomenon- and regime-specific, relying heavily on specialised domain knowledge and ad hoc upscaling \cite{Vescovi2016} or multi-scale coupling \cite{Guo2014,Cheng2023} algorithms. 
Recall that, unlike solids and fluids, granular materials do not have a unified theory.
As a result, the entire \emph{experiments–micro–macro–continuum–model} pipeline becomes an iterative and costly endeavour.
Addressing this challenge raises a new set of research questions:

\begin{enumerate}
\item Can data-driven approaches expedite the multi-step process of realising phenomenon-specific granular theories?
\item Can data-driven alternatives fully replace contemporary physics-based models by overcoming their inherent limitations?
\item Can data-driven methods help realise the unified theory that the GM community has long strived for?
\item More broadly, how can we best utilise the available GM data to complement our current approaches to modelling discrete media?
\end{enumerate}

Over the last two decades, machine learning techniques have evolved into powerful alternatives for computational mechanics, enabling researchers to process vast amounts of data and extract deeper insights from detailed simulations (see the recent review \cite{wang2024machine}). As a result, ML has enabled researchers to perform tasks like constitutive model development \cite{Bahmani_2024} and surrogate modelling \cite{qu2023data,mayr2023boundary}, model identification \cite{HARTMANN2022104491} and design optimisation \cite{FRANSEN2023118526}.
Deep neural network-based surrogates have recently attracted attention as a more universal methodology~\citep{thuerey2021physics,brunton2023machine}, transforming computational fluid dynamics~\citep{guo2016convolutional, kochkov2021,li2021fourier,gupta2022towards,alkin2024universal}, weather forecasting~\citep{rasp2021data, weyn2020improving, sonderby2020metnet, pathak2022fourcastnet, lam2022graphcast, nguyen2023climax,bodnar2024aurora}, and molecular modelling \citep{batzner2022e3, batatia2022mace,Zeni:23,Merchant:23} or protein folding predictions~\citep{jumper2021highly}. Beyond the ever-urgent topic of computational efficiency, neural surrogates -- models that approximate the underlying physics by learning patterns and relationships directly from data -- have the potential to generalise across various phenomena and characteristics,
such as boundary conditions or coefficients~\citep{brandstetter2022message,mccabe2023multiple,herde2024poseidon}.

One characteristic of classical physics-based numerical methods is their required conditions for numerically solving the equations~\citep{quarteroni2008numerical}. Although these conditions are mitigated in these deep ``neural surrogates'', they still exist when looking at deep neural network-based techniques, where different deep learning architectures are prevalent across applications. This prevalence is exemplified when contrasting deep learning approaches with particle- and grid-based simulations. 

Graph neural networks (GNNs)~\citep{Scarselli:08, Kipf:17} are a natural choice for surrogate modelling of particle-based dynamics in DEM simulations. Predicted node accelerations are often integrated numerically to advance particle dynamics in a hybrid ML-numerical fashion~\citep{Sanchez:20, Pfaff:20, Mayr:23}. 
Many recent deep learning-based approaches for granular materials adopt GNN-based simulators \citep{choi2023graph,kumar2023accelerating,choi2024inverse}.
For continuum models, typically solved on structured or unstructured grids, Fourier neural operator (FNO)-based~\citep{li2021fourier}, convolutional neural network (CNN)-based~\citep{gupta2022towards,raonic2024convolutional}, or Transformer-based~\citep{cao2021choose,Li:OFormer,alkin2024universal} architectures could be suitable alternatives -- assuming fixed grids for simplicity reasons.
Most recently, NeuralDEM \citep{alkin2024} has demonstrated the ability to model large-scale coupled DEM-computational fluid dynamics (CFD) simulations using transformer-based multi-branch neural operators.

Despite this promise, building large-scale neural surrogates or particle \emph{foundation} models introduces the following concerns:
\begin{itemize}
    \item Training such models demands extensive, high-quality datasets. Yet, the opaque nature of granular materials limits experimental data availability and quality, making physics-based models such as DEM or continuum simulations indispensable as sources of reliable training data. Moreover, industrial processes involve diverse granular phenomena, geometries, and material parameters, requiring unified training strategies to effectively capture the underlying physics. 
    \item Interpretability and explainability remain major concerns. Domain experts need to extract and understand the learned physics to trust and apply these models. However, ML models often function as black boxes, lacking guarantees on their predictive behaviour.
    \item Model training is computationally intensive, sometimes comparable to GPU-parallelised particle simulations. Additionally, as with physics-based models requiring calibration, ML models suffer from performance drift and must be regularly updated as new data becomes available, therefore requiring efficient training strategies like active and transfer learning.
    \item A final challenge concerns defining what constitutes a ``good'' model. Beyond accuracy and efficiency, models must maintain consistent performance under uncertainty, which is crucial when probing parameter spaces for calibration, uncertainty quantification, and design optimisation.
\end{itemize}

Nevertheless, the confluence of computational granular mechanics and machine learning follows a similar path---much like the development of classical theories---by integrating prior knowledge with patterns learnt from complex, high-dimensional data. This implies that machine learning surrogates should be seen as complementary to, rather than replacements for, physics-based modelling advances. 

\section{Aim and objectives}
\label{sec:aim}

Contemporary physics-based approaches towards modelling granular materials and systems face several challenges to provide a unified continuum theory for granular materials.
In this paper, we present seven key challenges that must be addressed to advance the field of granular modelling in \secref{sec:GM_challenges} and explore how state-of-the-art machine learning methodologies could provide potential solutions to these challenges in \secref{sec:MLSolution}. We aim to briefly analyse each challenge and identify corresponding ML models and algorithms that might offer viable solutions.
Furthermore, we present two illustrative examples in \secref{sec:examples} to demonstrate potential workflows and integration of numerical models and their ML surrogates. Here, the focus lies on data structures and unification of workflows, which can be used in engineering and research.
This paper is not a comprehensive review but reflects our collective viewpoint on the future direction of this emerging field. All the discussions took place during and after the Lorentz Center Workshop, ``Machine Learning for Discrete Granular Media'', from 29 April to 3 May 2024.

We address two primary audiences: (1) experts in computational mechanics, physics, and engineering working on granular materials and (2) ML researchers interested in developing data-driven surrogates for modelling particulate dynamics and related phenomena. The specific objectives of this paper are:

\begin{itemize}
	\item To identify and articulate the key challenges faced by the granular materials modelling community.
	\item To highlight recent ML developments capable of capturing both local and global spatio-temporal responses of granular materials, including their associated uncertainties..
	\item To act as a source of follow-up research directions, effectively establishing a collaborative workflow between computational granular mechanics and ML communities.
\end{itemize}

\section{Challenges in granular material simulations}

\label{sec:GM_scales}

\subsection{Micro, meso and macro scales of granular systems}

\begin{figure}[t!]
    \centering
    \resizebox{0.6\textwidth}{!}{
    \begin{minipage}{\textwidth}
    \centering
    \begin{minipage}{0.33\textwidth}
        \includegraphics[width=\textwidth]{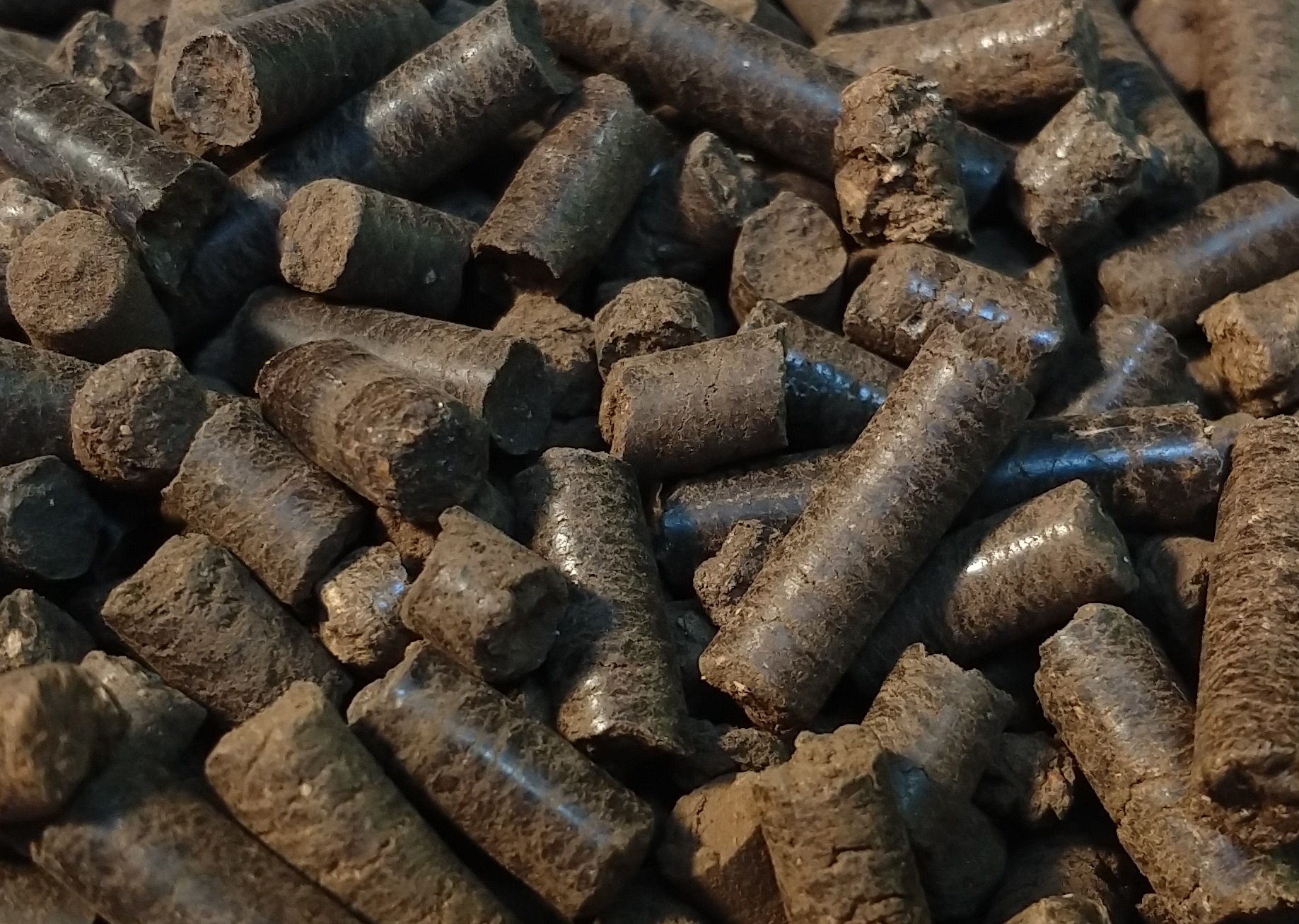}
    \end{minipage}
    \begin{minipage}{0.33\textwidth}
        \includegraphics[width=\textwidth]{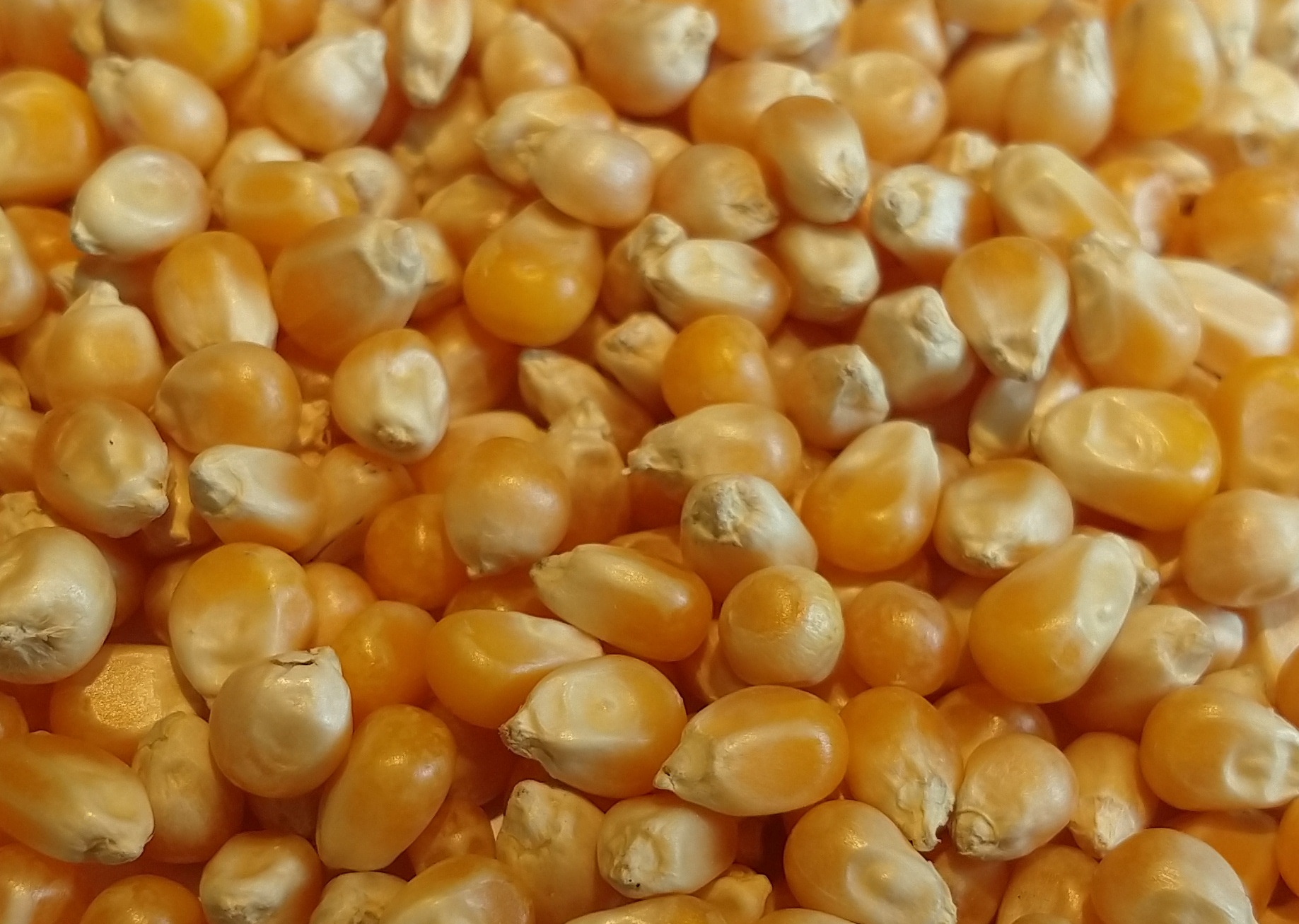}
    \end{minipage}
    \begin{minipage}{0.33\textwidth}
        \includegraphics[width=\textwidth]{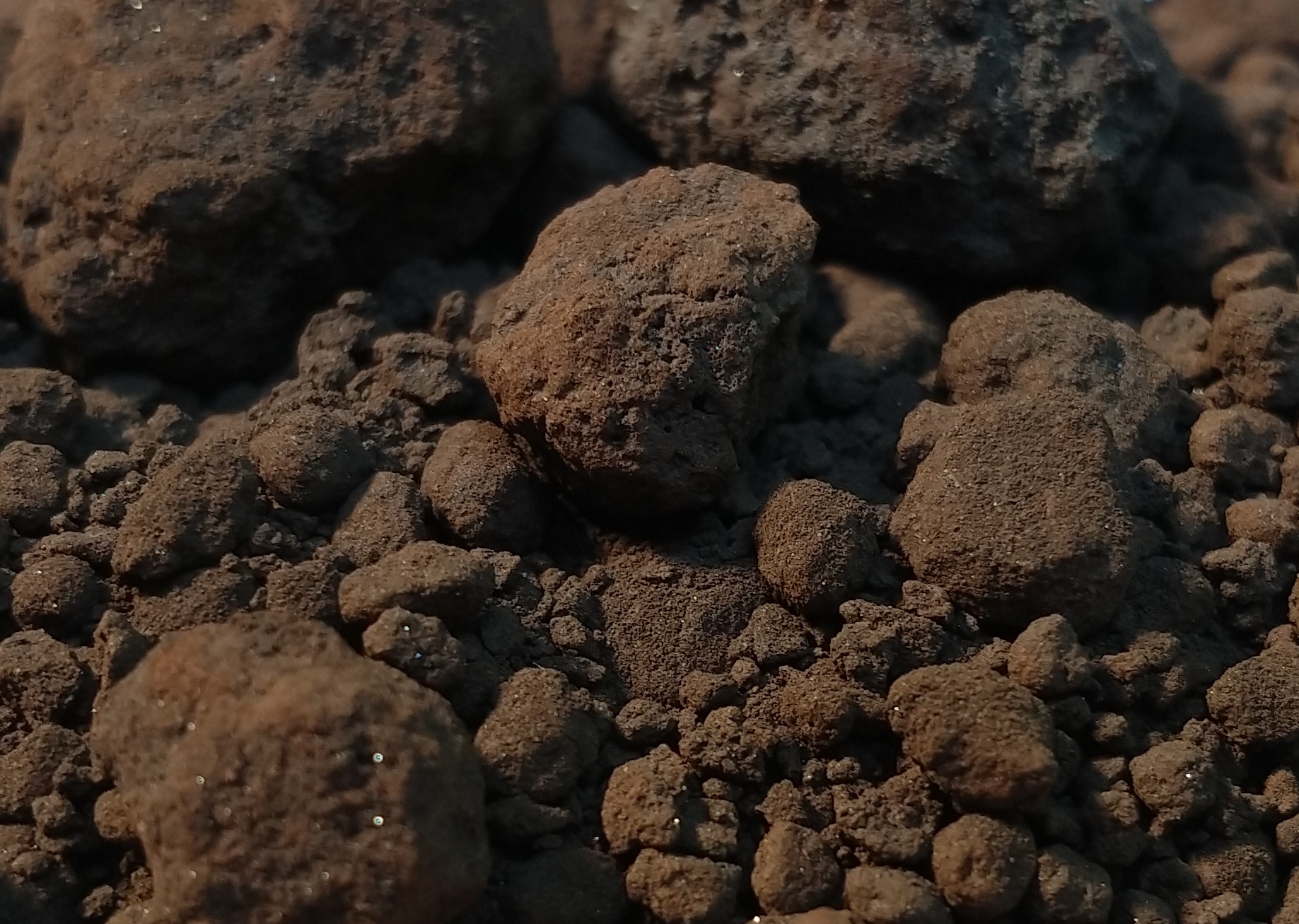}
    \end{minipage}
    \\[0.1em]
    \begin{minipage}{0.33\textwidth}
        \includegraphics[width=\textwidth]{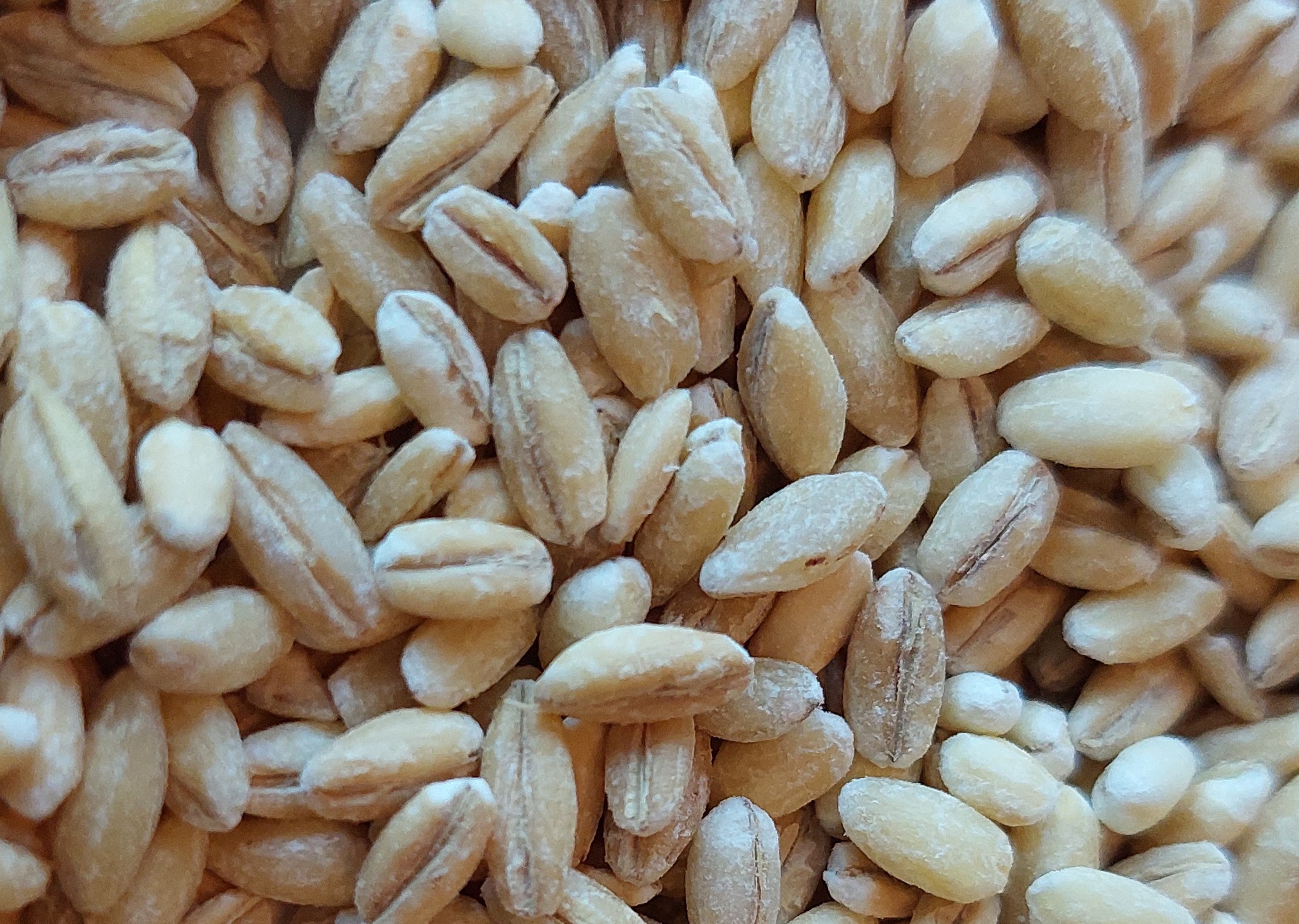}
    \end{minipage}
    \begin{minipage}{0.33\textwidth}
        \includegraphics[width=\textwidth]{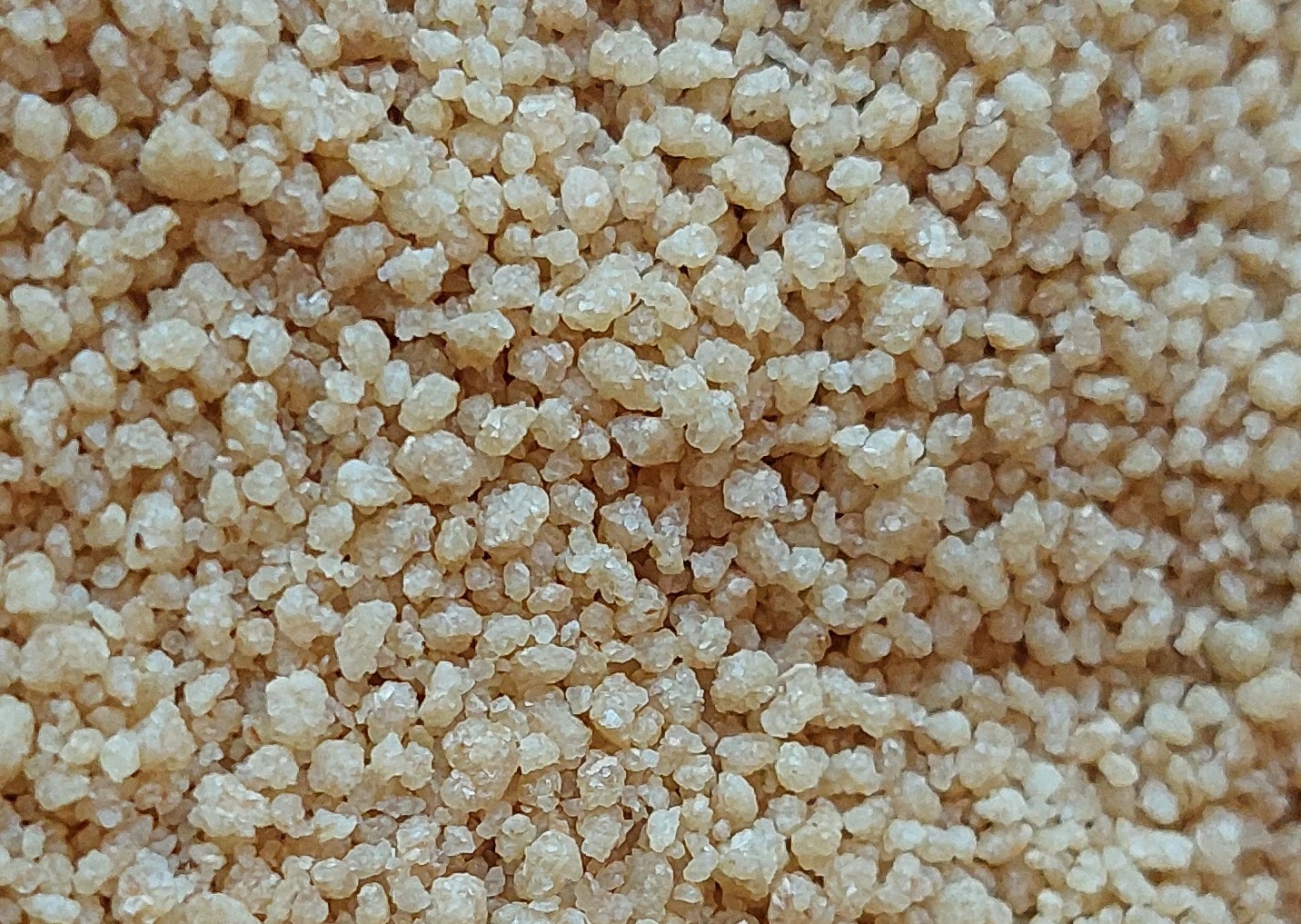}
    \end{minipage}
    \begin{minipage}{0.33\textwidth}
        \includegraphics[width=\textwidth]{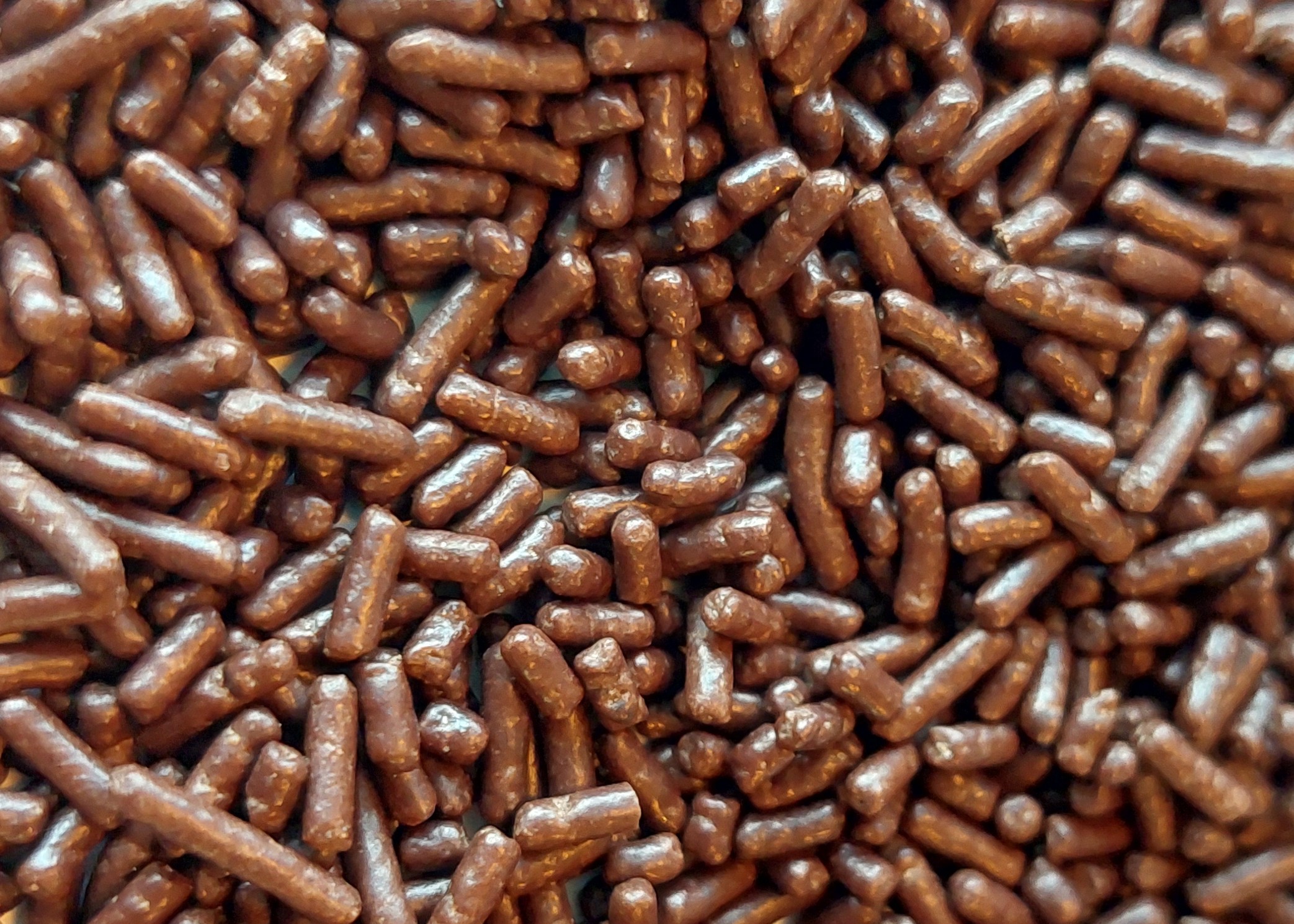}
    \end{minipage}
    \\[0.1em]
    \begin{minipage}{0.33\textwidth}
        \includegraphics[width=\textwidth]{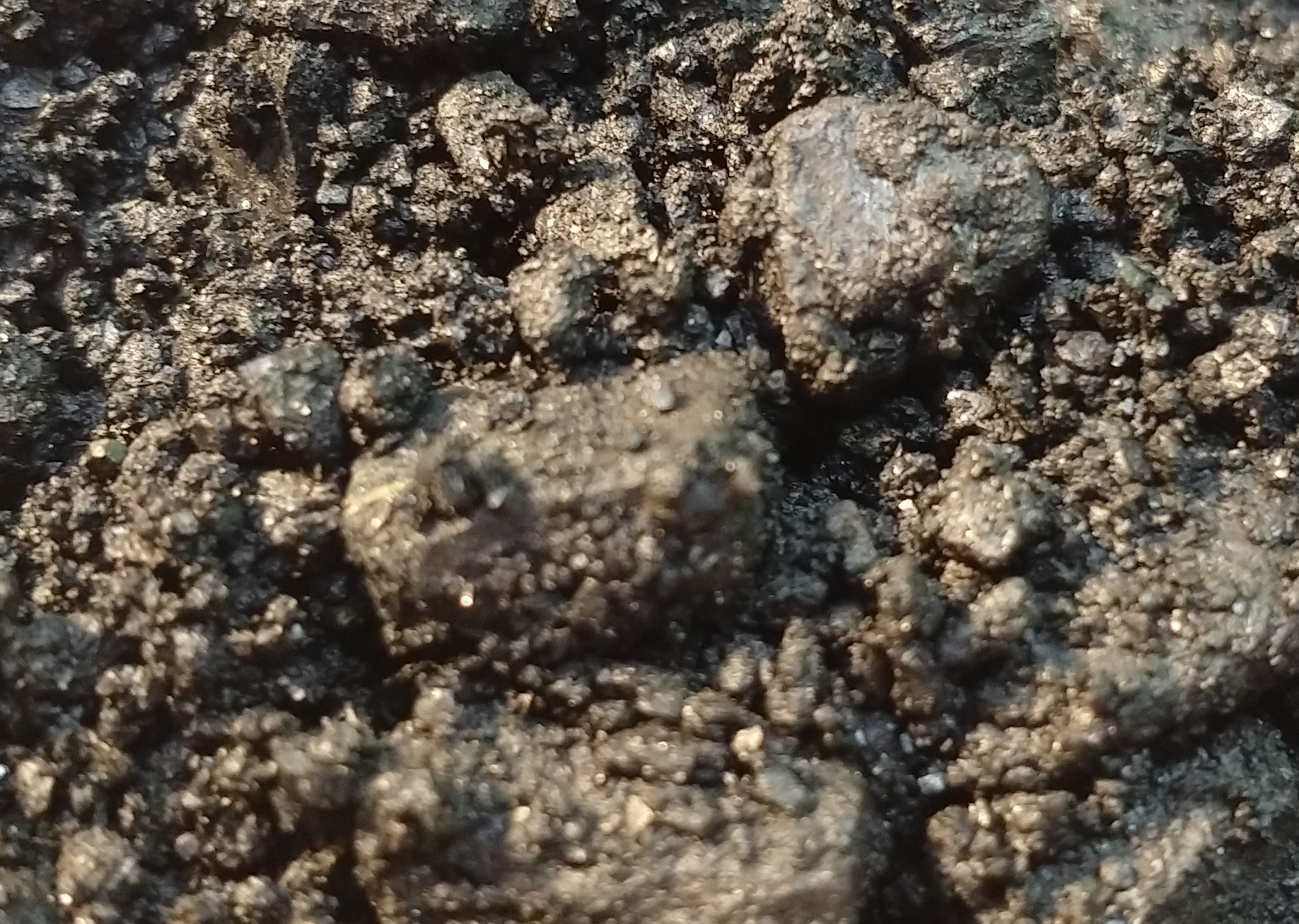}
    \end{minipage}
    \begin{minipage}{0.33\textwidth}
        \includegraphics[width=\textwidth]{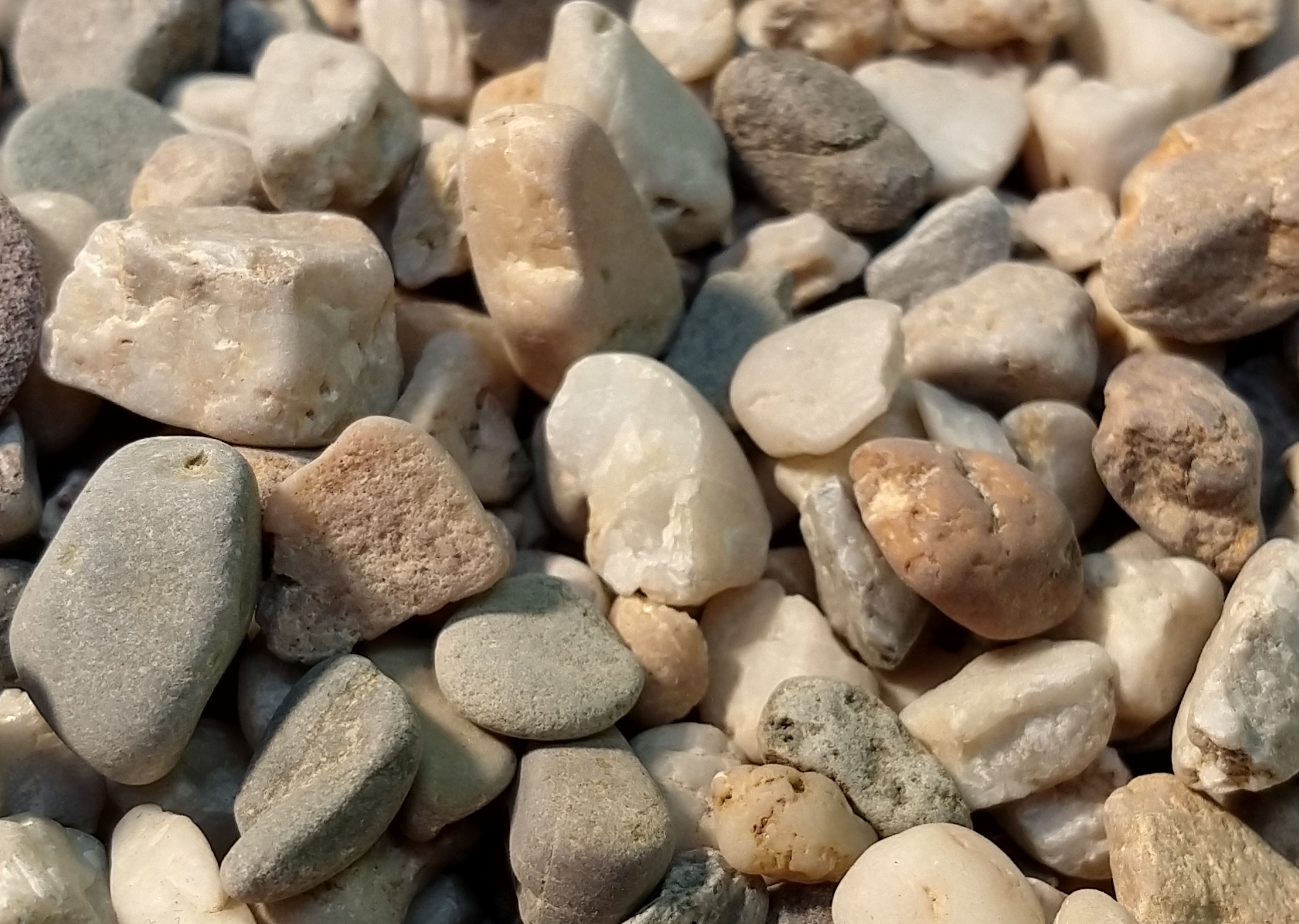}
    \end{minipage}
    \begin{minipage}{0.33\textwidth}
        \includegraphics[width=\textwidth]{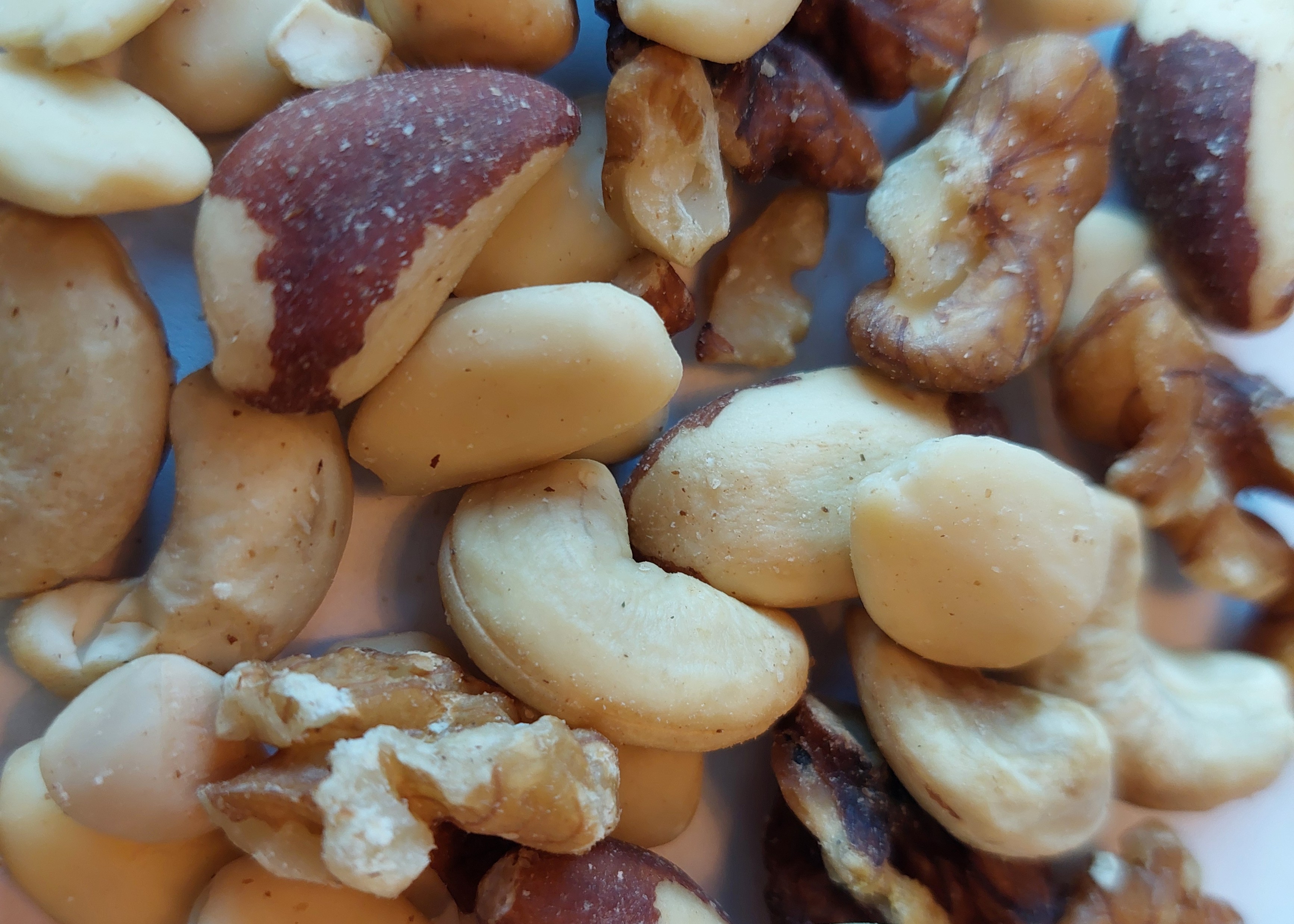}
    \end{minipage}
    \end{minipage}
    }
    \caption{Examples of granular materials of various sizes and shapes.}
    \label{fig:msmGM}
\end{figure}

The behaviour of granular systems is dominated by discrete micro-scale mechanisms, e.g.\ the interaction of particles through forces at contacts. DEM allows explicit modelling of these mechanisms across all regimes of granular materials. However, the computational cost is often prohibitive, especially when considering micro-scale complexities such as particle shape variations, wide particle size distributions, or complex interaction physics, as illustrated in \figref{fig:msmGM}.

\begin{figure}[b!]
    \centering
	\includegraphics[width=0.4\linewidth]{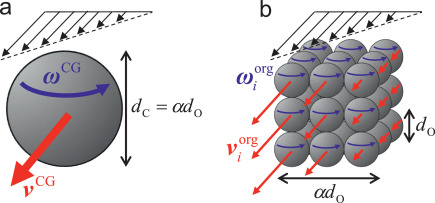}
	\caption{Schematic of (a) coarse-grained particles and (b) original particles. Reprinted with permission from \citep{Nakamura2020}.}
	\label{fig:mesoparticles}
\end{figure}

A common technique to reduce computational costs is \emph{Coarse-graining} (CG) where meso-particles---computational particles representing aggregates of real particles---are used within DEM simulations. While CG lowers computational demands, it introduces new challenges, particularly in identifying equivalent mesoscopic properties to ensure that the \emph{coarse-grained} system faithfully represents the original microstructure (\figref{fig:mesoparticles}). 

\begin{figure}[t!]
	\centering
	\includegraphics[width=\linewidth]{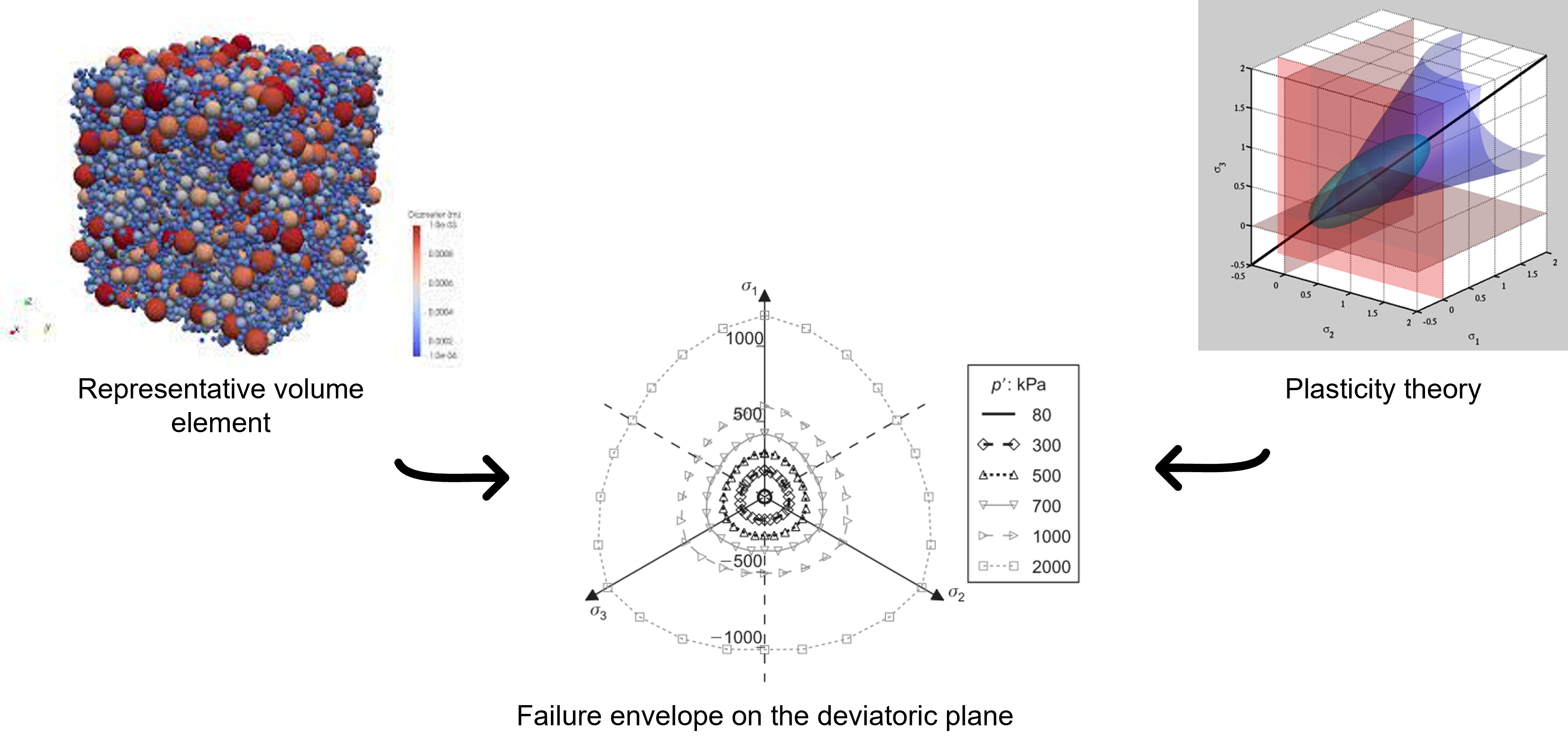}
	\caption{Micro-macro relationship between a representative volume element (DEM) and a phenomenological representation (constitutive law) of granular materials. Figure adapted from \cite{Zhao2013}.}
	\label{fig:macro_models}
\end{figure}

Modelling industrial-scale granular systems is difficult due to the vast amount of particles involved. Continuum methods, such as the finite element method (FEM), material point method (MPM) or smooth particle hydrodynamics (SPH), are often used to reduce computational costs by avoiding explicit micro-scale modelling. These methods represent granular materials through continuum fields, constitutive laws and partial differential equations (PDEs). These methods represent granular materials through continuum fields and partial differential equations, but this comes at the expense of phenomenological constitutive models, which may not fully capture discrete granular effects.
\figref{fig:macro_models} shows a micro-macro relationship between a representative volume element and a phenomenological representation of granular materials, which has been sought for decades within the GM community.

\begin{figure}[b!]
	\begin{minipage}[t]{0.5\textwidth}
	\centering
	\includegraphics[height=0.4\textwidth]{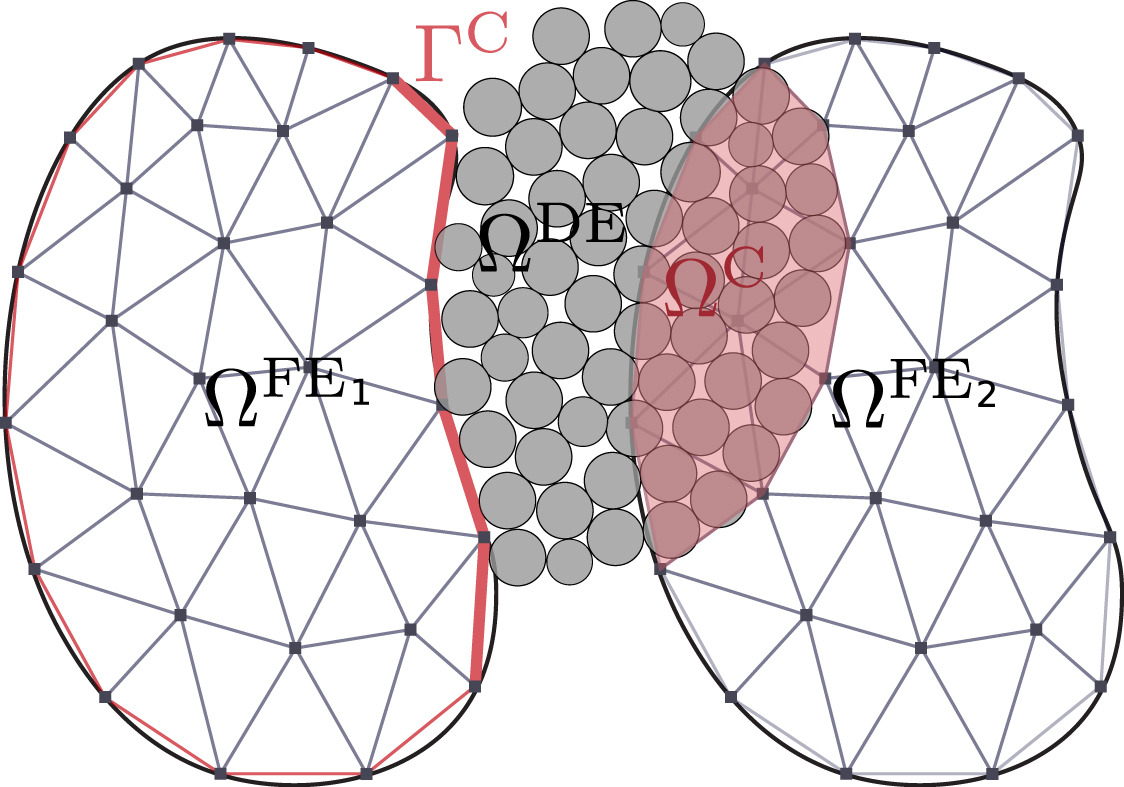}
        \\
        (a)
	\end{minipage}
	\begin{minipage}[t]{0.5\textwidth}
	\centering
	\includegraphics[height=0.6\textwidth]{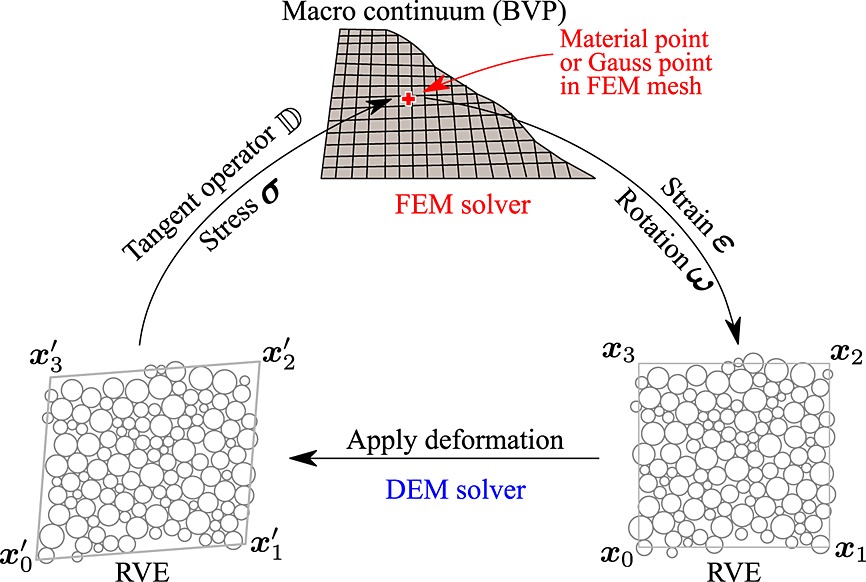}
        \\
        (b)
	\end{minipage}
	\caption{Multi-scale modelling for granular materials: (a) concurrent and (b) hierarchical methods. Reprinted with permission from \cite{Cheng2023} and \cite{Guo2014}.}
	\label{fig:msmmodels}
\end{figure}

Multi-scale methods attempt to bridge this gap by embedding discrete micro-scale simulations within macro-scale continuum models, for instance, through replacing constitutive laws or parts of the domain with DEM simulations. Although such multi-scale methods improve physical realism, the embedded micro-scale simulations still remain the bottleneck to improve the computational efficiency, see \figref{fig:msmmodels}.
When applying the methods above to multi-physical problems, different length/time scales are encountered, e.g.\ when the fluid between particles has to be modelled because the length scales corresponding to micro, meso and macro levels for each physical process often differ. For example, the length and time scales for heat transfer or pore fluid flow differ from plastic mechanical deformation where strain typically localizes in the granular materials.

\subsection{Physical regimes of granular materials}
\label{sec:GM_regimes}

\begin{figure}[t!]
    \centering
    \includegraphics[width=0.4\linewidth]{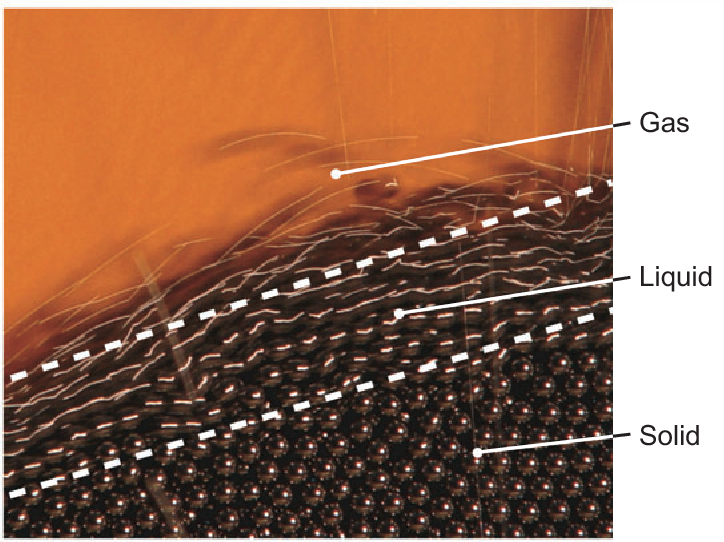}
    \caption{Three physical states of granular materials. Reprinted with permission from \cite{forterre2008flows}.}
    \label{fig:GMstates}
\end{figure}

On top of the challenges of determining appropriate length/time scales---critical for describing the relevant physics---the physical states of granular materials often vary, resulting from evolving internal length/time scales due to external loads.
This causes granular materials to exhibit quasi-static (solid-like), steady-state flow (fluid-like) and dynamic (transitional) responses, as shown in \figref{fig:GMstates}.
In the following, we briefly introduce these three physical regimes and their characteristic features, including kinetics, initial and boundary conditions, and typical geometries.

\subsubsection{Quasi-static behaviour of granular solids}
\label{sec:quasi}

\begin{figure}[b!]
    \centering
    \includegraphics[width=0.4\linewidth]{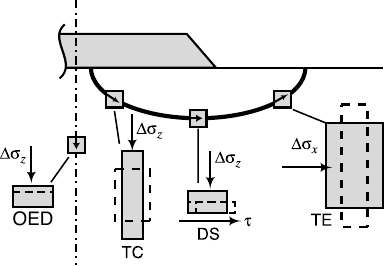}
    \caption{Typical examples of quasi-static processes in granular materials (soil) underneath an embankment. The deformation field of a granular sample (dashed line) is controlled by the acting stress field (i.e., $\Delta \sigma_i$, $\tau$), often captured by soil element tests such as the oedometer (OED), triaxial compression (TC), triaxial extension (TE) and direct shear (DS) tests. Figure adapted from \cite{budhu2010}.}
    \label{fig:sketch_quasiStatic}
\end{figure}

Quasi-static processes are characterised by slow loading conditions, so it can be assumed that the process is in equilibrium at a given instant in time. Under these conditions, any inertial effects present in other transient and steady states are neglected. Typical examples include oedometer, triaxial and direct shear tests (see Fig.~\ref{fig:sketch_quasiStatic}), which are essential for deriving material parameters later used in geotechnical design. 

In a quasi-static state, loads (or stresses) may be changed to estimate (or determine) the sample deformation (or strain), or vice-versa and boundary conditions are set to reflect the process under consideration. For example, soil under an embankment may experience different boundary and stress conditions. Directly underneath the centre of the embankment, soils experience one-dimensional compression, as in an oedometer test (OED), where a cylindrical soil sample is compressed vertically under constant stress (incrementally applied) whilst deformation in other directions remains constant. Different locations along the failure plane may experience triaxial compression (TC), triaxial extension (TE) or direct shear (DS) where the combination of stresses and boundary conditions differ.

\subsubsection{Steady-state flow of granular fluids}

Steady-state flow refers to conditions where the macroscopic properties of a granular material remain constant over time despite continuous internal movement and rearrangement of individual particles. In steady-state, properties such as shear rate, velocity profiles, and packing density reach equilibrium even though particles are still in motion and interacting. Examples of steady-state flow in granular materials include rotating drums, silos during discharge, and fluidised beds. In these systems, a stable flow is maintained under constant loading conditions. Modelling steady-state flows typically involves DEM to simulate particle interactions and, when fluids are present, a coupled DEM-CFD approach to capture particle-fluid dynamics. Such steady-state conditions are key to understanding the flowability of granular materials in industrial applications like bulk material handling, where a consistent flow is essential for the efficiency of the machines and manufacturing processes.

\subsubsection{Transient regimes between solid-like and fluid-like}
 
\begin{figure}
    \begin{minipage}[t]{0.5\textwidth}
        \centering
        \includegraphics[width=0.7\textwidth]
        {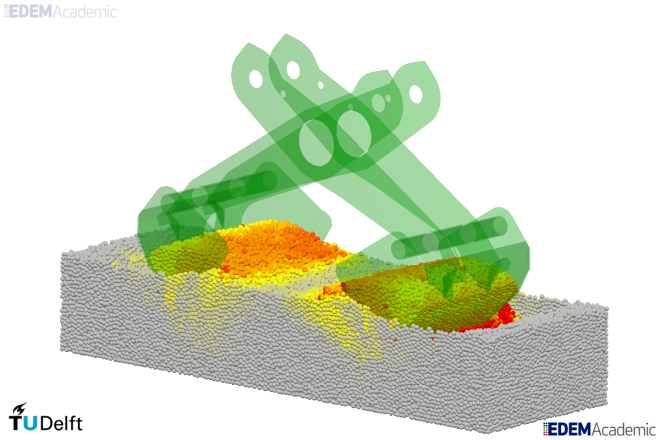}
        \subcaption{\centering}
    \end{minipage}
    \begin{minipage}[t]{0.5\textwidth}
        \centering
        \includegraphics[width=0.7\textwidth, height=0.5\textwidth]{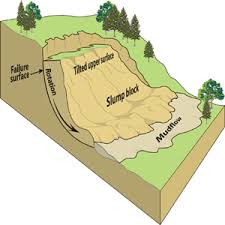}
        \subcaption{\centering}
    \end{minipage}
    \caption{Examples of dynamic granular flow processes: (a) grabbing and (b) landslide. Reprinted with permission from \cite{SCHOTT202129}.}
    \label{fig:voorbeelden}
\end{figure}

Dynamic granular flows are non-steady, and their evolution strongly depends on timescales. Grabbing processes and landslides are representative examples (\figref{fig:voorbeelden}). These processes often involve abrupt transitions between quasi-static and flowing states.

In grabbing processes, the non-steady nature arises from the rapid manipulation or displacement of granular materials. Here, timescales are crucial as they dictate the response of the material to external forces and vice versa, meaning that the grab response depends on packing density and interactions between material and equipment.
Abrupt events such as dike failures and landslides are examples of dynamic granular flows at a catastrophic scale, characterised by sudden and often unpredictable shifts in particle distributions and motion due to stress accumulation, geological instabilities, or external triggers. Understanding the relevant timescales is essential for effective risk assessment, hazard mitigation, and disaster management.

From a modelling perspective, the challenges that grabbing processes (cf. \figref{fig:voorbeelden}a) pose is accurate representation of granular material-machine interactions, including the effects of real particle sizes and shapes, leading to large or even unfeasible computational expenses. Additionally, granular materials exhibit varying degrees of cohesion which may fundamentally alter the rheological behaviour of the flows \cite{staron2023cohesive}.

Similarly, modelling landslide failures (\figref{fig:voorbeelden}b) requires simulating the initiation and progression of soil, rock, and debris movement down slopes. Challenges include determining material properties, identifying failure triggers, and accurately capturing the evolution of instabilities across multiple spatial and temporal scales, often requiring CFD-DEM approaches \cite{zhao2017coupled}. Such models have versatile applications, from predicting failure initiation to delineating impact zones, runout areas, and flow velocities, which are crucial to assessing risks and designing mitigation measures. 

\subsection{Main challenges in granular material simulations}
\label{sec:GM_challenges}

The peculiar behaviour of granular materials across scales and states, as described above, gives rise to the following main modelling challenges:
    
\begin{enumerate}
    \item \textbf{Prohibitive computational costs for large-scale DEM:} Modelling granular systems using DEM becomes computationally prohibitive as real-world problems become large.
    Problems with hundreds of millions of particles can be targeted with high-performance computing techniques. However, this comes at a significant computational resource cost and requires efficient parallelisation methods across CPU or GPU cores. Large-scale problems, however, can contain particles far exceeding this amount (e.g. granular flows during dike collapse or landslide and sediment transport in rivers), exceeding the bounds of current computational capabilities.
    \item \textbf{Identifying effective coarse-grained representation:} For design and optimisation tasks, faster models are needed.
    Ensuring that the models effectively capture relevant phenomena at a larger length/time scale without excessive computational cost is essential. This necessitates the identification of coarse-grained properties and structures for meso-particles, allowing for simplified yet effective modelling within the DEM modelling framework.
    \item \textbf{Expensive and repetitive DEM in multi-scale frameworks:} Multi-scale approaches require many micro-scale simulations that often undergo similar loading conditions within certain regions of the macroscopic problem. Although the difference is subtle, these simulations collectively contribute to the global phenomena like shear localisation and jamming/unjamming. Moreover, particle-scale characteristics, such as particle shape, surface roughness, and crushability, add extra complexity and computational cost to these micro-scale simulations.
    \item \textbf{Separation and evolution of length and time scales:} Model resolution is crucial, as macro-scale phenomena emerge from micro-scale interactions. Understanding the extent to which micro-scale details are needed involves elucidating the relationship between the macroscopic characteristics of granular materials, including anisotropy, dilatancy and stress-path dependency, and particle-scale and microstructural characteristics. 
    \item \textbf{Lack of a unified continuum theory across all granular regimes:} \secref{sec:GM_regimes} shows that many industrial applications involve moving structures and granular materials transitioning from quasi-static to free-flowing. Describing the transient behaviour of granular materials is an active area of research; DEM remains the fundamental tool for obtaining relevant data for discovering unified continuum theories for granular materials in gas, fluid and solid-like regimes.
    \item \textbf{Coupling with multi-physics and external structures:} Many problems require integrating multi-physics solvers and handling different material boundaries and domains. These include coupling DEM with CFD or SPH, addressing the interaction between grains and pore fluids, and accommodating changing geometry and loading conditions, which are often uncertain and application-dependent. This capability is essential for accurate system simulations with important multi-physical and grain-structure interactions.
    \item \textbf{Model calibration, validation, and uncertainty quantification:} Experimental data for granular materials is often limited to macro-scale quantities, making GM model calibration and validation challenging. Additionally, different types of granular materials, ranging from calcareous and siliceous sands to rocks, respond distinctively to external loads, complicating the development of generalised GM models. Therefore, quantifying and constraining parameter uncertainties across scales is critical for building robust GM digital twins.
    \end{enumerate}

\section{Machine learning methods for computational mechanics and engineering problems}
\label{sec:MLSolution}

\begin{figure}
    \centering
    \includegraphics[width=0.7\textwidth]{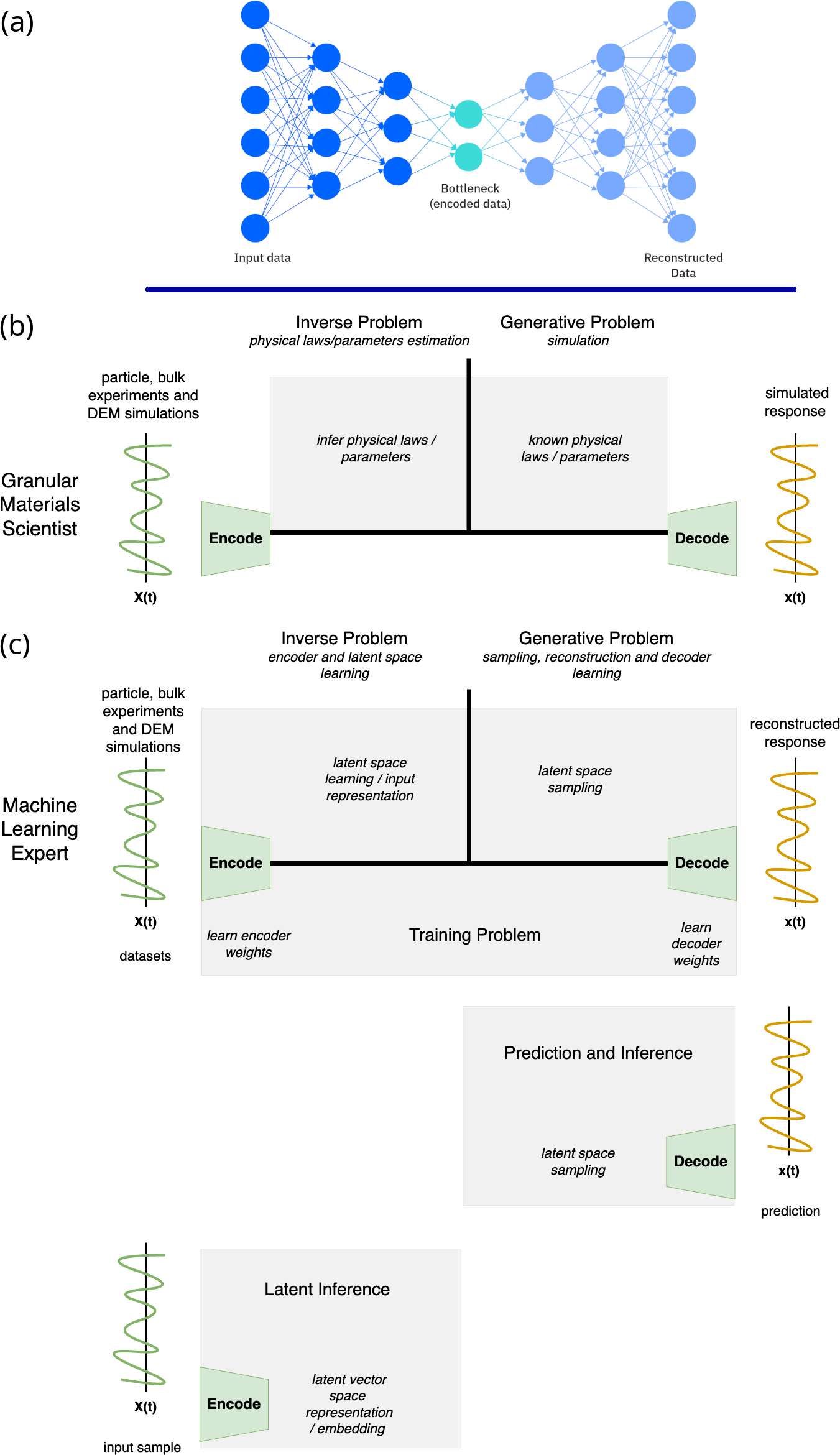}
\caption{A dual encode-decode perspective of granular material modelling and machine learning. (a) illustrates a simple autoencoder architecture, highlighting the \href{https://www.ibm.com/think/topics/latent-space}{encode-decode} philosophy in deep learning. (b) presents the granular material scientist's perspective, focusing on model parameter estimation from particle and bulk measurements, DEM simulations, $X(t)$, and generating simulated responses using known parameters. (c) the ML expert's perspective is showcased, emphasising the training problem through model weights and latent space learning from datasets, with response reconstruction through latent sampling and decoding. It further illustrates the prediction and inference problems, where encoded input samples are decoded to predict and understand the response $x(t)$, alongside latent inference that encodes input samples into a latent parameter space for interpretation.}
\label{fig:generative_model}
\end{figure}

GM experts traditionally focus on inferring physical laws from experiments and simulations to develop phenomenological models, typically relying on years of experience and a deep understanding of the underlying physics. In contrast, ML experts formulate algorithms that learn patterns and physical behaviours directly from data, aiming to uncover hidden structures and governing laws \cite{bishop2007}.
Therefore, given sufficient experimental and numerical data, both GM and ML communities are confronted with the same fundamental challenge: how to encode and infer hidden laws or models from data into abstract mathematical representations.
The data must contain meaningful instances of the relevant physics to be effectively learned. Once this knowledge is encoded, it can be generalized or decoded to predict unseen scenarios. 
\figref{fig:generative_model} illustrates this dual encoding-decoding perspective for both GM and ML experts.

Machine learning encompasses extensive methods to obtain hidden knowledge by training models from heterogeneous datasets.
Classical machine learning methods commonly used in computational mechanics include supervised learning, such as support vector machines and Gaussian processes for classification and regression tasks, and unsupervised learning, such as principal component analysis and manifold learning for feature extraction, dimensionality reduction and clustering.
Probabilistic approaches, including Bayesian inference and probabilistic distribution and stochastic process models, are often used to incorporate uncertainty explicitly, making them particularly valuable for uncertainty quantification.
While classical machine learning methods focus on structured, interpretable models, modern deep learning extends these capabilities by employing neural networks to extract knowledge from high-dimensional, often unstructured data.

\subsection{Encode-decode philosophy}

As already briefly stated in the introduction, recent deep neural network-based learning advances show promise in addressing the previously described granular material modelling challenges. Before exploring these developments in more detail, it is useful to first introduce the concept of \emph{latent space} in deep learning.

Paraphrasing \href{https://www.ibm.com/think/topics/latent-space}{IBM Think}, a latent space is a compact representation of data that retains only the essential aspects, capturing its underlying patterns and structure. The process of mapping data into this space is referred to as \emph{encoding}. By doing so, encoding expresses complex data efficiently and meaningfully, enhancing the ability of machine learning models to understand and manipulate it while reducing computational requirements. 

Typically, encoding data to a latent space involves some dimensionality reduction. That is compressing high-dimensional data to a lower-dimensional space and, as a result, omitting irrelevant or redundant information. For example, the famous \hyperlink{http://yann.lecun.com/exdb/mnist/}{MNIST} dataset contains tens of thousands of 28x28 grayscale images of handwritten digits, each represented as a 784-dimensional vector where each dimension corresponds to a pixel value between 0 (black) and 1 (white). The vectors would be 2,352-dimensional for colour images, with three values for each pixel (RGB). However, most of the image is an empty background, so reducing the vector to only the relevant dimensions (the latent space) can enhance a model’s ability to process the images efficiently and accurately.

A well-known neural network architecture for encoding/compressing/reducing data to a latent space is the autoencoder, see~\figref{fig:generative_model}(a). A self-supervised model aims to compress input data through dimensionality reduction and reconstruct the original data from this compressed form. In a typical autoencoder, the encoder consists of layers with progressively fewer nodes, compressing the input as it passes through each layer. The decoder then uses the compressed latent vector to reconstruct the original input. Hereby extracting the most important features of the data and effectively learning the latent space of the input. Modelling a latent space is integral to several state-of-the-art deep learning algorithms in generative computer vision, language modelling, etc.

\subsection{Guide to selected machine learning methods for granular material simulations}

The subsections below introduce key ML methods relevant to computational mechanics and granular material simulations.
\secref{subsec:lstm} covers standard neural network architectures for sequence learning, which are particularly suited to the history-dependent behaviour of granular materials, and recent advances on \emph{interpretable} data-driven constitutive modelling.
\secref{subsec:latent_models} expands on the encoding-decoding philosophy (\figref{fig:generative_model})(a) to represent high-dimensional data using lower-dimensional latent spaces, focusing on the interpretability of ML.
\secref{subsec:gnn} shows how graph neural networks (GNNs) have emerged as an efficient ML surrogate, assembling the geometric features of interacting particles over the so-called ``latent graphs''.
Moving beyond conventional ML surrogates, \secref{subsec:neural_surrogates} introduces recent advances in neural operator learning and their applications to CFD and DEM simulations. 
\secref{subsec:rom} considers high-dimensional systems parametrised by time and parameters and introduces a few reduced-order modelling from classical to deep learning-based for mapping between the full-order and the reduced-order space.
With these surrogate modelling options introduced, \secref{subsec:prob_learning_uq} returns to probabilistic learning and Bayesian uncertainty quantification, again in the context of parametrised problems, and demonstrate how surrogate model uncertainties can be incorporated.
We also discuss where appropriate how these methods and models address the GM challenges summarised in \secref{sec:GM_challenges} and then further synthesized in \secref{subsec:ML_GM_group}.

\subsection{Sequence-based learning for path-dependent constitutive behaviours}
\label{subsec:lstm}

The elastic-plastic behaviour of granular media is known for its strong loading path dependence, which can be perceived as a time series problem. From a machine learning perspective, a natural choice for solving these problems lies within the family of deep learning models like recurrent neural networks (RNNs). The core idea behind RNNs is to leverage the network's ability to maintain a "memory" of previous inputs, enabling it to capture the sequential and temporal dependencies inherent in time-varying data. The previous applications of time-series ML models into elastoplascity modelling of granular media include long short-term memory (LSTM) \cite{wang2018multiscale, ma2022predictive}, gated recurrent units (GRU) \cite{wang2019meta, qu2021towards} and temporal convolutional network (TCN) \cite{wang2022data}. 

Instead of directly inputting sequence data into a time-series ML model, an alternative is to encode the loading history experienced by granular media via internal variables. 
Non-sequential ML models like multi-layer perception (MLP) can also be employed to predict path-dependent stress responses  \cite{ghaboussi1991knowledge}. 
More recent studies with MLP include the accumulated absolute strain \cite{huang2020machine} and the Frobenius norm-based internal variable \cite{wang2024multi}.
However, interpretability \cite{Bahmani_2024, sun2022data}, robustness \citep{wang2021non} and speed \citep{phan2025hydra} of these black-box neural network models could be important issues for high-consequent applications. 

To improve the reliability of the constitutive models, incorporating physics or universally acknowledged laws in underpinning reliable data-driven constitutive modelling of granular materials 
is a viable option.
The principle of frame indifference, which states that material laws should be independent of the observer's frame of reference or motion, can be enforced via various approaches. 
For isotropic materials, frame indifference can be established by formulating the material behaviours via invariants. For anisotropic materials, one may enforce material frame indifference through data augmentation \cite{lefik2003artificial}, 
additional constraints in the loss function \cite{vlassis2022molecular}
or incorporating spectral representation of stress and strain in the learning problems \cite{heider2020so} or through equivariant neural network \cite{cai2023equivariant}.

\subsection{Interpretable data-driven constitutive law discovery for granular materials}
\label{subsec:constitutive_ml}

For highly complex path-dependent systems, such as granular materials, the expectation of enough data for training to unambiguously predict outcomes is unrealistic, whether the training data be physical experiments or computer simulations. As such, theory and model construction are an underdetermined inference problem \citep{golan2022information}. 
To ensure robustness, adversarial attacks and knowledge graphs can be helpful tools. In adversarial attacks, one may train an artificial intelligence agent to design experiments that expose the weakness of a material model. This can be done in a deep reinforcement learning framework where the adversarial agent is rewarded by designing loading paths that maximize discrepancy \cite{wang2021non}. The construction of knowledge graphs, on the other hand, can enhance the robustness of the models while also improving interoperability. Instead of mapping strain history to stress directly, knowledge graph models require the determination of relationships among state variables and represent it with directed graphs \citep{wang2019meta, sun2022data, he2022thermodynamically}. A direct graph is a two-tuple $\mathbb{G} = (\mathbb{V}, \mathbb{E})$ with node set $\mathbb{V}$ and directed edge set $\mathbb{E}$. In this case, relationships among state variables, such as void ratio, coordination numbers, and other measures, can be mathematically represented by connecting nodes with directed edges. This directed edge can represent causal relations \citep{sun2022data} or the order of predictions that optimizes the prediction accuracy \cite{wang2019meta, wang2019cooperative}. While the knowledge graph does not necessarily lead to an expression simply enough for a human user to comprehend, the connectivity of the graphs does offer a set of falisiable hypotheses that can be proved or disproved by third parties. 

Another popular approach to improve credibility of machine learning model is to enhance the interpretiability \cite{murdoch2019definitions} such that the learned models achieve sufficient descriptive accuracy and relevancy that gives confidence to the accurate predictions. 
\cite{murdoch2019definitions} defines two categories of tools to improve interoperability, i.e., model-based and posthoc interpretiability. In the first case, one may introduce modularity design on the machine learning models, providing insight into the inner workings of the models. For instance, \cite{vlassis2021sobolev} uses neural networks to parametrize components of elastoplasticity models. \cite{vlassis2020geometric} introduce semi-supervised learning to relate topological information of polycrystals to elastic stored energy. \cite{Bahmani_2024} leverages the neural additive method to enforce modularity by forcing the material models to be a linear combination of features of each strain component or strain invariants. 

Post hoc analysis can also improve the interpretability of machine learning models after they are trained. For instance, \cite{Bahmani_2024} replaces the learned feature of the neural additive model with symbolic equations through symbolic regression \cite{cranmer2020discovering}. 
\cite{phan2025hydra} improve the expressivity of the neural additive model by introducing projection steps such that the feature space may reproduce polynomials of arbitrary orders and dimensions. 

Finally, training data determines a supervised ML model's performance. 
Recent advancements in this area include applying active learning or deep reinforcement learning to quantify predictive uncertainty and prioritize the selection of informative strain paths which can maximize the predictive performance of a model \cite{wang2019cooperative, villarreal2023design, qu2023deep}. In addition, transfer learning is proposed to achieve cross-paradigm knowledge sharing. By fully using low-fidelity yet cost-effective data (e.g. numerical integration data from phenomenological models), the reliance on high-fidelity and expensive data, such as physical or virtual experimental data, can be minimized \cite{qu2023data}. A similar concept of multi-fidelity can be found in the works of \cite{su2023multifidelity} and \cite{zhang2024multifidelity}, but they adopted a tailored multi-fidelity network for their respective purposes.
If experiment data is not sufficient to train a predictive model, data obtained from simulations of synthesis granular assemblies may provide the necessary information to build a closure. Generating a synthesis of granular assemblies consistent with the observed data from micro-CT or SEM images can be done by generative AI, such as generative adversarial networks (GANs) or denoising diffusion probablistic models (DDPM), as shown in \cite{nguyen2022synthesizing} and \cite{vlassis2024synthesizing}.

\subsection{Generative latent variable models for interpretable representations}
\label{subsec:latent_models}
Generative latent variable models (LVMs) are mathematical frameworks used in computational mechanics to represent high-dimensional data using lower-dimensional latent spaces. These models establish mappings between latent variables and observable data, facilitating dimensionality reduction, data generation and analysis of mechanical systems. In computational mechanics, several categories of generative models have gained prominence.

Variational autoencoders (VAEs) are one such category, defining probabilistic encoders and decoders that map between data and latent spaces. VAEs have found applications in dimensionality reduction for fluid dynamics and materials science \cite{VAE_Application}. Another significant category is GANs, which consist of generator and discriminator networks trained adversarially. GANs have been successfully applied to generate microstructures, flow fields and material property distributions \cite{GAN_Application}. Flow-based Models represent a third category, employing sequences of invertible transformations between latent and data spaces. These models have proven useful in molecular dynamics and turbulence modelling \cite{Flow_Application}.

Creating interpretable latent spaces is a key focus in computational mechanics, with several approaches being developed. Disentanglement techniques \cite{10.1007/978-3-030-85584-0_5}, such as $\beta$-VAE, aim to learn latent variables corresponding to independent variation factors in the data \cite{Disentanglement_Study}. Physics-informed Latent Spaces take a different approach, incorporating physical constraints or prior knowledge into latent space representations \cite{Physics_Informed_Latent}. Manifold Learning methods explicitly model the underlying data manifold, including geometric deep learning approaches \cite{Manifold_Learning}. 
The importance of interpretable encoding has been demonstrated using statistical learning techniques to encode complex discrete element simulation data into more manageable and insightful representations \cite{Wilke2021kds}.
Considering time as an untangled latent dimension known as temporal-preserving latent variable models, \cite{Balshaw2023} highlights the importance of utilising some manifold coordinates rather than solely relying on data space indicators like reconstruction errors to enhance interpretable latent spaces.

\subsection{Graph neural network surrogates for particulate systems}
\label{subsec:gnn}

Deep learning surrogates have emerged as powerful frameworks for replicating complex physical systems, offering a computationally efficient alternative to traditional simulations \citep{zhang2024science}.
These models directly approximate the underlying physics by learning patterns and relationships from data obtained by simulations.
Once trained, surrogate models can provide predictions at significantly lower computational costs, circumventing the high expense associated with traditional solvers.
Graph neural networks (GNNs)~\citep{Scarselli:08, Kipf:17} provide an effective framework for modelling the dynamics of granular materials and particle systems. By representing granular assemblies as graphs, with nodes corresponding to particles and edges encoding their interactions, GNNs can learn the underlying interaction laws governing particle motion in a permutation-invariant manner.

One instance of applying GNNs for granular flow modelling is the Graph Neural Network-based Simulator (GNS) \citep{Sanchez:20}. 
GNS takes the current state of the granular domain (particle positions, velocities, boundaries) as input and predicts the acceleration of each particle, which is then numerically integrated to model the time evolution of the particles. Trained on high-fidelity simulations, GNS demonstrates accurate predictions while being orders of magnitude faster than traditional numerical methods.
Applications of GNS for particulate systems involve granular column collapse \citep{choi2023graph}, granular flows with complex boundaries \citep{mayr2023boundary} and solving inverse problems \citep{kumar2023accelerating,choi2024inverse,jiang2024integrating}.

In recent years, geometric deep learning~\citep{bronstein2021geometric} on graphs has investigated many directions to incorporate knowledge on domains, signals, and transformation groups into networks designs. 
For example, equivariance has emerged as a powerful inductive bias for GNNs when applied to dynamical systems~\citep{satorras2021n, brandstetter2021geometric, sharma2025dynamical}. Such equivariant GNNs leverage geometric information of nodes while maintaining equivariance w.r.t. to rotations, translations, and reflections, the fundamental symmetries that many physics laws are based upon.
However, a major drawback of GNNs is that their computational complexity grows with the number of particles as the number of nodes in the graph grows accordingly.
Thus, computational complexity becomes infeasible for an increasing number of particles \citep{alkin2024universal,musaelian2023}.
There are many approaches for scaling GNNs to larger particulate systems, including 
evaluating subgraphs and combining the solutions \citep{bonnet2022}, limiting the receptive field of the neural network and applying domain decomposition \citep{musaelian2023,kozinsky2023} and using a hierarchy of coarser graphs \citep{qi2017,fortunato2022,lino2022multi}.
Another approach is to reduce the problem's dimensionality, as demonstrated in \citep{haeri2024subspace}, which uses PCA on the data before training a GNS.

\subsection{Neural operator surrogates for particulate systems}
\label{subsec:neural_surrogates}

\begin{figure}[t]
    \centering
    \includegraphics[width=\linewidth,trim={0cm 0cm 0cm 0.5cm},clip]{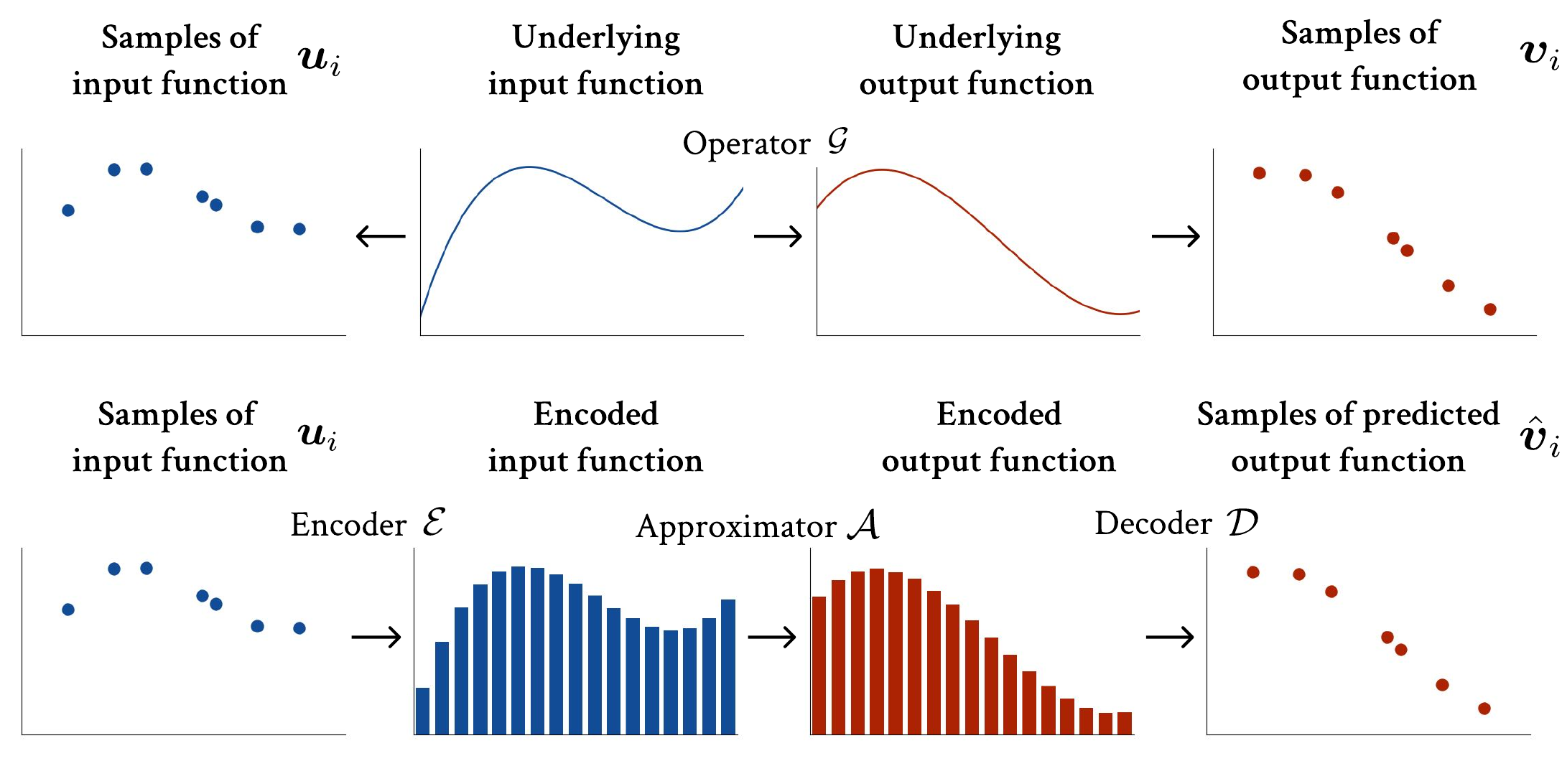}
    \caption{Illustration of neural operator learning. (Top row) The ground truth operator $\mathcal{G}$ maps an input function to an output function. However, we only have access to samples of these functions -- for example, particle velocities at their respective positions in a DEM simulation.
    (Bottom row) The neural operator $\hat{\mathcal{G}}$ approximates the ground truth operator $\mathcal{G}$ using three maps, an encoder $\mathcal{E}$, an approximator $\mathcal{A}$, and a decoder $\mathcal{D}$. 
    The approximation of $\mathcal{G}$ is ideally independent of the number of sampled input points and can approximate the output function for an arbitrary number of points. (Reprint with permission from \cite{alkin2024}.)
    }
    \label{fig:NO_sketch}
\end{figure}

Neural operator learning ~\citep{Lu2019LearningNO, Li2020NeuralOG, li2021fourier, kovachki2023neural,alkin2024universal} is a formulation that offers architecture designs for learning mappings between function spaces, making them well-suited for approximating solutions of PDEs. Neural operators have been formulated for graph neural network layers~\citep{Li2020NeuralOG}, Fourier neural operator layers~\citep{li2021fourier}, convolution neural layers~\citep{raonic2024convolutional}, or transformer layers~\citep{Li:OFormer,alkin2024universal}. Following~\cite{kovachki2023neural}, we assume $\mathcal{U}, \mathcal{V}$ to be Banach spaces of functions on compact domains $\mathcal{X} \subset \mathbb{R}^{d_x}$ or $\mathcal{Y} \subset \mathbb{R}^{d_y}$, mapping into $\mathbb{R}^{d_u}$ or $\mathbb{R}^{d_v}$, respectively. 
Operator learning is defined to approximate a ground truth operator $\mathcal{G}: \mathcal{U} \rightarrow \mathcal{V}$ via a neural network $\hat{\mathcal{G}}: \mathcal{U} \rightarrow \mathcal{V}$.
To train a neural operator $\hat{\mathcal{G}}$, a widely used approach is to construct a dataset of discrete data pairs $(\boldsymbol{u}_{i,j}, \boldsymbol{v}_{i,j'})$, 
which correspond to $\boldsymbol{u}_i$ and $\boldsymbol{v}_i$ evaluated at spatial locations $j=1,\ldots,K$ and $j'=1,\ldots,K'$, where $K$ and $K'$ can, but most not be equal. This is shown in the top part of Figure~\ref{fig:NO_sketch}.
On this dataset, $\hat{\mathcal{G}}$ is trained to map $\boldsymbol{u}_{i,j}$ to $\boldsymbol{v}_{i,j'}$ via supervised learning, as sketched in the bottom part of Figure~\ref{fig:NO_sketch}.
$\hat{\mathcal{G}}$ is composed of three maps~\citep{seidman2022,alkin2024universal}:
$\hat{\mathcal{G}} \coloneqq \mathcal{D} \circ \mathcal{A} \circ \ \mathcal{E}
$, the encoder $\mathcal{E}$, the approximator $\mathcal{A}$, 
and the decoder $\mathcal{D}$. First, the encoder $\mathcal{E}$ transforms the discrete function samples $\boldsymbol{u}_{i,j}$ to a latent representation of the input function. The approximator $\mathcal{A}$ then
maps the latent representation of the input function to a representation of the output function. Lastly, the decoder evaluates the output function at spatial locations $j'$. The neural network $\hat{\mathcal{G}}$ is then trained via gradient descent, using the gradient of, e.g., a mean squared error loss in the discretized space $\mathcal{L}_i = \frac{1}{K'} \sum_{j'}\lVert \hat{\boldsymbol{v}}_{i,j'} - \boldsymbol{v}_{i,j'}\rVert_2^2$, where $\rVert_2$ is the Euclidean norm. 

An important property of a neural operator is discretisation convergence, which ensures that the solutions produced by the neural operator remain invariant with respect to the number of input samples $K$ and the number of output function samples $K'$.
To ensure discretisation convergence, commonly used architectures for the encoder $\mathcal{E}$ include graph neural networks (GNNs) \cite{Li2020NeuralOG, li2023geometryinformed}, transformers \cite{cao2021choose,Li:OFormer,hao2023gnot,wang2024lno}, or a hybrid approach combining both, as demonstrated in \cite{alkin2024universal}.
Once the encoded input function has been sufficiently sampled for accurate representation, any further increase in the number of input samples $K$ should not affect it. 
This aligns with growing evidence that neural networks can effectively capture physical phenomena while requiring significantly less fine-grained discretisation compared to traditional numerical methods \cite{kochkov2021}.
For the decoder $\mathcal{D}$, recent work \cite{wang2024cvit} has shown that formulating the decoder as a conditional neural field allows for independent point-wise evaluation, thereby ensuring discretisation convergence.

Most state-of-the-art neural operator approaches are primarily designed for geometrically simple domains with regular grids, leaving neural operator formulations for particle-based dynamics relatively under explored.
Recently, Universal Physics Transformers (UPT) \cite{alkin2024universal} introduced a flexible and scalable neural operator designed to work with irregular grids and/or particle systems.
Operating without grid- or particle-based latent structures enables flexibility and scalability across meshes and particles.
The diverse applicability and efficacy of UPTs have been demonstrated in both mesh-based fluid simulations and Lagrangian-based fluid dynamics.
Building upon the UPT framework, \cite{alkin2024} proposes a multi-branch neural operator capable of processing multi-physics quantities. Their approach demonstrates strong performance in hopper simulations with 250k particles and fluidised bed reactors comprising 500k particles and 160k CFD cells, achieving faithful modelling over trajectories of up to 28 seconds, equivalent to 2800 machine learning time steps.

\subsection{Data-driven reduced-order modelling}
\label{subsec:rom}

Reduced-order modelling (ROM) \cite{hesthaven2016certified,quarteroni2015reduced,BGW2015pmorSurvery}, alternatively termed model order reduction, is a promising strategy for accelerating numerical simulations of parametrised problems in computational mechanics and hence to enable multi-query, real-time computations. Generally speaking, ROM techniques reduce problem complexity by intelligently representing high-dimensional dynamical systems in low-dimensional latent spaces with controlled accuracy, so that the reduced dimensionality substantially improves computational efficiency. A ROM scheme typically features an offline/online decomposition, in which a reduced model is constructed offline from a collection of full-order solution data covering a time-parameter domain of interest. Such a low-dimensional system is evaluated online for new time instances and parameter configurations. The parameters of ROM are not limited to the physical features of the system, e.g. material properties or boundary/initial conditions \cite{quarteroni2015reduced,Lee2020autoencoder,hesthaven2018non,guo2021learning}. Still, the geometric parameters that describe the configuration of the computational domain \cite{quarteroni2015reduced,KABALAN2025115929,YE2024112639,MATRAY2024117243}.  

Consider a high-dimensional system $\dot{\mathbf{u}} = f(\mathbf{u};\boldsymbol{\mu})$ for the full-order state vector $\mathbf{u}\in \mathbb{R}^N$ characterised by $d$ parameters $\boldsymbol{\mu}\in \mathcal{D}\subset \mathbb{R}^d$ varying in a parameter space $\mathcal{D}$, the offline construction of an $r$-dimensional reduced-order model ($r\ll N$) mainly entails two steps. The first step is to learn a time/parameter-independent encoding-decoding structure \cite{champion2019sindy,yu2019non} for dimensionality reduction (or manifold learning) \cite{bishop2007,murphy2012machine}. In particular, such a structure is written as follows
\begin{equation}
    \mathbf{u}\approx \mathcal{F}_d ~\circ~ \mathcal{F}_e (\mathbf{u})\,, 
\end{equation}
in which $\mathcal{F}_e:\mathbb{R}^N \to \mathbb{R}^r$ is an encoder that maps to the reduced space and $\mathcal{F}_d:\mathbb{R}^r \to \mathbb{R}^N$ is a decoder recovering back to the full space. A classical treatment for this is the proper orthogonal decomposition \cite{Berkooz1993,liang2002proper,quarteroni2015reduced} that takes advantage of linear principal component analysis \cite{murphy2012machine}. More recently, deep autoencoders are becoming popular for the discovery of low-dimensional latent spaces \cite{champion2019sindy,Lee2020autoencoder,fresca2022pod,botteghi2022deep} and they are powerful tools for compressing high-dimensional solution data that represent complex mechanical behaviours in granular modelling. 

\begin{figure}
    \centering
    \includegraphics[width=0.5\linewidth]{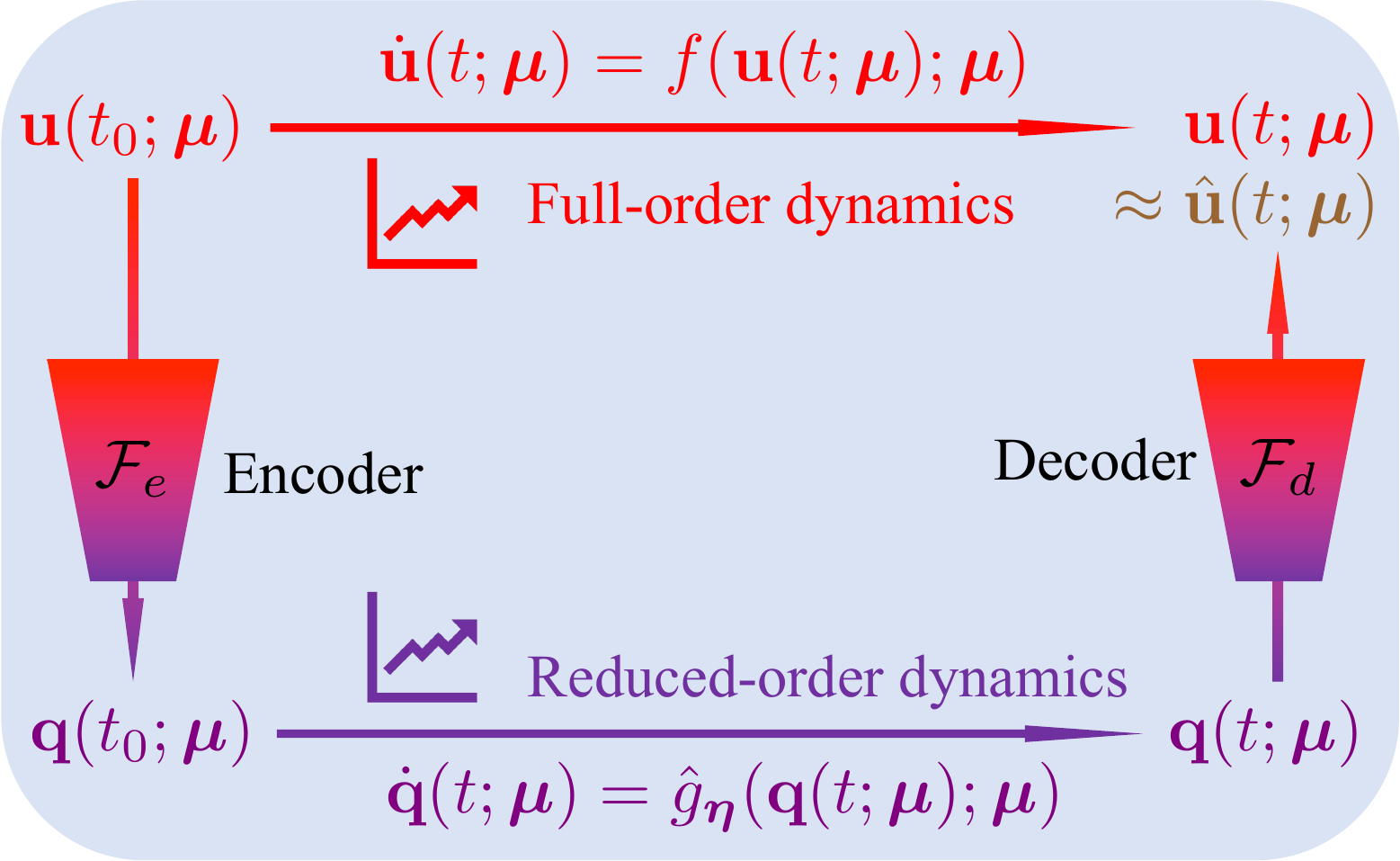}
    \caption{A conceptual diagram for data-driven reduced-order modelling.}
    \label{fig:rom}
\end{figure}

The second step aims to represent reduced-order dynamics in the low-dimensional latent space. Towards this end, a `projection' of the full-order system onto the reduced space can be written as
\begin{equation}
    \dot{\mathbf{q}} = [\nabla \mathcal{F}_d(\mathbf{q})]^\dagger f(\mathcal{F}_d(\mathbf{q});\boldsymbol{\mu})\,.
\end{equation}
However, the assembly of such a reduced system is inconvenient in granular simulations, primarily because its formulation requires intrusive access to the full-order code that often involves complex, nonlinear constitutive laws. To overcome this difficulty, a straightforward strategy is to construct generic surrogate models directly for the reduced-state solution $\mathbf{q}(t;\boldsymbol{\mu})$ via, e.g. neural networks \cite{hesthaven2018non,chen2021physics,fresca2022pod}, polynomial response surfaces \cite{swischuk2019MOR_ML}, radial basis function interpolation \cite{xiao2015non}, or Gaussian processes \cite{guo2018reduced,guo2019data,kast2020non,cicci2023uncertainty}. An example of this reduced-state solution is illustrated in \figref{fig:rom}. Such a `black-box' treatment can conveniently handle the interpolation of parameter dependency within the time domain covered by training data. Still, there is no guarantee of generalisation performance in predicting for future states. Alternatively, one can approximate the right-hand side of the reduced system with certain model parametrisation $\hat{g}_{\boldsymbol{\eta}}$, i.e.,
\begin{equation}
    \dot{\mathbf{q}} = \hat{g}_{\boldsymbol{\eta}}(\mathbf{q};\boldsymbol{\mu})\,,
\end{equation}
and pose the data-driven learning of reduced-order equations as the parameter estimation for $\boldsymbol{\eta}$ --- an inverse problem \cite{GW2021learning}. When prior knowledge about the granular dynamics and/or their numerical implementation is incorporated into the model parametrisation $\hat{g}_{\boldsymbol{\eta}}$, improved temporal predictiveness can be seen when the learned low-dimensional system is evaluated beyond training coverage. The parameter estimation in reduced-order dynamics can be accomplished with various strategies, including data-driven operator inference \cite{PW2016operatorInference,QKPW2020liftAndLearn,guo2022bayesian}, sparse identification of latent dynamics \cite{brunton2016sindy,champion2019sindy,schaeffer2017learning}, neural ordinary differential equations \cite{chen2018neural}, surrogate modelling of Runge-Kutta schemes \cite{zhuang2021model} and recurrent learning for reduced-state time series \cite{conti2024multi,botteghi2024recurrent}.

\subsection{Probabilistic learning for uncertainty quantification}\label{subsec:prob_learning_uq}

Almost all natural and industrial processes show a large variability.
This variability introduces uncertainties in real-world modelling and observations, hindering our understanding of these processes.
In computational science and engineering, we frequently explore how these uncertainties propagate from observational data and characterising input features typically defined within mathematical/numerical models to the predictions produced by these models.
This is particularly challenging for granular materials due to their evolving microstructure, representative scales and physical regimes (see \secref{sec:GM_challenges}).
Such complexities restrict observations to a single scale or regime at any given time.

Adding uncertainties to the state $\mathbf{u}$ and parameters $\boldsymbol{\mu}$ of a suitable mathematical/numerical model $ \mathbb{F} $ describing the physical process results in the following state-parameter space model,
\begin{equation}
	\begin{aligned}
		\mathbf{u}_t &= \mathbb{F}(\mathbf{u}_{t-1}, \boldsymbol{\mu}_t) + \boldsymbol{\nu}_t, \\
		\mathbf{v}_t &= \mathbb{H}(\mathbf{u}_t) + \boldsymbol{\omega}_t.
	\end{aligned}
	\label{eq:dynamic_system}
\end{equation}
where $ \boldsymbol{\mu}_t $, $ \mathbf{u}_t $ and $ \mathbf{v}_t $ are random variables describing uncertainties of the model parameters, states and observations, respectively.
$ \mathbb{H} $ represents a mapping between hidden state and observable vectors,
and $ \boldsymbol{\nu}_t $ and $ \boldsymbol{\omega}_t $ are the modelling and measurement uncertainties, respectively,  often defined to be zero-mean normal distributions. 
\eqref{eq:dynamic_system} describes the time evolution of the state of the model $\mathbf{u}_t$ and the state of the observables $\mathbf{v}_t$. The state/parameter estimations for $\mathbf{u}_t$ and $\boldsymbol{\mu}_t$ are often performed using Bayesian inference.

Bayes' rule provides a general framework for estimating the hidden state of a dynamical system from partial observations using a predictive model of the system dynamics $ \mathbb{F} $.
Bayesian statistics offers a methodology for quantifying the impact of various uncertainty sources and enables a probabilistic integration of prior knowledge about physics and mechanics into the inference process.
The state, often augmented by the system's parameters as shown in \eqiref{eq:dynamic_system}, changes in time according to an iterative process of predicting and updating.
Within the context of Bayesian uncertainty quantification \cite{Sarkka2013}, the probability distribution of the system's state is updated whenever new observations become available, using the recursive Bayes' theorem.
In material uncertainties, the system dynamics are predicted by a material model, and the observations are experimental data. The goal is to estimate the probability distribution of the model's parameters from the experimental data using simulations of material responses.

\begin{figure}
    \centering
    \includegraphics[width=0.75\linewidth]{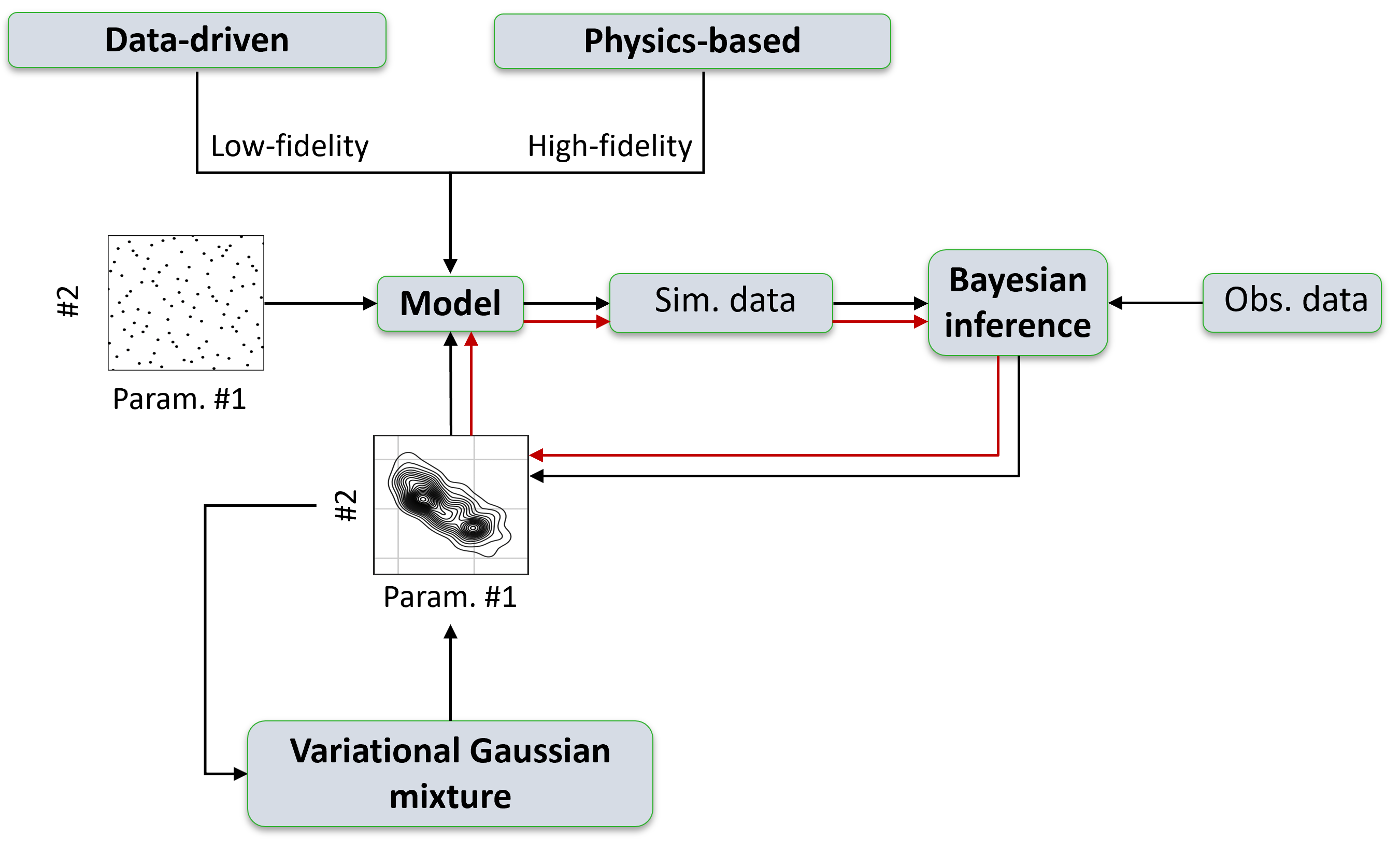}
    \caption{Integration of data-driven and physics-based models for multi-level sampling-based Bayesian uncertainty quantification \cite{Cheng2024}.}
    \label{fig:gl_flowchart}
\end{figure}

Probabilistic and Bayesian learning are powerful frameworks that provide a structured way to integrate model prediction and real-world observations or estimate the model uncertainties based on limited observation data. They are critical to risk assessment, uncertainty quantification, and optimisation of granular material processes and design.
However, because granular systems show strong nonlinearity and discontinuity, the estimation of the uncertainties are usually achieved via extensive sampling of the model solutions, e.g. via Markov-Chain Monte Carlo or other variations of the Monte Carlo sampling schemes \cite{Andrieu2010, Giles_2015}.
The purpose of using machine learning for these problems is two-fold: (1) to reduce the computational cost of the model evaluations via surrogate modelling (e.g. Gaussian process regression and responses surfaces) and (2) to learn the underlying conditional probability distribution via cluster algorithms \cite{bishop2007} and making use of it to improve the efficiency of the sampling scheme.
These data-driven surrogates can be integrated within the Bayesian framework in conjunction with physics-based models, as schematized in \figref{fig:gl_flowchart} and implemented in the open-source Bayesian uncertainty quantification software package \href{https://grainlearning.readthedocs.io/en/latest/rnn.html}{GrainLearning}.

In particular, Gaussian processes (GP) emulation is a powerful tool of probabilistic supervised learning for surrogate modelling with uncertainty quantification (UQ) \cite{murphy2012machine,williams2006gaussian}. It is a generally applicable regression method in approximating maps between characterising input features and output quantities of interest (QoI) \cite{fuhg2022local,guan2023machine,guo2018reduced,guo2021learning}. GP regression is a non-parametric Bayesian approach that considers a GP prior over regression functions and then estimates the posterior predictions by conditioning on the observational data. The posterior mean function can be viewed as an approximation in a reproducing kernel Hilbert space \cite{williams2006gaussian}, which enables effective treatment of nonlinearities. Additionally, the predictive distribution facilitates the function approximation with quantifiable uncertainties, hence certifying the data-driven model's credibility. Knowledge about governing physics can be incorporated into GP surrogate models by leveraging the property that Gaussianity can be preserved through linear operations. This allows to integrate differential equation constraints into the regression problem and has been applied to forward and inverse problems governed by linear equations \cite{raissi2017machine,pfortner2022physics} and extended to nonlinear problems \cite{chen2021solving,ye2024gaussian}. 

\subsection{Identifying and classifying machine learning solutions for granular material challenges}
\label{subsec:ML_GM_group}

Given the strong prospect for ML in modelling granular materials, \secref{subsec:lstm} - \secref{subsec:prob_learning_uq}, below we attempt to present possible ML-based solutions for the earlier stated challenges, \secref{sec:GM_challenges}, as follows.

\begin{enumerate}
	\item \textbf{Prohibitive computational cost for large-scale problems with DEM:} 
	A typical DEM computation cycle involves: (1) contact detection, (2) contact force determination, and (3) particle position updates. 
	These computations are expensive due to particle interactions' highly nonlinear and discontinuous nature. 
	Machine learning offers a potential solution by creating surrogate models that learn a lower-dimensional representation of particle dynamics. 
	Neighbor-aware methods, such as Convolutional neural networks (CNNs) and graph neural networks (GNNs), can learn local interactions between particles and boundaries \cite{xu2022improved}. 
	However, integrating these approaches with DEM results in hybrid models that trade off accuracy for speed. 
	For example, CNNs and GNNs struggle with capturing dynamic contact changes, which requires careful validation against full DEM models to quantify accuracy.
	Another promising approach are GNN-based neural surrogates \cite{Sanchez:20,choi2023graph} and neural operators \cite{alkin2024universal,alkin2024}, which learn end-to-end mappings from DEM simulation data, bypassing the need for hand-crafted features, but may require substantial datasets to achieve generalisation.
	\item \textbf{Identification of coarse-grained particles and structures for meso-scale representation:} 
	Coarse-graining reduces computational complexity by representing clusters of particles or microstructures at a higher level, akin to molecular dynamics. 
	Techniques like graph coarsening with GNNs can preserve essential mechanical properties by intelligently aggregating nodes while maintaining the connectivity patterns \cite{Cai2021, Husic2020}. 
	However, achieving a balance between simplification and mechanical fidelity is challenging. 
	Validation frameworks must be in place to assess whether the mechanical properties, such as effective stiffness or yielding, are maintained across the coarsened scale representations.
	
	\item \textbf{Repetitive, still expensive micro-scale simulations within a multi-scale modelling framework:} 
	In multi-scale modelling, the repetitive nature of micro-scale simulations can be a significant bottleneck. 
	To address this, recurrent neural networks and simple multi-layer perceptrons have been utilised as surrogate models to represent the macroscopic response derived from DEM or FEM simulations \cite{2024IJNAM..48.1372G, guan2023machine, Maia2023}. 
	However, these models often ignore grain-scale variations, limiting their utility in dynamic environments with complex and nonlinear interactions.
        One promising direction is the integration of GNN surrogates within a multi-scale modelling framework, which allows capturing the intricate micromechanics leading to plasticity while overcoming GNN's scalability issue \cite{Storm2024}.
    Additionally, clustering and classification algorithms \cite{Chaouch2024} can also be employed to identify regions of the materials that undergo similar load conditions, significantly reducing the number of RVE problems to be solved.
	\item \textbf{Separation and evolution of length and time scales:} 
	Granular materials exhibit complex behaviour with evolving length and time scales as their microstructure changes in response to external loads. 
	Representation learning and other pattern recognition methods can help extract relevant scales from data \cite{Linden2016, Tian2018, Tordesillas2022}. 
	However, capturing these scales' dynamic evolution requires real-time models, such as transfer learning techniques, which adjust to new data without forgetting previous knowledge. 
	
	\item \textbf{Lack of a unified continuum theory for granular materials in quasi-static, rapid, and transient motions:} 
	At the local level, ML methods can help discover constitutive laws for granular materials \cite{Bahmani_2024, 10.1115/1.4052684, vlassis2021sobolev}, but these models are limited to a solid mechanics framework. 
	The lack of established theoretical guidance for ML-based optimisation is a significant gap for fluid-like regimes. 
	Hybrid approaches, integrating ML with computational fluid dynamics (CFD) solvers, could provide a path forward by enabling a consistent continuum representation across flow regimes. 
	Data-driven neural operators such as UPT \cite{alkin2024universal} and geometry-informed neural operators (GINO) \cite{li23gino} are promising alternatives that bypass fixed theoretical assumptions by directly learning interactions from data.
	
	\item \textbf{Coupling to multi-physics and external rigid/deformable structures:} 
	In many industrial applications, granular materials interact with other physical systems, making the data highly heterogeneous. 
	Multi-modal deep learning methods can integrate diverse data streams, such as fluid pressures and solid stresses, into cohesive models. 
	A major technical challenge is ensuring consistency across these data types with different time and spatial scales. 
	Attention mechanisms in neural networks could be particularly helpful in focusing on the most relevant interactions in specific physical domains. 
	Additionally, the increased computational requirements for such models demand distributed computing approaches for real-time or large-scale applications.
	
	\item \textbf{Model calibration, validation, uncertainty quantification, and optimisation:} 
	Incorporating model or parameter uncertainties is critical in ML-driven granular material simulations. 
	To achieve computational feasibility, advanced techniques like sparse Gaussian Processes or variational inference should be considered to handle the high dimensionality of granular systems \cite{cicci2023uncertainty}. 
	Moreover, given the noisy nature of real-world data, pre-processing steps are essential to filter outliers and reduce overfitting. 
	Active learning can be used to select the most informative data points for model retraining, and robust acquisition functions can further enhance model reliability by prioritising scenarios that contribute to reducing overall model uncertainty \cite{CHENG2019268}.
\end{enumerate}

\subsection{Proposed machine learning workflow for granular material simulations (GranML)}\label{subsec:ml_workflow}

\begin{figure}
\includegraphics[width=\textwidth]{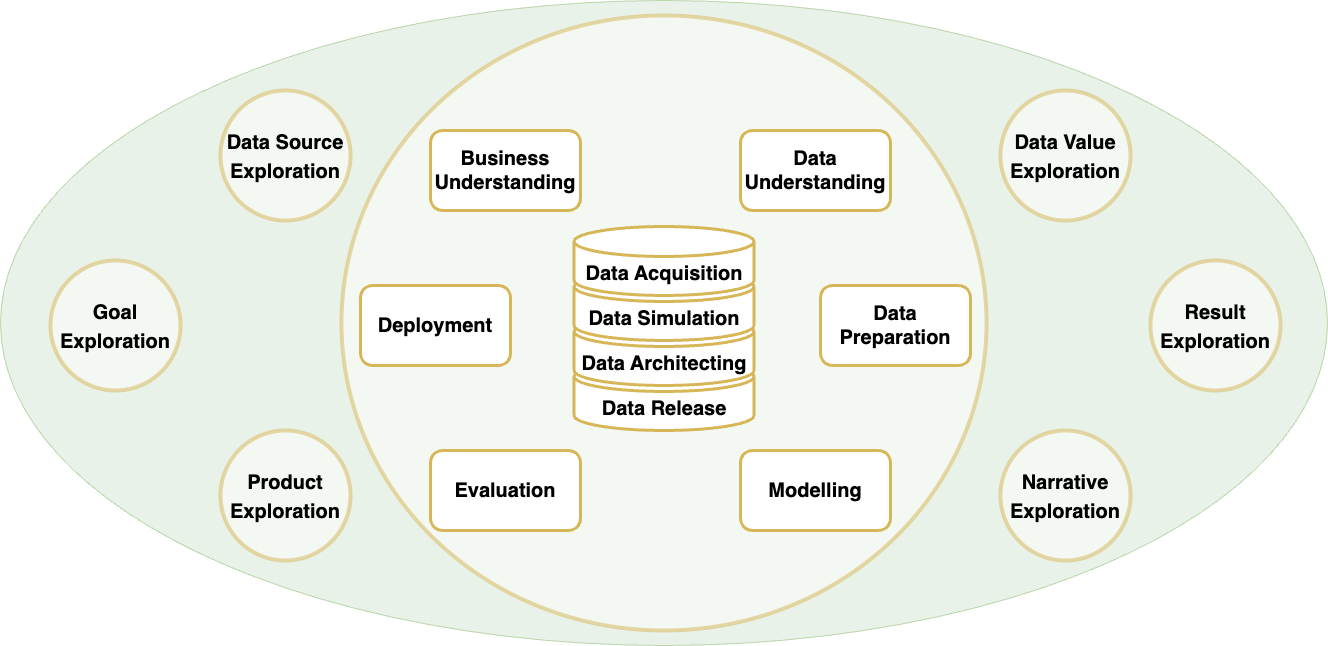}
\caption{Illustrates a variety of components of a data science trajectory (DST) \cite{martinez2019crisp} map, essential to any data science project.}
\label{fig:dst_map}
\end{figure}


The previous subsection shows that a matchmaking between ML methods and GM challenges should lead to an extraction and utilisation of knowledge (either universal or domain/application-specific), driven by the underlying research questions.
The two main reasons motivating this extensive academic endeavour are (i) a dramatic increase in the availability and the amounts of data and (ii) a boom in the variety of machine learning techniques. Given ML's potential, we also see new methodologies being proposed to extract this knowledge systematically \cite{martinez2019crisp}. \figref{fig:dst_map} illustrates a data science trajectory (DST) map, which is an extension of CRISP-DM \cite{chapman2000crisp} -- a \textit{de facto} standard for data mining and knowledge discovery projects.
Based on the definitions of each component of the data science trajectory (DST) (Fig. 4 in \cite{martinez2019crisp}), \secref{sec:examples} showcases concept machine learning workflows for a few example applications. We use the word \textit{concept} because of the multiple possible arrangements of the DST components. 

Our concept workflow mainly encompasses three stages, namely \textbf{data simulation, understanding and preparation}, \textbf{modelling} and \textbf{evaluation}. Each stage is critical to the overall model development and deployment process. For a complete digital twin, the process should operate through two parallel tracks: one focused on modelling and the other on sensing.
To assist readers, particularly those with an engineering background, we briefly describe the key components of this workflow as follows.


\begin{itemize}
\item 
The simulation workflow begins with \textbf{data preparation and understanding} using physics-based granular material models. This step typically includes a selection of appropriate spatial and temporal resolutions, data preprocessing to obtain quantities of interest, and sensitivity analyses across the sampling domain (e.g., time, parameters, boundary conditions) to generate datasets.
\item 
The \textbf{modelling} stage involves selecting and training suitable ML models (as discussed in \secref{sec:MLSolution}), including hyper-parameter tuning and retraining as new data becomes available.
Techniques such as transfer learning and active learning can improve efficiency. Typically, datasets are split into training, testing, and validation subsets to ensure good generalisation to unseen data.
\item 
In the \textbf{evaluation} stage, trained models are scored and assessed, leading to the \textbf{deployment} phase. Deployment involves integrating ML models with real-world sensing data to perform digital twinning tasks such as uncertainty quantification, optimisation, and data assimilation. The significant speed advantage of ML surrogates over physics-based simulations makes gradient- and sampling-based techniques practically viable within this workflow.
\end{itemize}

\section{Demonstrations of the GranML workflow for granular material simulations}
\label{sec:examples}

The main challenges in granular material simulations, summarised in \secref{sec:GM_challenges}, are conventionally tackled through advanced numerical methods and theories. By integrating numerical simulations with ML techniques, “hybrid” approaches are developed (see \secref{subsec:ML_GM_group}). These are exemplified by the following cases aimed to address at least Challenges 1, 2, 5, and 7 of \secref{sec:GM_challenges}: the quasi-static compression and shearing of sand in the solid-like regime (\secref{sec:exampleSolid}) and the transient behaviour of granular column collapse in the fluid-like regime (\secref{sec:exampleFluid}). The section concludes with a set of recommendations to facilitate the use of ML in granular material simulations.

\subsection{Modelling quasi-static, solid-like behaviour of granular materials}\label{sec:exampleSolid}

\subsubsection{Compression and shearing of densely packed grains}

As discussed in \secref{sec:quasi}, granular materials at rest or moving slowly can often be considered in a quasi-static regime. Common engineering examples include powders, soils, and rocks compaction, provided the external loads are not rapidly applied. In such cases, the representative volume element (RVE) scale is of interest, where individual grains can be simulated as rigid bodies interacting via contact forces, Coulomb friction, cohesion, and other mechanisms.

\begin{figure}
    \centering
    \includegraphics[width=0.4\linewidth]{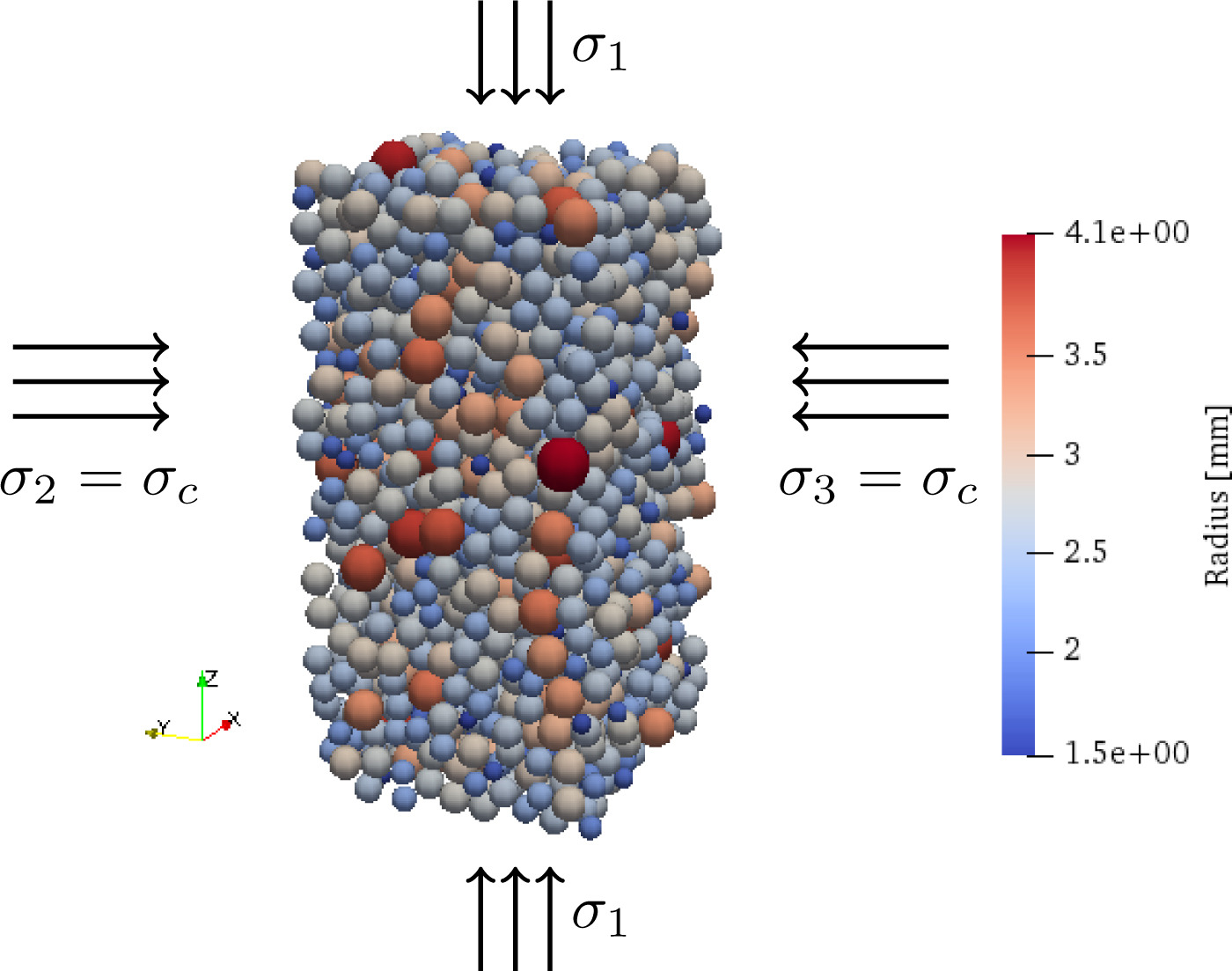}
    \caption{Illustration of a typical DEM simulation of triaxial compression. Reprinted with permission from \cite{HARTMANN2022104491}.}
    \label{fig:triax_DEM}
\end{figure}

A typical scenario is triaxial compression (see \figref{fig:triax_DEM}). Here, a granular assembly comprising several thousand particles (often with varying shapes and sizes) is compressed to a confining pressure of a few kPa, mirroring subsurface conditions. Despite many studies still employing spherical particles for simplicity, advanced algorithms also accommodate non-spherical shapes \cite{Feng2023} and more sophisticated contact laws \cite{Weinhart2020,Duriez2021a,Zhao2021a,Wang2022b}.

These quasi-static simulations record numerous snapshots of a system’s microstructural evolution (e.g., particle rearrangements, creation and loss of contacts). The data allow researchers to derive macroscopic properties—such as bulk anisotropy, dilatancy, and excess pore-pressure dynamics homogenized over the RVE. Capturing these variables as time-series data naturally lends itself to training relatively simple recurrent neural networks (RNNs), as exemplified by early work from Sun et al. \cite{wang2019meta}.

\subsubsection{LSTM-based surrogate modelling of stress--strain response}
\label{sec:example_LSTM}

\begin{figure}[b!]
    \centering
    \includegraphics[width=0.75\textwidth]{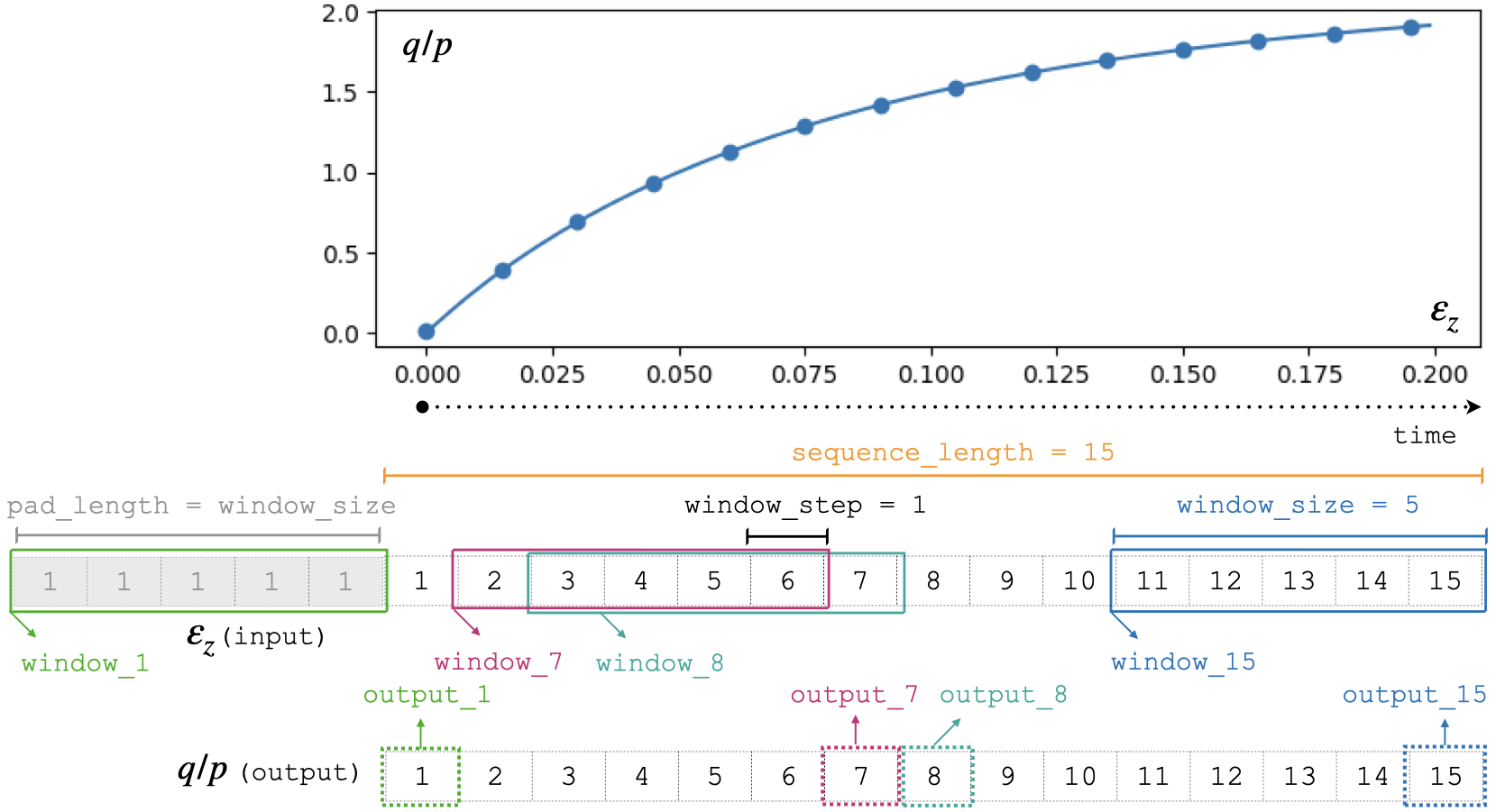}
    \caption{LSTM-DEM workflow to predict stress--strain response.}
    \label{fig:lstm_DEM}
\end{figure}

This section presents a concrete example of the machine learning workflow using long short-term memory (LSTM) networks to predict the stress-strain behaviour of granular materials based on simulation data generated from DEM \cite{Cheng2021a}. The schematic in \figref{fig:lstm_DEM} illustrates how the generic framework from \secref{subsec:ml_workflow} can be applied in a solid-like regime. A more detailed tutorial can be found in the \href{https://grainlearning.readthedocs.io/en/latest/rnn.html}{GrainLearning} documentation.

\textbf{Data simulation, understanding and preparation}: The Discrete Element Method (DEM) is used to simulate the compression of a RVE filled with grains to a certain porosity.
DEM simulations of triaxial compressions are performed using the open-source DEM package YADE to generate training data and identify Quantities of Interest (QoI) sequences of stress and strain measures. These simulations vary in contact parameters and are saved at regular time intervals. The primary QoIs include mesoscopic material behaviour homogenized over the RVE, including sequences of stress, strain, and anisotropy measures,  under various conditions (e.g., drained and undrained). A training dataset is then created, covering the evolution of these QoIs over the chosen loading paths for a wide range of contact parameters \cite{CHENG2019268} and microstructures \cite{Lubbe2022}.

\textbf{Modelling} (LSTM model training and hyperparameter tuning): The training phase involves transforming the data from HDF5 format into TensorFlow datasets suitable for LSTM models. Key transformations include merging arrays from different HDF5 groups, standardising the data and splitting it into training, validation, and test sets. An LSTM model does not require an encoder and decoder. However, in this particular example, a sliding window technique \cite{Cheng2024} is used to divide the sequences into smaller lengths (see \figref{fig:lstm_DEM}). 
Hyperparameter optimisation (including the sliding window size) uses the Weights and Biases (wandb) platform. This platform allows for comprehensive tracking of training metrics, model configurations and system performance across different runs. Users create configuration files to define the parameters and methods for optimisation, enabling efficient exploration of the hyperparameter space.
These configurations ensure the model is well-tuned, enhancing its predictive accuracy and generalizability. The optimised model and training metrics are saved for future use, facilitating reproducibility and further analysis.

During \textbf{evaluation}, the trained LSTM model accepts new sets of contact parameters and predicts the resulting stress–strain response over time. The predictions are evaluated against additional simulations to assess the model's accuracy and generalisability, such as DEM simulations performed with more particles. When \textbf{deployed}, LSTM-based surrogate models can efficiently yield physics-informed predictions for granular materials under quasi-static loading.
One significant advantage of LSTM-based surrogates is their integration with their physics-based counterparts in an uncertainty quantification or optimisation framework because the ML-based surrogates are much cheaper to run, and the error estimates of these surrogates can be computed for the Bayesian or probabilistic variants of these ML models.

\subsection{Modelling dynamic, fluid-like granular flows}
\label{sec:exampleFluid}

\subsubsection{Collapse of dense granular columns}

\begin{figure}
    \centering
    \includegraphics[width=0.8\linewidth]{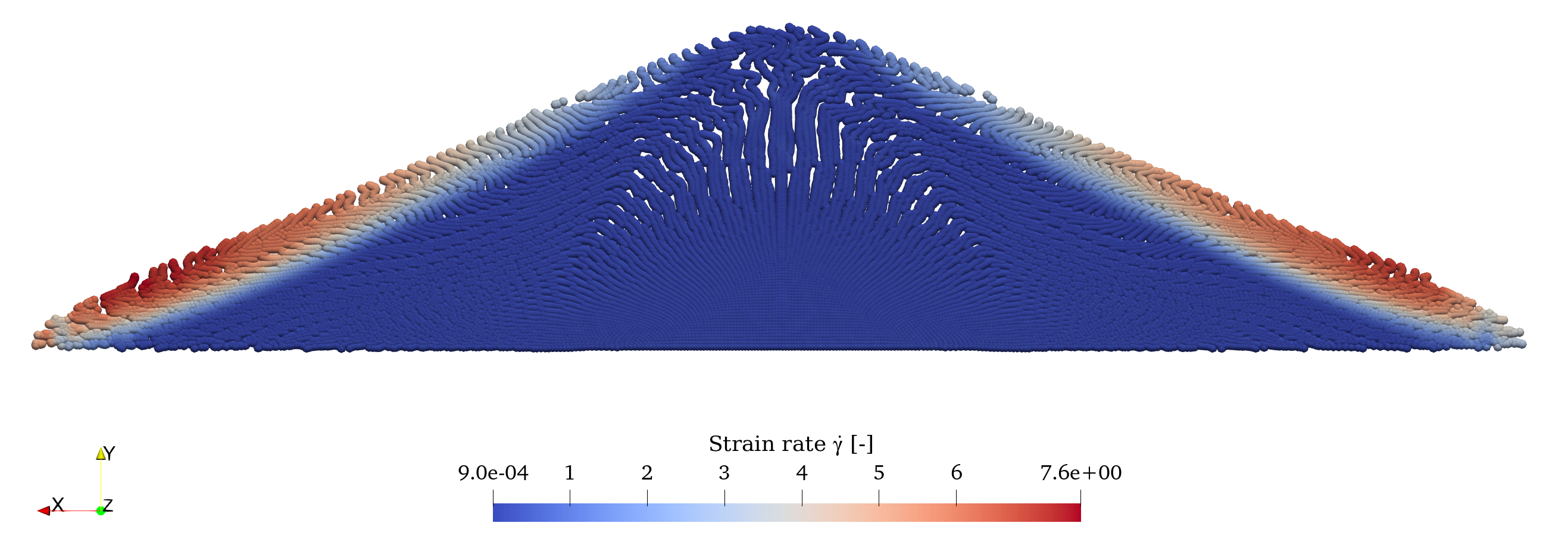}
    \caption{Illustration of a typical MPM simulation of granular column collapse using the $\mu(I)$ rheology model \cite{Lubbe2024}.}
    \label{fig:collapse_MPM}
\end{figure}

Fluid-like granular regimes occur when materials flow rapidly---often down slopes or when columns collapse. Chute flows, for example, arise naturally in landslides and industrial transport processes and can vary from dense ``liquid-like'' to dilute ``gas-like'' states \cite{gdr2004dense,forterre2008flows} (see \figref{fig:collapse_MPM}). Key factors such as channel inclination, particle shape, surface friction, and ambient fluid effects (e.g., air or water) all influence flow dynamics.

Contrary to the chute flows, the granular column collapse problem involves the sudden collapse of a vertical column of granular material, like sand, onto a flat surface, transitioning from a static to a dynamic state, as shown in \figref{fig:collapse_MPM}. The scientific community studies this configuration \cite{cabrera2019granular, sarlin2021collapse, staron2023cohesive, polania2024polydispersity} to understand how granular materials transition to a rapid deformation and then to a deposit. The collapse sequence reveals how granular materials spread and settle, which is crucial for understanding the extension of natural hazards, like landslides, avalanches, and debris flows, and designing safe and efficient storage and transport systems. 

Modelling the granular column collapse \cite{webb2024continuum,gans2023collapse} is challenging due to the complex and nonlinear behaviour of granular materials and their interaction with the surrounding media. Accurate simulations must capture intricate particle-particle and particle-fluid interactions, leading to non-trivial energy dissipation. Additionally, the influence of the surface on which the material collapses adds further complexity.

\subsubsection{GNS-based surrogate modelling  of granular column collapses}

This section presents a concrete example of the machine learning workflow using the Graph Neural Network-based Simulator (GNS) for predicting granular column collapse trajectories and their runout distances. This example demonstrates how the generic framework outlined in \secref{subsec:ml_workflow} can be applied to fluid-like regimes in geotechnical contexts.
Different from \secref{sec:example_LSTM}, the numerical method used to generate training data is the Material Point Method (MPM) \cite{kumar2023accelerating} and a GNS surrogate is trained on MPM data \cite{Joseph2022} to learn and predict granular flow dynamics.
Interested readers are referred to the examples in the \href{https://github.com/geoelements/gns}{GNS repository}.

\begin{figure}[b!]
    \centering
    \includegraphics[width=0.8\textwidth]{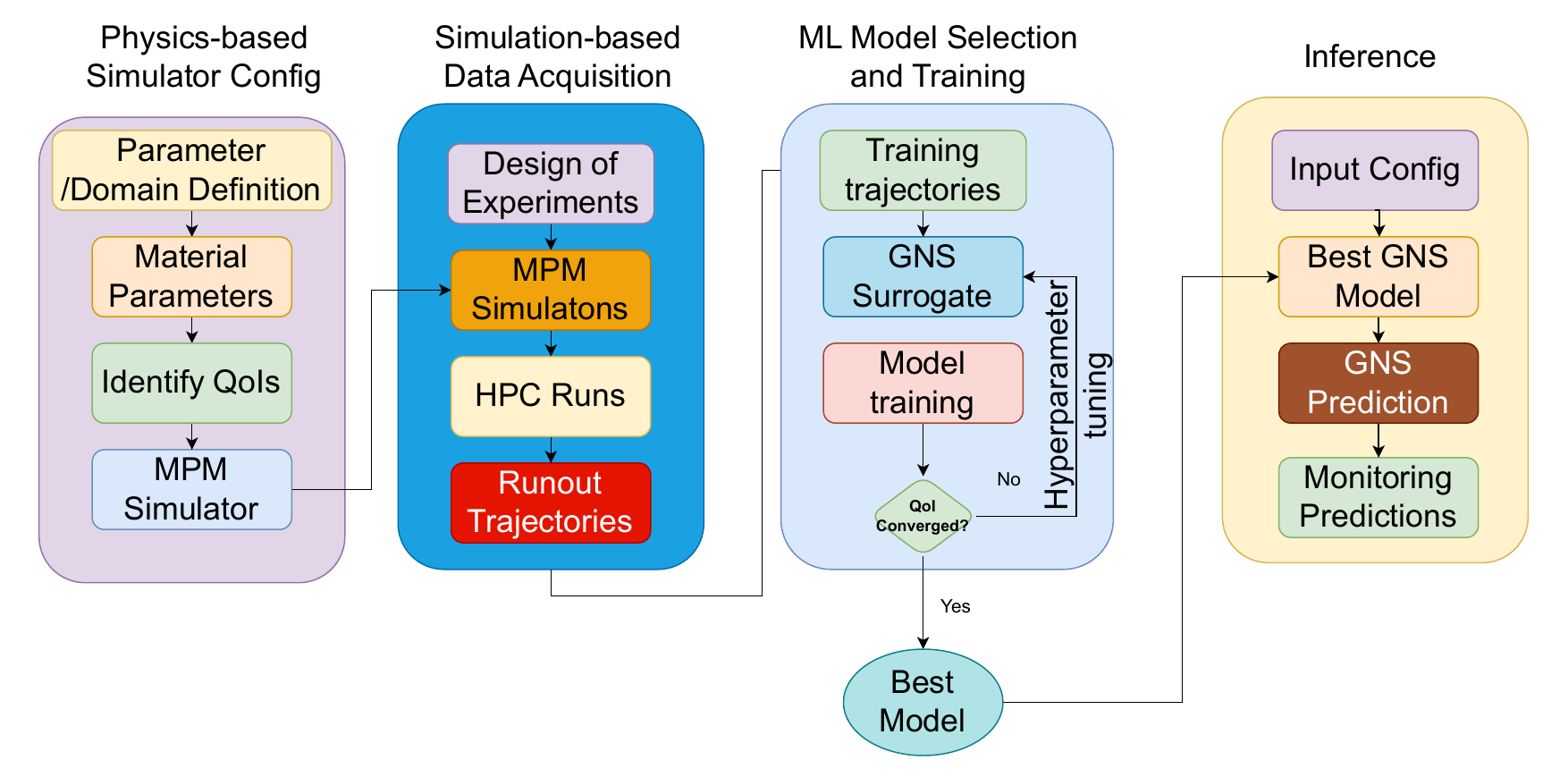}
    \caption{GNS-MPM workflow to predict granular column collapse evolution.}
    \label{fig:gns-mpm}
\end{figure}

\textbf{Data simulation, understanding and preparation}: This phase utilizes MPM simulations to generate training data and identify Quantities of Interest (QoI). The input parameters include granular column geometry (e.g. aspect ratios from 0.5 to 4.0) and material properties (e.g. friction angles from 15$^\circ$ to 45$^\circ$). A parametric sweep across these variables produces diverse granular flow trajectories. The primary Quantities of Interest (QoI) is the runout distance, with material point positions tracked over time. The simulations capture complex behaviours such as column collapse dynamics and barrier interactions. This phase establishes the ground truth data for training and validation, ensuring a comprehensive representation of granular flow physics across various configurations.

\textbf{Modelling} (GNS ML training): The GNS surrogate is trained on the MPM-generated trajectories. The GNS architecture incorporates strong inductive biases through graph representation and message passing operations. The model consists of an encoder, a processor with $M=10$ message passing steps and a decoder. Domain knowledge is injected through an updater function that applies Euler integration for position updates, forcing the model to focus on learning unknown dynamics. Key hyperparameters include the connectivity radius $R$ (e.g. 0.030 m for 2D, 0.025 m for 3D) and the number of message-passing steps. Training data is augmented with Gaussian noise to improve long-term prediction stability. The model is trained using an Adam optimiser with learning rate decay on a multi-GPU setup. Training progress is monitored using mean squared error loss on both training and validation sets, with the process continuing until convergence or a maximum of 5 million steps.

During \textbf{evaluation}, the trained GNS model, known for its adaptability, takes as input the initial geometry of the granular mass, material properties and boundary conditions. This input is encoded into a graph representation with $N$ vertices representing material points. The GNS then performs $M=10$ message-passing steps to predict particle dynamics. The updater function computes the next state based on these predictions. This process is repeated for $K$ timesteps to simulate the entire granular flow evolution. Post-processing extracts relevant information such as runout distance, flow height and energy evolution. The GNS predictions are evaluated against MPM simulations or experimental data, assessing performance on various problem configurations, including different aspect ratios and upscaled domains. Metrics such as normalized runout distance and energy conservation are used to quantify the model's accuracy and generalizability. Once \textbf{deployed}, the GNS model's ability to generalize to unseen configurations and scales, leveraging the physics-informed architecture to make accurate predictions efficiently.

\subsection{Open infrastructure and best practices for machine learning in granular material simulations}
\label{sec:recommendationsOpenInfra}

Data requirements vary widely depending on the scientific challenges and applications presented in \secref{sec:examples}. Regardless of the specific context, well-structured workflows, rigorous software engineering, and consistent data management principles enhance the reproducibility and interoperability of ML approaches for modelling granular materials. In particular, applying FAIR (Findable, Accessible, Interoperable, Reusable) data and software practices and leveraging existing frameworks like the Open Science Framework (\href{https://osf.io/}{OSF}) helps ensure that ML tools remain transparent, reliable, and reusable across diverse research groups.
 can help organise and share data effectively.
It is essential to establish a structured process for defining and transferring data from physics-based simulations to surrogate modelling and referencing benchmarks to ensure reproducibility and clarity.

\paragraph{Data management for granular material simulations}
Granular material simulations typically generate large data sets ranging from particle positions and velocities to boundary conditions and material parameters. Effectively handling these datasets requires robust infrastructure (e.g., high-performance or parallel file systems), efficient and portable file formats (such as HDF5 or NetCDF), and comprehensive metadata schemas. Metadata should thoroughly record simulation parameters, temporal information, and the relationships between different data files or snapshots (\appref{app:data_management}).

\paragraph{Best software practices}

High-quality software is crucial for reproducible research in physics-informed data-driven modelling. By employing version control (e.g., Git), adopting a transparent release process, and publishing well-documented code in accessible repositories, developers enable the wider research community to verify results and build upon their prior work. Rigorous software testing, including unit and integration tests, helps prevent errors during collaborative development. Continuous integration that is deployable on many open platforms (e.g., GitHub and GitLab) can automate this testing. Lastly, it is important to follow packaging conventions (e.g., PyPI for Python software) and account for code maintenance or deprecation strategies (\appref{app:best_software}).

\paragraph{Data preparation for machine learning models}

Raw simulation data often needs cleaning and normalisation to remove outliers or rescale variables before being fed into ML pipelines. Feature engineering transforms or extracts relevant attributes (e.g., packing density, velocity gradients) that help the ML model recognize underlying patterns. Appropriate data partitioning into training, validation, and test sets is critical: it prevents overfitting, makes model tuning more reliable, and provides an unbiased performance estimate.
Additionally, large and heterogeneous datasets may benefit from efficient data-loading and batching strategies—essential, for instance, when dealing with millions of particles in a single simulation. In some cases, data augmentation can increase the variety of training samples and improve model generalisation (\appref{app:data_prep_ML}). 

\paragraph{Training machine learning models}

Model training should be approached systematically to ensure clear documentation, reproducibility, and verifiable results. Key aspects include specifying the model architecture (e.g., the number and type of neural network layers) and clarifying input–output formats. Frequent monitoring of loss functions and metrics (e.g., mean squared error or accuracy) helps guide hyperparameter tuning, which can benefit from advanced methods like grid search, Bayesian optimisation, or other automated approaches.
Techniques such as early stopping, cross-validation, and learning-rate scheduling bolster model reliability (\appref{app:ML_training}).

\section{Conclusions and outlook}
\label{sec:conclude}

The Lorentz Center workshop on ``Machine Learning for Discrete Granular Media'', brought together researchers from the granular materials and machine learning communities to discuss the state-of-the-art and challenges in an emerging research field that integrates machine learning and numerical simulations of granular materials.
The workshop participants collectively created a knowledge base, which resulted in this position paper. First, the challenges in granular material simulations were reviewed, followed by a selective overview of ML approaches that proved useful in developing constitutive laws (theory), solving governing equations (numerical), and facilitating digital twinning, e.g., with uncertainty quantification and optimisation.
Synthesising on the current state-of-the-art and the discussions centered around the application, adaptation, and efficacy of ML in modelling granular materials, this section outlines future directions that touch on constitutive theory, neural operator learning, uncertainty quantification for granular materials, and the key issues on interpretability, scalability, and hybrid approaches within ML fields.
The vision presented in this paper aims to serve research communities within granular mechanics and physics and broadly in various engineering applications.

\paragraph{Learning material's constitutive laws}

Materials' constitutive laws are largely based on classical theories of elasticity, plasticity and rheology, where physical constraints (e.g. thermodynamics) can be considered in handcrafting constitutive laws for granular materials.
Advances in particle-scale methods now allow for the simulation of all the intricate features of granular material behaviour, including stress path dependency, fabric anisotropy (microstructure) and particle shapes, which can be expressed in stress and strain sequences.
A major challenge here is to create ML-based constitutive laws based on particle-scale information that are more efficient and accurate (compared to experimental evidence) and provide insights into constitutive theories familiar to engineers.
``Recurrent'' black-box architectures (e.g., GRU, and LSTM) have shown their effectiveness in predicting stress and strain pairs, which are the macroscopic responses of granular materials.
A future direction for learning constitutive models lies in the development of hybrid approaches, where ML models are guided by physical constraints and scarce experimental data and complemented by abundant simulation data.
For example, ML can assist in defining appropriate strain energy density functions or yield surfaces in stress space (augmented by other variables), rather than relying on handcrafted models developed by domain experts.
Symbolic regression can derive mathematical formulations directly from data, providing a middle ground between black-box ML models and fully physics-based approaches.
These constitutive model surrogates should be general enough to cover both the solid-like (small strain, rate-independent) and fluid-like (large strain, rate-dependent) regimes of granular materials, incorporating physical principles (or biases) through regularisation.
Lastly, it is essential to systematically verify ML-based constitutive surrogates by degenerating them into limiting cases, such as elastic solid and Newtonian fluid behaviour and benchmarking them against them.
            
\paragraph{Learning field evolutions of underlying governing equations}

Neural operators are capable of learning mappings between functional spaces.
They are promising tools for bypassing conventional numerical methods to solve governing equations that have strong nonlinearity and discontinuity without the issues of numerical instabilities due to spatial/temporal discretisation.
These methods allow generalisation across different problem geometries and material parameters without needing handcrafted theories at a specific length or time scale.
However, when applied to granular materials, neural operators face scalability issues due to the dynamic particle-scale interactions that exhibit macroscopic material behaviour transitioning between fluid-like and solid-like.
Future directions should focus on refining encoder/decoder architectures to express spatial and temporal patterns in both the physical regime and on utilising transfer learning techniques to specific engineering tasks, e.g., geometries, materials, etc.
Learning the multi-scale features of granular systems relies on latent state representations that can handle strong discontinuity and nonlinearity at the lower scales while capturing macroscopic fields of interest under various initial and boundary conditions.
One example of circumventing the scalability issue is training on the underlying fields instead of fine-grained particle kinematics and interaction features in graph neural networks.
However, the physically admissible microstructures, ready to be plugged again into numerical solvers like DEM, are difficult to reconstruct.

\paragraph{Learning uncertainties from limited real-world data}

Many engineering problems in industry involve risk assessment and/or design optimisation, which integrate real-world uncertainty (although limited) into the physical processes represented by their predictive models.
Given the stochastic nature of granular systems and the variability in material properties and conditions, uncertainty quantification is crucial in granular simulations and practical engineering applications.
Data-driven surrogates, either based on classical machine learning or deep learning (see \secref{sec:MLSolution}), are essential in constraining uncertainties with real-world measurement data or design objectives (e.g. through Bayesian updating and optimisation).
Although deep learning models can learn to generalize across unseen parameter combinations, their probability landscape over the parameters are very likely to be different from the counterpart of the numerical models, adding additional uncertainties that are difficult to trace.
A promising direction would be adding model uncertainties to deep learning models (e.g. Bayesian neural networks).
The stochastic behaviour in real-world applications could be approached through ensembles: running multiple predictions with different parameters and averaging the results.
Future works should also include the trade-off between large and small neural networks for general or specific engineering tasks to better use compute resources. 
Techniques such as Bayesian inference and Gaussian processes can be instrumental in understanding the robustness of ML-based constitutive models and identifying part of ML predictions that might no longer be reliable.
By incorporating uncertainty quantification via the combined use of physics-based models and their ML surrogates into digital twin applications, researchers can enable more informed decision-making for real-world engineering processes.

\paragraph{Open science infrastructure to facilitate ``GranML''}

The importance of data quality, sharing and reproducibility was discussed extensively in the workshop.
For ML models to be broadly applicable and trusted by the community, they must be developed using standardized datasets and workflows that ensure reproducibility against well-known engineering benchmark problems.
Open science practices can provide the infrastructure necessary for developing models and data resources collaboratively.
To facilitate progress, future efforts should establish community-wide standards for data sharing and model benchmarking, such as the Open Network on DEM Simulations (ON-DEM) funded by the European Cooperation in Science and Technology and DesignSafe by the US National Science Foundation.
An open-access platform akin to ImageNet could serve as a repository for training datasets, pre-trained models, and evaluation metrics. Such a platform would not only improve reproducibility but also foster collaboration, enabling researchers from different disciplines to contribute to and benefit from shared data and methodologies.

In addition to the topics addressed during the workshop and in this position paper, several broader challenges emerged that are particularly relevant to machine learning for granular materials.
These include issues of scalability when dealing with large-scale systems involving millions of particles, the development of hybrid approaches in which physics-based and data-driven models are used and the need for improved interpretability to obtain engineering insights.
The potential applications of ML models for granular materials extend far beyond civil and mechanical engineering.
More accurate, scalable, and interpretable ML-based approaches would also benefit fields such as pharmaceutical manufacturing, food processing, particle technology, and agriculture.
Progress in the research area addressed in this position paper and those beyond will enable transitioning ML from being a novel, exploratory tool to becoming an integral part of the assessment and design toolbox across multiple industries.
Ultimately, the key to progress lies in fostering collaboration across domains—bringing together data scientists, engineers, and domain experts—to develop trustworthy, scalable models capable of addressing the pressing challenges of granular material applications.

\section{Acknowledgments}

All authors contributed equally to the first draft. D. Barreto, M.A. Cabrera, H. Cheng, D. Daniel, M. Fransen, B. Kieskamp, J. Nuttall, J. Ooi, S. Papanicolopulos, D. Schott and T. Weinhart wrote the first draft of \secref{sec:GM_scales}. B. Alkin, J. Brandstetter, H. Cheng, A. Fürst, M. Guo, K. Kumar, T. Qu, T. Shuku, W. Sun, D. R. Tunuguntla, D.N. Wilke, and D. Ye wrote the first draft of \secref{sec:MLSolution}. X. Fan, J. Morrissey, and L. Orozco wrote the first draft of \secref{sec:examples}. H. Cheng, D. R. Tunuguntla, and D.N. Wilke wrote the first draft of \secref{sec:intro}. H. Cheng wrote the first draft of \secref{sec:aim} and \secref{sec:conclude}. D. R. Tunuguntla led the revision of \secref{sec:intro}. M. Fransen led the revision of \secref{sec:GM_scales}, A. Fürst and D. R. Tunuguntla led the revision of \secref{sec:MLSolution}, H. Cheng, D. R. Tunuguntla and D.N. Wilke led the revision of \secref{sec:examples}. X. Fan and J. Morrissey wrote and reviewed \secref{sec:app}. H. Cheng carried out the final edits of all the sections.

The workshop organisers, Hongyang Cheng, Marc Fransen, and Bojana Rosic, and the participants would like to thank the Lorentz Center for providing the venue and facilitating the organisation. The main organiser, Hongyang Cheng, would like to acknowledge the financial support from the Netherlands eScience Center under grant number NLESC.OEC.2021.032.

\begin{appendices}
\section{Suggestions for open infrastructure and best practices for material learning in granular material simulations}
\label{sec:app}

Best practices regarding FAIR data and software in this context include the creation of a standardised pre-processing pipeline to produce input for ML models, the establishment of a reproducible software environment that enables users to replicate workflows using identical computational setups, and the adoption of domain metadata standards covering vocabularies, formats, and data structures, as discussed in \secref{sec:recommendationsOpenInfra}.
Our practices demonstrated through the examples in \secref{sec:examples} align with these principles by ensuring that ready-to-use input data for ML models is standardised, shared, and well-documented according to domain standards. The following appendices expand on these areas with more detailed recommendations and specific strategies.

\subsection{Data management for granular material simulations}
\label{app:data_management}
\begin{itemize}
    \item Describe the storage infrastructure used for handling large datasets, emphasising the use of high-performance systems such as parallel file systems.
    \item Specify the data formats employed (e.g. HDF5, NetCDF), justifying their selection based on factors like portability and efficiency.
    \item Detail the metadata schema, focusing on how it captures simulation parameters, particle-scale material properties, and spatio-temporal information in a structured manner.
    \item Outline the version control strategy for simulation codes and datasets, emphasising its role in ensuring reproducibility.
    \item If applicable, discuss any data compression techniques used, focusing on how they balance storage efficiency with data integrity.
\end{itemize}

\subsection{Best software practices}
\label{app:best_software}

It is important to recognize that different types of software vary in complexity. For instance, a DEM (Discrete Element Method) code contains multiple modules and greater complexity compared to a set of scripts designed for running simulations or machine learning model training. The following recommendations apply broadly to different types of software, though the extent and specifics may differ based on the software's nature and purpose \cite{software_management_plan}.

\begin{itemize}
	\item Use version control
	\begin{itemize}
		\item Adopt a version control system (e.g. Git) to track changes in code, ensure collaboration and manage releases effectively.
		\item Use tags to mark specific releases according to a clear versioning scheme (e.g. semantic versioning). This helps users understand compatibility between software versions and enhances reproducibility by enabling users to replicate results with a particular version.
	\end{itemize}
	\item Repository publication
	\begin{itemize}
		\item Host the software in a publicly accessible remote repository (e.g. GitHub, GitLab) to promote reusability, transparency and collaboration.
		\item Include license (e.g. MIT, Apache2, GPL) to specify terms of use and reusability.
		\item For scholarly recognition, publish your software in a service that will store snapshots of your software and assign a DOI (e.g Zenodo, Software Heritage, OSF - Open Science Framework, SourceForge) and ensure citation files (e.g. Citation File Format - CFF) are available so that your software can be correctly credited.
	\end{itemize}
	\item Documentation
	\begin{itemize}
		\item Provide user documentation that explains what the software does and how to use it, ideally in a Markdown format (e.g. README.md). This reduces barriers for new users and makes the software more approachable.
		\item Developer documentation (e.g. docstrings, inline comments, contributing guidelines) should explain how to extend or modify the software.
		\item Deployment documentation should clarify system requirements, dependencies and detailed installation instructions.
	\end{itemize}
	\item Software testing
	\begin{itemize}
		\item Implement a rigorous testing strategy, including unit tests, integration tests and regression tests, to ensure the software functions as intended in different environments.
		\item Use code quality tools (e.g. linters) and coverage tools to assess how much of the code is being tested.
		\item Set up CI tools (e.g. Github Actions, GitLab CI, Travis CI) to automatically run tests on each commit. This ensures that new changes do not introduce errors and maintain software quality over time.
	\end{itemize}
	\item Modular code and software engineering standards:
	\begin{itemize}
		\item Ensure code adheres to best practices in terms of modularity, readability and code reusability. Encourage collaboration by setting guidelines for contributing and reviewing code (e.g. pull requests, code reviews).
		\item Follow recognized software engineering standards for packaging and distributing software using package managers like PyPI (Python), so that users can easily install the software.
	\end{itemize}
	\item Maintenance and Sustainability:
	\begin{itemize}
		\item Plan for the long-term maintenance of the software by ensuring there are sufficient resources and support structures. This could involve creating a developer community or ensuring future projects continue the software’s maintenance.
		\item Develop a retirement strategy if the software is no longer actively maintained, ensuring that users are aware of its status.
	\end{itemize}
	\item Ensure that the software is secure against potential vulnerabilities such as dependency management, vendor lock-in and cross-platform compatibility.
\end{itemize}

\subsection{Data preparation for machine learning models}
\label{app:data_prep_ML}

\begin{itemize}
    \item Describe the preprocessing pipeline for particle data, including cleaning, normalisation and augmentation methods.
    \item Explain the feature engineering process, detailing how relevant features are extracted and selected from raw particle data.
    \item Present the methodology for partitioning data into training, validation and test sets, justifying the chosen approach.
    \item Discuss efficient data loading and batching strategies, particularly for handling large-scale particle datasets during model training.
    \item If used, explain any data augmentation techniques applied to enhance model generalisation, providing the rationale for their selection.
\end{itemize}

\subsection{Training machine learning models}

\label{app:ML_training}

\begin{itemize}
\item Model description
    \begin{itemize}
    \item Architecture: Document the model architecture (e.g. neural network layers, types of layers, recurrent/convolutional, etc.)
    \item Inputs and outputs: Define the format, type and sizes of input data and the expected outputs.
    \item Model size: Specify the number of parameters, memory footprint and compute requirements.
    \item Regularisation techniques: Document the techniques used (e.g. L1/L2, dropout, batch normalisation) and their place in the pipeline.
    \end{itemize}
\item Loss functions and error metrics: Describe the loss and error metrics used and how are they monitored (i.e. per batch, epoch).
\item Hyperparameter tuning
    \begin{itemize}
    \item Document which techniques (e.g grid search, Bayesian optimisation, automated hyperparameter tuning tools) and in which order are they applied.
    \item Report on which additional techniques such as: early stopping, cross-validation, learning rate schedules used and their parameters.
    \end{itemize}
\item Training–validation–test workflow:
Document the training-validation-test workflow applied and the splitting of the data for each stage.
\end{itemize}

\end{appendices}


\begin{thebibliography}{100}

\bibitem{ABOUSLEIMAN2020123}
R.~Abousleiman, G.~Walton, and S.~Sinha.
\newblock Understanding roof deformation mechanics and parametric sensitivities
  of coal mine entries using the discrete element method.
\newblock {\em International Journal of Mining Science and Technology},
  30(1):123--129, 2020.
\newblock Special issue on ground control in mining in 2019.

\bibitem{Lopez2018}
J.~P. Aguilar-L{\'{o}}pez, J.~J. Warmink, R.~M. Schielen, and S.~J. Hulscher.
\newblock {Piping erosion safety assessment of flood defences founded over
  sewer pipes}.
\newblock {\em Eur. J. Environ. Civ. Eng.}, 22(6):707--735, 2018.

\bibitem{alkin2024universal}
B.~Alkin, A.~Fürst, S.~Schmid, L.~Gruber, M.~Holzleitner, and J.~Brandstetter.
\newblock Universal physics transformers, 2024.

\bibitem{alkin2024}
B.~Alkin, T.~Kronlachner, S.~Papa, S.~Pirker, T.~Lichtenegger, and
  J.~Brandstetter.
\newblock Neural{DEM} -- real-time simulation of industrial particulate flows,
  2024.

\bibitem{Alvarez2024}
J.~E. Alvarez, H.~Cheng, S.~Luding, and T.~Weinhart.
\newblock {Densification of visco-elastic powders during free and
  pressure-assisted sintering}.
\newblock {\em Int. J. Solids Struct.}, 294(March):112786, 2024.

\bibitem{Andrieu2010}
C.~Andrieu, A.~Doucet, and R.~Holenstein.
\newblock Particle markov chain monte carlo methods.
\newblock {\em Journal of the Royal Statistical Society: Series B (Statistical
  Methodology)}, 72(3):269--342, 2010.

\bibitem{Bahmani_2024}
B.~Bahmani, H.~S. Suh, and W.~Sun.
\newblock Discovering interpretable elastoplasticity models via the neural
  polynomial method enabled symbolic regressions.
\newblock {\em Computer Methods in Applied Mechanics and Engineering},
  422:116827, Mar. 2024.

\bibitem{Balshaw2023}
R.~Balshaw, P.~S. Heyns, D.~N. Wilke, and S.~Schmidt.
\newblock Latent indicators for temporal-preserving latent variable models in
  vibration-based condition monitoring under non-stationary conditions.
\newblock {\em Mechanical Systems and Signal Processing}, 199:110446, 9 2023.

\bibitem{batatia2022mace}
I.~Batatia, D.~P. Kovacs, G.~Simm, C.~Ortner, and G.~Cs{\'a}nyi.
\newblock Mace: Higher order equivariant message passing neural networks for
  fast and accurate force fields.
\newblock {\em Advances in Neural Information Processing Systems},
  35:11423--11436, 2022.

\bibitem{batzner2022e3}
S.~Batzner, A.~Musaelian, L.~Sun, M.~Geiger, J.~P. Mailoa, M.~Kornbluth,
  N.~Molinari, T.~E. Smidt, and B.~Kozinsky.
\newblock E(3)-equivariant graph neural networks for data-efficient and
  accurate interatomic potentials.
\newblock {\em Nature communications}, 13(1):2453, 2022.

\bibitem{BGW2015pmorSurvery}
P.~Benner, S.~Gugercin, and K.~Willcox.
\newblock A survey of projection-based model reduction methods for parametric
  dynamical systems.
\newblock {\em SIAM Review}, 57(4):483--531, 2015.

\bibitem{Berkooz1993}
G.~Berkooz, P.~Holmes, and J.~L. Lumley.
\newblock The proper orthogonal decomposition in the analysis of turbulent
  flows.
\newblock {\em Annual Review of Fluid Mechanics}, 25:539--575, 1993.

\bibitem{bishop2007}
C.~M. Bishop.
\newblock {\em Pattern Recognition and Machine Learning (Information Science
  and Statistics)}.
\newblock Springer, 1 edition, 2007.

\bibitem{bodnar2024aurora}
C.~Bodnar, W.~P. Bruinsma, A.~Lucic, M.~Stanley, J.~Brandstetter, P.~Garvan,
  M.~Riechert, J.~Weyn, H.~Dong, A.~Vaughan, et~al.
\newblock Aurora: A foundation model of the atmosphere.
\newblock {\em arXiv preprint arXiv:2405.13063}, 2024.

\bibitem{bonnet2022}
F.~Bonnet, J.~Mazari, P.~Cinnella, and P.~Gallinari.
\newblock Airfrans: High fidelity computational fluid dynamics dataset for
  approximating reynolds-averaged navier--stokes solutions.
\newblock {\em Advances in Neural Information Processing Systems},
  35:23463--23478, 2022.

\bibitem{botteghi2022deep}
N.~Botteghi, M.~Guo, and C.~Brune.
\newblock Deep kernel learning of dynamical models from high-dimensional noisy
  data.
\newblock {\em Scientific Reports}, 12:21530, 2022.

\bibitem{botteghi2024recurrent}
N.~Botteghi, P.~Motta, A.~Manzoni, P.~Zunino, and M.~Guo.
\newblock Recurrent deep kernel learning of dynamical systems.
\newblock {\em arXiv preprint arXiv:2405.19785}, 2024.

\bibitem{brandstetter2021geometric}
J.~Brandstetter, R.~Hesselink, E.~van~der Pol, E.~J. Bekkers, and M.~Welling.
\newblock Geometric and physical quantities improve e (3) equivariant message
  passing.
\newblock {\em arXiv preprint arXiv:2110.02905}, 2021.

\bibitem{brandstetter2022message}
J.~Brandstetter, D.~Worrall, and M.~Welling.
\newblock Message passing neural pde solvers.
\newblock {\em arXiv preprint arXiv:2202.03376}, 2022.

\bibitem{bronstein2021geometric}
M.~M. Bronstein, J.~Bruna, T.~Cohen, and P.~Veli{\v{c}}kovi{\'c}.
\newblock Geometric deep learning: Grids, groups, graphs, geodesics, and
  gauges.
\newblock {\em arXiv preprint arXiv:2104.13478}, 2021.

\bibitem{Manifold_Learning}
M.~M. Bronstein, J.~Bruna, Y.~LeCun, A.~Szlam, and P.~Vandergheynst.
\newblock Geometric deep learning: going beyond euclidean data.
\newblock {\em IEEE Signal Processing Magazine}, 34(4):18--42, 2017.

\bibitem{brunton2023machine}
S.~L. Brunton and J.~N. Kutz.
\newblock Machine learning for partial differential equations, 2023.

\bibitem{brunton2016sindy}
S.~L. Brunton, J.~L. Proctor, and J.~N. Kutz.
\newblock Discovering governing equations from data by sparse identification of
  nonlinear dynamical systems.
\newblock {\em Proceedings of the National Aca{DEM}y of Sciences},
  113(15):3932--3937, 2016.

\bibitem{budhu2010}
M.~Budhu.
\newblock {\em Soil mechanics and foundations}.
\newblock John Wiley and Sons, 2010.

\bibitem{cabrera2019granular}
M.~Cabrera and N.~Estrada.
\newblock Granular column collapse: Analysis of grain-size effects.
\newblock {\em Physical Review E}, 99(1):012905, 2019.

\bibitem{cai2023equivariant}
C.~Cai, N.~Vlassis, L.~Magee, R.~Ma, Z.~Xiong, B.~Bahmani, T.-F. Wong, Y.~Wang,
  and W.~Sun.
\newblock Equivariant geometric learning for digital rock physics: estimating
  formation factor and effective permeability tensors from morse graph.
\newblock {\em International Journal for Multiscale Computational Engineering},
  21(5), 2023.

\bibitem{Cai2021}
C.~Cai, D.~Wang, and Y.~Wang.
\newblock {Graph Coarsening with Neural Networks}.
\newblock pages 1--26, 2021.

\bibitem{cao2021choose}
S.~Cao.
\newblock Choose a transformer: Fourier or galerkin.
\newblock {\em Advances in neural information processing systems},
  34:24924--24940, 2021.

\bibitem{champion2019sindy}
K.~Champion, B.~Lusch, J.~N. Kutz, and S.~L. Brunton.
\newblock Data-driven discovery of coordinates and governing equations.
\newblock {\em Proceedings of the National Aca{DEM}y of Sciences},
  116(45):22445--22451, 2019.

\bibitem{Chaouch2024}
S.~Chaouch and J.~Yvonnet.
\newblock Unsupervised machine learning classification for accelerating fe2
  multiscale fracture simulations.
\newblock {\em Computer Methods in Applied Mechanics and Engineering},
  432:117278, 2024.

\bibitem{chapman2000crisp}
P.~Chapman, J.~Clinton, R.~Kerber, T.~Khabaza, T.~Reinartz, C.~Shearer,
  R.~Wirth, et~al.
\newblock Crisp-dm 1.0: Step-by-step data mining guide.
\newblock {\em SPSS inc}, 9(13):1--73, 2000.

\bibitem{chen2018neural}
R.~T.~Q. Chen, Y.~Rubanova, J.~Bettencourt, and D.~K. Duvenaud.
\newblock Neural ordinary differential equations.
\newblock {\em Advances in neural information processing systems}, 31, 2018.

\bibitem{chen2021physics}
W.~Chen, Q.~Wang, J.~S. Hesthaven, and C.~Zhang.
\newblock Physics-informed machine learning for reduced-order modeling of
  nonlinear problems.
\newblock {\em Journal of Computational Physics}, 446:110666, 2021.

\bibitem{chen2021solving}
Y.~Chen, B.~Hosseini, H.~Owhadi, and A.~M. Stuart.
\newblock Solving and learning nonlinear pdes with gaussian processes.
\newblock {\em Journal of Computational Physics}, 447:110668, 2021.

\bibitem{Cheng2021a}
H.~Cheng.
\newblock {DEM} simulation data containing micro- and macro-scale quantities of
  granular materials under triaxial compression. dataset. version 1. available
  at https://doi.org/10.4121/16632559.v1, 2021.

\bibitem{Cheng2024}
H.~Cheng, L.~Orozco, R.~Lubbe, A.~Jansen, P.~Hartmann, and K.~Thoeni.
\newblock Grainlearning: A {Bayesian} uncertainty quantification toolbox for
  discrete and continuum numerical models of granular materials.
\newblock {\em Journal of Open Source Software}, 9(97):6338, 2024.

\bibitem{CHENG2019268}
H.~Cheng, T.~Shuku, K.~Thoeni, P.~Tempone, S.~Luding, and V.~Magnanimo.
\newblock An iterative {Bayesian} filtering framework for fast and automated
  calibration of {DEM} models.
\newblock {\em Computer Methods in Applied Mechanics and Engineering},
  350:268--294, 2019.

\bibitem{Cheng2023}
H.~Cheng, A.~R. Thornton, S.~Luding, A.~L. Hazel, and T.~Weinhart.
\newblock Concurrent multi-scale modeling of granular materials: Role of
  coarse-graining in {{FEM}}-{DEM} coupling.
\newblock {\em Computer Methods in Applied Mechanics and Engineering},
  403:115651, 2023.

\bibitem{choi2023graph}
Y.~Choi and K.~Kumar.
\newblock Graph neural network-based surrogate model for granular flows, 2023.

\bibitem{choi2024inverse}
Y.~Choi and K.~Kumar.
\newblock Inverse analysis of granular flows using differentiable graph neural
  network simulator.
\newblock {\em arXiv preprint arXiv:2401.13695}, 2024.

\bibitem{cicci2023uncertainty}
L.~Cicci, S.~Fresca, M.~Guo, A.~Manzoni, and P.~Zunino.
\newblock Uncertainty quantification for nonlinear solid mechanics using
  reduced order models with gaussian process regression.
\newblock {\em Computers \& Mathematics with Applications}, 149:1--23, 2023.

\bibitem{conti2024multi}
P.~Conti, M.~Guo, A.~Manzoni, A.~Frangi, S.~L. Brunton, and J.~Nathan~Kutz.
\newblock Multi-fidelity reduced-order surrogate modelling.
\newblock {\em Proceedings of the Royal Society A}, 480(2283):20230655, 2024.

\bibitem{cranmer2020discovering}
M.~Cranmer, A.~Sanchez~Gonzalez, P.~Battaglia, R.~Xu, K.~Cranmer, D.~Spergel,
  and S.~Ho.
\newblock Discovering symbolic models from deep learning with inductive biases.
\newblock {\em Advances in neural information processing systems},
  33:17429--17442, 2020.

\bibitem{Cundall1979ADN}
P.~A. Cundall and O.~D.~L. Strack.
\newblock A discrete numerical model for granular assemblies.
\newblock {\em Geotechnique}, 29:47--65, 1979.

\bibitem{Duriez2021a}
J.~Duriez and C.~Galusinski.
\newblock {A Level Set-Discrete Element Method in YADE for numerical,
  micro-scale, geomechanics with refined grain shapes}.
\newblock {\em Comput. Geosci.}, 157(September):104936, 2021.

\bibitem{Feng2023}
Y.~T. Feng.
\newblock {Thirty years of developments in contact modelling of non-spherical
  particles in {DEM}: a selective review}.
\newblock {\em Acta Mech. Sin. Xuebao}, 39(1), 2023.

\bibitem{forterre2008flows}
Y.~Forterre and O.~Pouliquen.
\newblock Flows of dense granular media.
\newblock {\em Annu. Rev. Fluid Mech.}, 40(1):1--24, 2008.

\bibitem{fortunato2022}
M.~Fortunato, T.~Pfaff, P.~Wirnsberger, A.~Pritzel, and P.~Battaglia.
\newblock Multiscale meshgraphnets.
\newblock {\em arXiv preprint arXiv:2210.00612}, 2022.

\bibitem{Fransen2025}
M.~Fransen.
\newblock {\em Unravelling gravel: Including stochastic behaviour of granular
  materials in design of bulk handling equipment}.
\newblock Dissertation (tu delft), Delft University of Technology, 2024.

\bibitem{Fransen2022}
M.~P. Fransen, M.~Langelaar, and D.~Schott.
\newblock Including stochastics in metamodel-based {DEM} model calibration.
\newblock {\em Powder Technology}, 406:117400, 2022.

\bibitem{FRANSEN2023118526}
M.~P. Fransen, M.~Langelaar, and D.~L. Schott.
\newblock Deterministic vs. robust design optimization using {DEM}-based
  metamodels.
\newblock {\em Powder Technology}, 425:118526, 2023.

\bibitem{fresca2022pod}
S.~Fresca and A.~Manzoni.
\newblock {POD-DL-ROM}: enhancing deep learning-based reduced order models for
  nonlinear parametrized pdes by proper orthogonal decomposition.
\newblock {\em Computer Methods in Applied Mechanics and Engineering},
  388:114181, 2022.

\bibitem{fuhg2022local}
J.~N. Fuhg, M.~Marino, and N.~Bouklas.
\newblock Local approximate gaussian process regression for data-driven
  constitutive models: development and comparison with neural networks.
\newblock {\em Computer Methods in Applied Mechanics and Engineering},
  388:114217, 2022.

\bibitem{gans2023collapse}
A.~Gans, A.~Abramian, P.-Y. Lagr{\'e}e, M.~Gong, A.~Sauret, O.~Pouliquen, and
  M.~Nicolas.
\newblock Collapse of a cohesive granular column.
\newblock {\em Journal of Fluid Mechanics}, 959:A41, 2023.

\bibitem{Gao2014}
Z.~Gao, J.~Zhao, X.~S. Li, and Y.~F. Dafalias.
\newblock {A critical state sand plasticity model accounting for fabric
  evolution}.
\newblock {\em Int. J. Numer. Anal. Methods Geomech.}, 38(4):370--390, mar
  2014.

\bibitem{gdr2004dense}
G.~M. gdrmidi@ polytech. univ-mrs. fr http://www. lmgc. univ-montp2. fr/MIDI/.
\newblock On dense granular flows.
\newblock {\em The European Physical Journal E}, 14:341--365, 2004.

\bibitem{GW2021learning}
O.~Ghattas and K.~Willcox.
\newblock Learning physics-based models from data: perspectives from inverse
  problems and model reduction.
\newblock {\em Acta Numerica}, 30:445--554, 2021.

\bibitem{Giles_2015}
M.~B. Giles.
\newblock Multilevel monte carlo methods.
\newblock {\em Acta Numerica}, 24:259–328, 2015.

\bibitem{golan2022information}
A.~Golan and J.~Harte.
\newblock Information theory: A foundation for complexity science.
\newblock {\em Proceedings of the National Aca{DEM}y of Sciences},
  119(33):e2119089119, 2022.

\bibitem{Golshan2020}
S.~Golshan, R.~Sotudeh-Gharebagh, R.~Zarghami, N.~Mostoufi, B.~Blais, and
  J.~Kuipers.
\newblock Review and implementation of {CFD}-{DEM} applied to chemical process
  systems.
\newblock {\em Chemical Engineering Science}, 221:115646, 2020.

\bibitem{Flow_Application}
A.~Grover, M.~Dhar, and S.~Ermon.
\newblock Flow-gan: Combining maximum likelihood and adversarial learning in
  generative models.
\newblock In {\em Proceedings of the AAAI Conference on Artificial
  Intelligence}, volume~32, 2018.

\bibitem{2024IJNAM..48.1372G}
Q.~{Guan}, Z.~{Yang}, N.~{Guo}, and L.~{Chen}.
\newblock {Deep learning‑accelerated multiscale approach for granular
  material modeling}.
\newblock {\em International Journal for Numerical and Analytical Methods in
  Geomechanics}, 48(5):1372--1389, Apr. 2024.

\bibitem{guan2023machine}
S.~Guan, T.~Qu, Y.~Feng, G.~Ma, and W.~Zhou.
\newblock A machine learning-based multi-scale computational framework for
  granular materials.
\newblock {\em Acta Geotechnica}, 18(4):1699--1720, 2023.

\bibitem{guo2018reduced}
M.~Guo and J.~S. Hesthaven.
\newblock Reduced order modeling for nonlinear structural analysis using
  gaussian process regression.
\newblock {\em Computer Methods in Applied Mechanics and Engineering},
  341:807--826, 2018.

\bibitem{guo2019data}
M.~Guo and J.~S. Hesthaven.
\newblock Data-driven reduced order modeling for time-dependent problems.
\newblock {\em Computer Methods in Applied Mechanics and Engineering},
  345:75--99, 2019.

\bibitem{guo2022bayesian}
M.~Guo, S.~A. McQuarrie, and K.~E. Willcox.
\newblock {Bayesian} operator inference for data-driven reduced-order modeling.
\newblock {\em Computer Methods in Applied Mechanics and Engineering},
  402:115336, 2022.

\bibitem{Guo2014}
N.~Guo and J.~Zhao.
\newblock A coupled {FEM}/{DEM} approach for hierarchical multiscale modelling
  of granular media.
\newblock {\em International Journal for Numerical Methods in Engineering},
  99(11):789--818, 2014.

\bibitem{guo2021learning}
T.~Guo, O.~Roko{\v{s}}, and K.~Veroy.
\newblock Learning constitutive models from microstructural simulations via a
  non-intrusive reduced basis method.
\newblock {\em Computer Methods in Applied Mechanics and Engineering},
  384:113924, 2021.

\bibitem{guo2016convolutional}
X.~Guo, W.~Li, and F.~Iorio.
\newblock {Convolutional neural networks for steady flow approximation}.
\newblock In {\em Proceedings of the 22nd ACM SIGKDD International Conference
  on Knowledge Discovery and Data Mining}, pages 481--490, 2016.

\bibitem{gupta2022towards}
J.~K. Gupta and J.~Brandstetter.
\newblock Towards multi-spatiotemporal-scale generalized pde modeling.
\newblock {\em arXiv preprint arXiv:2209.15616}, 2022.

\bibitem{haeri2024subspace}
A.~Haeri, D.~Holz, and K.~Skonieczny.
\newblock Subspace graph networks for real-time granular flow simulation with
  applications to machine-terrain interactions.
\newblock {\em Engineering Applications of Artificial Intelligence},
  135:108765, 2024.

\bibitem{hao2023gnot}
Z.~Hao, Z.~Wang, H.~Su, C.~Ying, Y.~Dong, S.~Liu, Z.~Cheng, J.~Song, and
  J.~Zhu.
\newblock Gnot: A general neural operator transformer for operator learning.
\newblock In {\em International Conference on Machine Learning}, pages
  12556--12569. PMLR, 2023.

\bibitem{HARTMANN2022104491}
P.~Hartmann, H.~Cheng, and K.~Thoeni.
\newblock Performance study of iterative {Bayesian} filtering to develop an
  efficient calibration framework for {DEM}.
\newblock {\em Computers and Geotechnics}, 141:104491, 2022.

\bibitem{he2022thermodynamically}
X.~He and J.-S. Chen.
\newblock Thermodynamically consistent machine-learned internal state variable
  approach for data-driven modeling of path-dependent materials.
\newblock {\em Computer Methods in Applied Mechanics and Engineering},
  402:115348, 2022.

\bibitem{heider2020so}
Y.~Heider, K.~Wang, and W.~Sun.
\newblock So (3)-invariance of informed-graph-based deep neural network for
  anisotropic elastoplastic materials.
\newblock {\em Computer Methods in Applied Mechanics and Engineering},
  363:112875, 2020.

\bibitem{herde2024poseidon}
M.~Herde, B.~Raoni{\'c}, T.~Rohner, R.~K{\"a}ppeli, R.~Molinaro,
  E.~de~B{\'e}zenac, and S.~Mishra.
\newblock Poseidon: Efficient foundation models for pdes.
\newblock {\em arXiv preprint arXiv:2405.19101}, 2024.

\bibitem{hesthaven2016certified}
J.~S. Hesthaven, G.~Rozza, and B.~Stamm.
\newblock {\em Certified Reduced Basis Methods for Parametrized Partial
  Differential Equations}, volume 590.
\newblock Springer, 2016.

\bibitem{hesthaven2018non}
J.~S. Hesthaven and S.~Ubbiali.
\newblock Non-intrusive reduced order modeling of nonlinear problems using
  neural networks.
\newblock {\em Journal of Computational Physics}, 363:55--78, 2018.

\bibitem{Disentanglement_Study}
I.~Higgins, L.~Matthey, A.~Pal, C.~Burgess, X.~Glorot, M.~Botvinick,
  S.~Mohamed, and A.~Lerchner.
\newblock beta-vae: Learning basic visual concepts with a constrained
  variational framework.
\newblock In {\em International Conference on Learning Representations}, 2017.

\bibitem{HORABIK2016206}
J.~Horabik and M.~Molenda.
\newblock Parameters and contact models for {DEM} simulations of agricultural
  granular materials: A review.
\newblock {\em Biosystems Engineering}, 147:206--225, 2016.

\bibitem{huang2020machine}
D.~Huang, J.~N. Fuhg, C.~Wei{\ss}enfels, and P.~Wriggers.
\newblock A machine learning based plasticity model using proper orthogonal
  decomposition.
\newblock {\em Computer Methods in Applied Mechanics and Engineering},
  365:113008, 2020.

\bibitem{Husic2020}
B.~E. Husic, N.~E. Charron, D.~Lemm, J.~Wang, A.~P{\'{e}}rez, M.~Majewski,
  A.~Kr{\"{a}}mer, Y.~Chen, S.~Olsson, G.~{De Fabritiis}, F.~No{\'{e}}, and
  C.~Clementi.
\newblock {Coarse graining molecular dynamics with graph neural networks}.
\newblock {\em J. Chem. Phys.}, 153(19):241722, nov 2020.

\bibitem{jiang2024integrating}
Y.~Jiang, E.~Byrne, J.~Glassey, and X.~Chen.
\newblock Integrating graph neural network-based surrogate modeling with
  inverse design for granular flows.
\newblock {\em Industrial \& Engineering Chemistry Research},
  63(20):9225--9235, 2024.

\bibitem{jumper2021highly}
J.~Jumper, R.~Evans, A.~Pritzel, T.~Green, M.~Figurnov, O.~Ronneberger,
  K.~Tunyasuvunakool, R.~Bates, A.~{\v{Z}}{\'\i}dek, A.~Potapenko, et~al.
\newblock Highly accurate protein structure prediction with alphafold.
\newblock {\em Nature}, 596(7873):583--589, 2021.

\bibitem{KABALAN2025115929}
A.~Kabalan, F.~Casenave, F.~Bordeu, V.~Ehrlacher, and A.~Ern.
\newblock Elasticity-based morphing technique and application to reduced-order
  modeling.
\newblock {\em Applied Mathematical Modelling}, 141:115929, 2025.

\bibitem{kamrin2024advances}
K.~Kamrin, K.~M. Hill, D.~I. Goldman, and J.~E. Andrade.
\newblock Advances in modeling dense granular media.
\newblock {\em Annual Review of Fluid Mechanics}, 56(1):215--240, 2024.

\bibitem{kast2020non}
M.~Kast, M.~Guo, and J.~S. Hesthaven.
\newblock A non-intrusive multifidelity method for the reduced order modeling
  of nonlinear problems.
\newblock {\em Computer Methods in Applied Mechanics and Engineering},
  364:112947, 2020.

\bibitem{Khosravi2020}
A.~Khosravi, A.~Martinez, and J.~T. Dejong.
\newblock {Discrete element model ({DEM}) simulations of cone penetration test
  (CPT) measurements and soil classification}.
\newblock {\em Can. Geotech. J.}, 57(9):1369--1387, 2020.

\bibitem{Kieckhefen2020}
P.~Kieckhefen, S.~Pietsch, M.~Dosta, and S.~Heinrich.
\newblock Possibilities and limits of computational fluid dynamics–discrete
  element method simulations in process engineering: A review of recent
  advancements and future trends.
\newblock {\em Annual Review of Chemical and Biomolecular Engineering},
  11(Volume 11, 2020):397--422, 2020.

\bibitem{Kipf:17}
T.~N. Kipf and M.~Welling.
\newblock Semi-supervised classification with graph convolutional networks.
\newblock In {\em International Conference on Learning Representations}, 2017.

\bibitem{kochkov2021}
D.~Kochkov, J.~A. Smith, A.~Alieva, Q.~Wang, M.~P. Brenner, and S.~Hoyer.
\newblock Machine learning–accelerated computational fluid dynamics.
\newblock {\em Proceedings of the National Aca{DEM}y of Sciences},
  118(21):e2101784118, 2021.

\bibitem{kovachki2023neural}
N.~Kovachki, Z.~Li, B.~Liu, K.~Azizzadenesheli, K.~Bhattacharya, A.~Stuart, and
  A.~Anandkumar.
\newblock Neural operator: Learning maps between function spaces with
  applications to pdes.
\newblock {\em Journal of Machine Learning Research}, 24(89):1--97, 2023.

\bibitem{kozinsky2023}
B.~Kozinsky, A.~Musaelian, A.~Johansson, and S.~Batzner.
\newblock Scaling the leading accuracy of deep equivariant models to
  biomolecular simulations of realistic size.
\newblock In {\em Proceedings of the International Conference for High
  Performance Computing, Networking, Storage and Analysis}, pages 1--12, 2023.

\bibitem{kumar2023accelerating}
K.~Kumar and Y.~Choi.
\newblock Accelerating particle and fluid simulations with differentiable graph
  networks for solving forward and inverse problems.
\newblock In {\em Proceedings of the SC'23 Workshops of The International
  Conference on High Performance Computing, Network, Storage, and Analysis},
  pages 60--65, 2023.

\bibitem{lam2022graphcast}
R.~Lam, A.~Sanchez-Gonzalez, M.~Willson, P.~Wirnsberger, M.~Fortunato,
  A.~Pritzel, S.~Ravuri, T.~Ewalds, F.~Alet, Z.~Eaton-Rosen, et~al.
\newblock {GraphCast: Learning skillful medium-range global weather
  forecasting}.
\newblock {\em arXiv preprint arXiv:2212.12794}, 2022.

\bibitem{Lee2020autoencoder}
K.~Lee and K.~T. Carlberg.
\newblock Model reduction of dynamical systems on nonlinear manifolds using
  deep convolutional autoencoders.
\newblock {\em Journal of Computational Physics}, 404:108973, 2020.

\bibitem{lefik2003artificial}
M.~Lefik and B.~A. Schrefler.
\newblock Artificial neural network as an incremental non-linear constitutive
  model for a finite element code.
\newblock {\em Computer methods in applied mechanics and engineering},
  192(28-30):3265--3283, 2003.

\bibitem{Li2009a}
C.~Li, P.~B. Umbanhowar, H.~Komsuoglu, D.~E. Koditschek, and D.~I. Goldman.
\newblock {Sensitive dependence of the motion of a legged robot on granular
  media}.
\newblock {\em Proc. Natl. Acad. Sci. U. S. A.}, 106(9):3029--3034, 2009.

\bibitem{Li2002}
X.~S. Li and Y.~F. Dafalias.
\newblock {Constitutive modeling of inherently anisotropic sand behavior}.
\newblock {\em J. Geotech. Geoenvironmental Eng.}, 128(10):868--880, oct 2002.

\bibitem{li2021fourier}
Z.~Li, N.~Kovachki, K.~Azizzadenesheli, B.~Liu, K.~Bhattacharya, A.~Stuart, and
  A.~Anandkumar.
\newblock Fourier neural operator for parametric partial differential
  equations, 2021.

\bibitem{li2023geometryinformed}
Z.~Li, N.~B. Kovachki, C.~Choy, B.~Li, J.~Kossaifi, S.~P. Otta, M.~A. Nabian,
  M.~Stadler, C.~Hundt, K.~Azizzadenesheli, and A.~Anandkumar.
\newblock Geometry-informed neural operator for large-scale 3d pdes, 2023.

\bibitem{li23gino}
Z.~Li, N.~B. Kovachki, C.~Choy, B.~Li, J.~Kossaifi, S.~P. Otta, M.~A. Nabian,
  M.~Stadler, C.~Hundt, K.~Azizzadenesheli, et~al.
\newblock Geometry-informed neural operator for large-scale 3d pdes.
\newblock {\em arXiv preprint arXiv:2309.00583}, 2023.

\bibitem{Li:OFormer}
Z.~Li, K.~Meidani, and A.~B. Farimani.
\newblock Transformer for partial differential equations' operator learning.
\newblock {\em Trans. Mach. Learn. Res.}, 2023, 2023.

\bibitem{Li2020NeuralOG}
Z.-Y. Li, N.~B. Kovachki, K.~Azizzadenesheli, B.~Liu, K.~Bhattacharya, A.~M.
  Stuart, and A.~Anandkumar.
\newblock Neural operator: Graph kernel network for partial differential
  equations.
\newblock {\em ArXiv}, abs/2003.03485, 2020.

\bibitem{liang2002proper}
Y.~C. Liang, H.~P. Lee, S.~P. Lim, W.~Z. Lin, K.~H. Lee, and C.~G. Wu.
\newblock Proper orthogonal decomposition and its applications---{Part I}:
  {T}heory.
\newblock {\em Journal of Sound and Vibration}, 252(3):527--544, 2002.

\bibitem{lino2022multi}
M.~Lino, S.~Fotiadis, A.~A. Bharath, and C.~D. Cantwell.
\newblock Multi-scale rotation-equivariant graph neural networks for unsteady
  eulerian fluid dynamics.
\newblock {\em Physics of Fluids}, 34(8), 2022.

\bibitem{LOMMEN2019273}
S.~Lommen, M.~Mohajeri, G.~Lodewijks, and D.~Schott.
\newblock {DEM} particle upscaling for large-scale bulk handling equipment and
  material interaction.
\newblock {\em Powder Technology}, 352:273--282, 2019.

\bibitem{Lu2019LearningNO}
L.~Lu, P.~Jin, G.~Pang, Z.~Zhang, and G.~E. Karniadakis.
\newblock Learning nonlinear operators via deeponet based on the universal
  approximation theorem of operators.
\newblock {\em Nature Machine Intelligence}, 3:218 -- 229, 2019.

\bibitem{Lubbe2024}
R.~Lubbe, H.~Cheng, P.~Gupta, S.~Luding, and V.~Magnanimo.
\newblock {A comparison of regime-specific continuum models for granular
  flows}.
\newblock In {\em Proceedings of the 11th International Conference on Conveying
  and Handling of Particulate Solids (CHoPS 2024), Edinburgh, UK}, 2024.

\bibitem{Lubbe2022}
R.~Lubbe, W.-J. Xu, Q.~Zhou, and H.~Cheng.
\newblock {Bayesian} calibration of {GPU}–based {DEM} meso-mechanics part ii:
  Calibration of the granular meso-structure.
\newblock {\em Powder Technology}, 407:117666, 2022.

\bibitem{ma2022predictive}
G.~Ma, S.~Guan, Q.~Wang, Y.~Feng, and W.~Zhou.
\newblock A predictive deep learning framework for path-dependent mechanical
  behavior of granular materials.
\newblock {\em Acta Geotechnica}, 17(8):3463--3478, 2022.

\bibitem{Maia2023}
M.~Maia, I.~Rocha, P.~Kerfriden, and F.~{van der Meer}.
\newblock Physically recurrent neural networks for path-dependent heterogeneous
  materials: Embedding constitutive models in a data-driven surrogate.
\newblock {\em Computer Methods in Applied Mechanics and Engineering},
  407:115934, 2023.

\bibitem{software_management_plan}
C.~Martinez-Ortiz, P.~Martinez~Lavanchy, L.~Sesink, B.~G. Olivier, J.~Meakin,
  M.~de~Jong, and M.~Cruz.
\newblock Practical guide to software management plans, Oct. 2022.

\bibitem{martinez2019crisp}
F.~Mart{\'\i}nez-Plumed, L.~Contreras-Ochando, C.~Ferri,
  J.~Hern{\'a}ndez-Orallo, M.~Kull, N.~Lachiche, M.~J. Ram{\'\i}rez-Quintana,
  and P.~Flach.
\newblock Crisp-dm twenty years later: From data mining processes to data
  science trajectories.
\newblock {\em IEEE transactions on knowledge and data engineering},
  33(8):3048--3061, 2019.

\bibitem{Dieter2020}
C.~Matl, Y.~Narang, R.~Bajcsy, F.~Ramos, and D.~Fox.
\newblock Inferring the material properties of granular media for robotic
  tasks.
\newblock In {\em 2020 IEEE International Conference on Robotics and Automation
  (ICRA)}, pages 2770--2777, 2020.

\bibitem{MATRAY2024117243}
V.~Matray, F.~Amlani, F.~Feyel, and D.~Néron.
\newblock A hybrid numerical methodology coupling reduced order modeling and
  graph neural networks for non-parametric geometries: Applications to
  structural dynamics problems.
\newblock {\em Computer Methods in Applied Mechanics and Engineering},
  430:117243, 2024.

\bibitem{mayr2023boundary}
A.~Mayr, S.~Lehner, A.~Mayrhofer, C.~Kloss, S.~Hochreiter, and J.~Brandstetter.
\newblock Boundary graph neural networks for 3d simulations, 2023.

\bibitem{Mayr:23}
A.~Mayr, S.~Lehner, A.~Mayrhofer, C.~Kloss, S.~Hochreiter, and J.~Brandstetter.
\newblock Boundary graph neural networks for 3d simulations.
\newblock In {\em Proceedings of the AAAI Conference on Artificial
  Intelligence}, volume~37, pages 9099--9107, 2023.

\bibitem{mccabe2023multiple}
M.~McCabe, B.~R.-S. Blancard, L.~H. Parker, R.~Ohana, M.~Cranmer, A.~Bietti,
  M.~Eickenberg, S.~Golkar, G.~Krawezik, F.~Lanusse, et~al.
\newblock Multiple physics pretraining for physical surrogate models.
\newblock {\em arXiv preprint arXiv:2310.02994}, 2023.

\bibitem{Merchant:23}
A.~Merchant, S.~Batzner, S.~S. Schoenholz, M.~Aykol, G.~Cheon, and E.~D. Cubuk.
\newblock Scaling deep learning for materials discovery.
\newblock {\em Nature}, pages 1--6, 2023.

\bibitem{Montella2023}
E.~Montellà, J.~Chauchat, C.~Bonamy, D.~Weij, G.~Keetels, and T.~Hsu.
\newblock Numerical investigation of mode failures in submerged granular
  columns.
\newblock {\em Flow}, 3:E28, 2023.

\bibitem{murdoch2019definitions}
W.~J. Murdoch, C.~Singh, K.~Kumbier, R.~Abbasi-Asl, and B.~Yu.
\newblock Definitions, methods, and applications in interpretable machine
  learning.
\newblock {\em Proceedings of the National Aca{DEM}y of Sciences},
  116(44):22071--22080, 2019.

\bibitem{murphy2012machine}
K.~P. Murphy.
\newblock {\em Machine Learning: A Probabilistic Perspective}.
\newblock MIT press, 2012.

\bibitem{musaelian2023}
A.~Musaelian, S.~Batzner, A.~Johansson, L.~Sun, C.~J. Owen, M.~Kornbluth, and
  B.~Kozinsky.
\newblock Learning local equivariant representations for large-scale atomistic
  dynamics.
\newblock {\em Nature Communications}, 14(1):579, 2023.

\bibitem{Nakamura2020}
H.~Nakamura, H.~Takimoto, N.~Kishida, S.~Ohsaki, and S.~Watano.
\newblock {Coarse-grained discrete element method for granular shear flow}.
\newblock {\em Chem. Eng. J. Adv.}, 4(September):100050, 2020.

\bibitem{nguyen2022synthesizing}
P.~C. Nguyen, N.~N. Vlassis, B.~Bahmani, W.~Sun, H.~Udaykumar, and S.~S. Baek.
\newblock Synthesizing controlled microstructures of porous media using
  generative adversarial networks and reinforcement learning.
\newblock {\em Scientific reports}, 12(1):9034, 2022.

\bibitem{nguyen2023climax}
T.~Nguyen, J.~Brandstetter, A.~Kapoor, J.~K. Gupta, and A.~Grover.
\newblock Climax: A foundation model for weather and climate, 2023.

\bibitem{OSullivan2011}
C.~O'Sullivan.
\newblock {\em {Particulate discrete element modelling: a geomechanics
  perspective}}.
\newblock Taylor {\&} Francis, 2011.

\bibitem{PARTELI201696}
E.~J. Parteli and T.~Pöschel.
\newblock Particle-based simulation of powder application in additive
  manufacturing.
\newblock {\em Powder Technology}, 288:96--102, 2016.

\bibitem{pathak2022fourcastnet}
J.~Pathak, S.~Subramanian, P.~Harrington, S.~Raja, A.~Chattopadhyay,
  M.~Mardani, T.~Kurth, D.~Hall, Z.~Li, K.~Azizzadenesheli, P.~Hassanzadeh,
  K.~Kashinath, and A.~Anandkumar.
\newblock {FourCastNet: A Global Data-driven High-resolution Weather Model
  using Adaptive Fourier Neural Operators}.
\newblock {\em arXiv preprint arXiv:2202.11214}, 2022.

\bibitem{PW2016operatorInference}
B.~Peherstorfer and K.~Willcox.
\newblock Data-driven operator inference for nonintrusive projection-based
  model reduction.
\newblock {\em Computer Methods in Applied Mechanics and Engineering},
  306:196--215, 2016.

\bibitem{Pfaff:20}
T.~Pfaff, M.~Fortunato, A.~Sanchez-Gonzalez, and P.~W. Battaglia.
\newblock Learning mesh-based simulation with graph networks.
\newblock {\em arXiv preprint arXiv:2010.03409}, 2020.

\bibitem{pfortner2022physics}
M.~Pf{\"o}rtner, I.~Steinwart, P.~Hennig, and J.~Wenger.
\newblock Physics-informed gaussian process regression generalizes linear pde
  solvers.
\newblock {\em arXiv preprint arXiv:2212.12474}, 2022.

\bibitem{phan2025hydra}
N.~N. Phan, W.~Sun, and J.~D. Clayton.
\newblock Hydra: Symbolic feature engineering of overparameterized eulerian
  hyperelasticity models for fast inference time.
\newblock {\em Computer Methods in Applied Mechanics and Engineering},
  437:117792, 2025.

\bibitem{polania2024polydispersity}
O.~Polan{\'\i}a, N.~Estrada, E.~Az{\'e}ma, M.~Renouf, and M.~Cabrera.
\newblock Polydispersity effect on dry and immersed granular collapses: an
  experimental study.
\newblock {\em Journal of Fluid Mechanics}, 983:A40, 2024.

\bibitem{qi2017}
C.~R. Qi, L.~Yi, H.~Su, and L.~J. Guibas.
\newblock Pointnet++: Deep hierarchical feature learning on point sets in a
  metric space.
\newblock {\em Advances in neural information processing systems}, 30, 2017.

\bibitem{QKPW2020liftAndLearn}
E.~Qian, B.~Kramer, B.~Peherstorfer, and K.~Willcox.
\newblock Lift \& {L}earn: {P}hysics-informed machine learning for large-scale
  nonlinear dynamical systems.
\newblock {\em Physica {D}: {N}onlinear {P}henomena}, 406:132401, 2020.

\bibitem{qu2021towards}
T.~Qu, S.~Di, Y.~Feng, M.~Wang, and T.~Zhao.
\newblock Towards data-driven constitutive modelling for granular materials via
  micromechanics-informed deep learning.
\newblock {\em International Journal of Plasticity}, 144:103046, 2021.

\bibitem{qu2023deep}
T.~Qu, S.~Guan, Y.~Feng, G.~Ma, W.~Zhou, and J.~Zhao.
\newblock Deep active learning for constitutive modelling of granular
  materials: From representative volume elements to implicit finite element
  modelling.
\newblock {\em International Journal of Plasticity}, 164:103576, 2023.

\bibitem{qu2023data}
T.~Qu, J.~Zhao, S.~Guan, and Y.~Feng.
\newblock Data-driven multiscale modelling of granular materials via knowledge
  transfer and sharing.
\newblock {\em International Journal of Plasticity}, 171:103786, 2023.

\bibitem{quarteroni2015reduced}
A.~Quarteroni, A.~Manzoni, and F.~Negri.
\newblock {\em Reduced Basis Methods for Partial Differential Equations: An
  Introduction}.
\newblock Springer, 2015.

\bibitem{quarteroni2008numerical}
A.~Quarteroni and A.~Valli.
\newblock {\em Numerical approximation of partial differential equations},
  volume~23.
\newblock Springer Science \& Business Media, 2008.

\bibitem{raissi2017machine}
M.~Raissi, P.~Perdikaris, and G.~E. Karniadakis.
\newblock Machine learning of linear differential equations using gaussian
  processes.
\newblock {\em Journal of Computational Physics}, 348:683--693, 2017.

\bibitem{Physics_Informed_Latent}
M.~Raissi, P.~Perdikaris, and G.~E. Karniadakis.
\newblock Physics-informed neural networks: A deep learning framework for
  solving forward and inverse problems involving nonlinear partial differential
  equations.
\newblock {\em Journal of Computational Physics}, 378:686--707, 2019.

\bibitem{raonic2024convolutional}
B.~Raonic, R.~Molinaro, T.~De~Ryck, T.~Rohner, F.~Bartolucci, R.~Alaifari,
  S.~Mishra, and E.~de~B{\'e}zenac.
\newblock Convolutional neural operators for robust and accurate learning of
  pdes.
\newblock {\em Advances in Neural Information Processing Systems}, 36, 2024.

\bibitem{williams2006gaussian}
C.~E. Rasmussen and C.~K. Williams.
\newblock {\em Gaussian processes for machine learning}, volume~2.
\newblock MIT press Cambridge, MA, 2006.

\bibitem{rasp2021data}
S.~Rasp and N.~Thuerey.
\newblock {Data-driven medium-range weather prediction with a resnet pretrained
  on climate simulations: A new model for weatherbench}.
\newblock {\em Journal of Advances in Modeling Earth Systems},
  13(2):e2020MS002405, 2021.

\bibitem{roy2024role}
S.~Roy and T.~Weinhart.
\newblock The role of granular matter in additive manufacturing.
\newblock {\em Granular Matter}, 26(4):102, 2024.

\bibitem{Sanchez:20}
A.~Sanchez-Gonzalez, J.~Godwin, T.~Pfaff, R.~Ying, J.~Leskovec, and
  P.~Battaglia.
\newblock Learning to simulate complex physics with graph networks.
\newblock In {\em International conference on machine learning}, pages
  8459--8468. PMLR, 2020.

\bibitem{sarlin2021collapse}
W.~Sarlin, C.~Morize, A.~Sauret, and P.~Gondret.
\newblock Collapse dynamics of dry granular columns: From free-fall to
  quasistatic flow.
\newblock {\em Physical Review E}, 104(6):064904, 2021.

\bibitem{satorras2021n}
V.~G. Satorras, E.~Hoogeboom, and M.~Welling.
\newblock E (n) equivariant graph neural networks.
\newblock In {\em International conference on machine learning}, pages
  9323--9332. PMLR, 2021.

\bibitem{Scarselli:08}
F.~Scarselli, M.~Gori, A.~C. Tsoi, M.~Hagenbuchner, and G.~Monfardini.
\newblock The graph neural network model.
\newblock {\em IEEE transactions on neural networks}, 20(1):61--80, 2008.

\bibitem{schaeffer2017learning}
H.~Schaeffer.
\newblock Learning partial differential equations via data discovery and sparse
  optimization.
\newblock {\em Proceedings of the Royal Society A: Mathematical, Physical and
  Engineering Sciences}, 473(2197):20160446, 2017.

\bibitem{SCHOTT202129}
D.~Schott, J.~Mohajeri, J.~Jovanova, S.~Lommen, and W.~{de Kluijver}.
\newblock Design framework for {DEM}-supported prototyping of grabs including
  full-scale validation.
\newblock {\em Journal of Terramechanics}, 96:29--43, 2021.

\bibitem{Schott2023}
D.~L. Schott and J.~Mohajeri.
\newblock {\em Multibody Dynamics and Discrete Element Method Co-Simulations
  for Large-Scale Industrial Equipment}, chapter~4, pages 107--143.
\newblock John Wiley \& Sons, Ltd, 2023.

\bibitem{seidman2022}
J.~Seidman, G.~Kissas, P.~Perdikaris, and G.~J. Pappas.
\newblock Nomad: Nonlinear manifold decoders for operator learning.
\newblock {\em Advances in Neural Information Processing Systems},
  35:5601--5613, 2022.

\bibitem{shaheen2021influence}
M.~Y. Shaheen, A.~R. Thornton, S.~Luding, and T.~Weinhart.
\newblock The influence of material and process parameters on powder spreading
  in additive manufacturing.
\newblock {\em Powder technology}, 383:564--583, 2021.

\bibitem{sharma2025dynamical}
V.~Sharma and O.~Fink.
\newblock Dynami-cal graphnet: A physics-informed graph neural network
  conserving linear and angular momentum for dynamical systems, 2025.

\bibitem{SHMULEVICH200737}
I.~Shmulevich, Z.~Asaf, and D.~Rubinstein.
\newblock Interaction between soil and a wide cutting blade using the discrete
  element method.
\newblock {\em Soil and Tillage Research}, 97(1):37--50, 2007.

\bibitem{sonderby2020metnet}
C.~K. S{\o}nderby, L.~Espeholt, J.~Heek, M.~Dehghani, A.~Oliver, T.~Salimans,
  S.~Agrawal, J.~Hickey, and N.~Kalchbrenner.
\newblock Metnet: A neural weather model for precipitation forecasting.
\newblock {\em arXiv preprint arXiv:2003.12140}, 2020.

\bibitem{staron2023cohesive}
L.~Staron, L.~Duchemin, and P.-Y. Lagr{\'e}e.
\newblock Cohesive granular columns collapsing: Numerics questioning failure,
  cohesion, and friction.
\newblock {\em Journal of Rheology}, 67(5):1061--1072, 2023.

\bibitem{Storm2024}
J.~Storm, I.~Rocha, and F.~{van der Meer}.
\newblock A microstructure-based graph neural network for accelerating
  multiscale simulations.
\newblock {\em Computer Methods in Applied Mechanics and Engineering},
  427:117001, 2024.

\bibitem{su2023multifidelity}
M.~Su, N.~Guo, and Z.~Yang.
\newblock A multifidelity neural network (mfnn) for constitutive modeling of
  complex soil behaviors.
\newblock {\em International Journal for Numerical and Analytical Methods in
  Geomechanics}, 47(18):3269--3289, 2023.

\bibitem{sun2022data}
X.~Sun, B.~Bahmani, N.~N. Vlassis, W.~Sun, and Y.~Xu.
\newblock Data-driven discovery of interpretable causal relations for deep
  learning material laws with uncertainty propagation.
\newblock {\em Granular Matter}, 24:1--32, 2022.

\bibitem{swischuk2019MOR_ML}
R.~Swischuk, L.~Mainini, B.~Peherstorfer, and K.~Willcox.
\newblock Projection-based model reduction: Formulations for physics-based
  machine learning.
\newblock {\em Computers \& Fluids}, 179:704--717, 2019.

\bibitem{Sarkka2013}
S.~Särkkä.
\newblock {\em {Bayesian} Filtering and Smoothing}, volume~3 of {\em Institute
  of Mathematical Statistics Textbooks}.
\newblock Cambridge University Press, Cambridge, UK, 2013.

\bibitem{thuerey2021physics}
N.~Thuerey, P.~Holl, M.~Mueller, P.~Schnell, F.~Trost, and K.~Um.
\newblock {Physics-based Deep Learning}.
\newblock {\em arXiv preprint arXiv:2109.05237}, 2021.

\bibitem{VAE_Application}
A.~Tong, J.~Huang, G.~Wolf, D.~van Dijk, and S.~Krishnaswamy.
\newblock Trajectorynet: A dynamic optimal transport network for modeling
  cellular dynamics.
\newblock In {\em International Conference on Machine Learning}, pages
  9526--9536. PMLR, 2020.

\bibitem{Tordesillas2022}
A.~Tordesillas, S.~Zhou, J.~Bailey, and H.~Bondell.
\newblock A representation learning framework for detection and
  characterization of dead versus strain localization zones from pre- to
  post-failure.
\newblock {\em Granular Matter}, 24(75), 2022.

\bibitem{tunuguntla2014mixture}
D.~R. Tunuguntla, O.~Bokhove, and A.~R. Thornton.
\newblock A mixture theory for size and density segregation in shallow granular
  free-surface flows.
\newblock {\em Journal of fluid mechanics}, 749:99--112, 2014.

\bibitem{Linden2016}
J.~H. van~der Linden, G.~A. Narsilio, and A.~Tordesillas.
\newblock Machine learning framework for analysis of transport through complex
  networks in porous, granular media: A focus on permeability.
\newblock {\em Phys. Rev. E}, 94:022904, Aug 2016.

\bibitem{Joseph2022}
J.~Vantassel and K.~Kumar.
\newblock Graph network simulator datasets, 2022.

\bibitem{Vescovi2016}
D.~Vescovi and S.~Luding.
\newblock Merging fluid and solid granular behavior.
\newblock {\em Soft Matter}, 12:8616--8628, 2016.

\bibitem{villarreal2023design}
R.~Villarreal, N.~N. Vlassis, N.~N. Phan, T.~A. Catanach, R.~E. Jones, N.~A.
  Trask, S.~L. Kramer, and W.~Sun.
\newblock Design of experiments for the calibration of history-dependent models
  via deep reinforcement learning and an enhanced kalman filter.
\newblock {\em Computational Mechanics}, 72(1):95--124, 2023.

\bibitem{Vlahinic2016}
I.~Vlahini{\'{c}}, R.~Kawamoto, E.~And{\`{o}}, G.~Viggiani, and J.~E. Andrade.
\newblock {From computed tomography to mechanics of granular materials via
  level set bridge}.
\newblock {\em Acta Geotechnica}, pages 1--11, 10 2016.

\bibitem{vlassis2020geometric}
N.~N. Vlassis, R.~Ma, and W.~Sun.
\newblock Geometric deep learning for computational mechanics part i:
  Anisotropic hyperelasticity.
\newblock {\em Computer Methods in Applied Mechanics and Engineering},
  371:113299, 2020.

\bibitem{10.1115/1.4052684}
N.~N. Vlassis and W.~Sun.
\newblock {Component-Based Machine Learning Paradigm for Discovering
  Rate-Dependent and Pressure-Sensitive Level-Set Plasticity Models}.
\newblock {\em Journal of Applied Mechanics}, 89(2):021003, 11 2021.

\bibitem{vlassis2021sobolev}
N.~N. Vlassis and W.~Sun.
\newblock Sobolev training of thermodynamic-informed neural networks for
  interpretable elasto-plasticity models with level set hardening.
\newblock {\em Computer Methods in Applied Mechanics and Engineering},
  377:113695, 2021.

\bibitem{vlassis2024synthesizing}
N.~N. Vlassis, W.~Sun, K.~A. Alshibli, and R.~A. Regueiro.
\newblock Synthesizing realistic sand assemblies with denoising diffusion in
  latent space.
\newblock {\em International Journal for Numerical and Analytical Methods in
  Geomechanics}, 48(16):3933--3956, 2024.

\bibitem{vlassis2022molecular}
N.~N. Vlassis, P.~Zhao, R.~Ma, T.~Sewell, and W.~Sun.
\newblock Molecular dynamics inferred transfer learning models for
  finite-strain hyperelasticity of monoclinic crystals: Sobolev training and
  validations against physical constraints.
\newblock {\em International Journal for Numerical Methods in Engineering},
  123(17):3922--3949, 2022.

\bibitem{wang2018multiscale}
K.~Wang and W.~Sun.
\newblock A multiscale multi-permeability poroplasticity model linked by
  recursive homogenizations and deep learning.
\newblock {\em Computer Methods in Applied Mechanics and Engineering},
  334:337--380, 2018.

\bibitem{wang2019meta}
K.~Wang and W.~Sun.
\newblock Meta-modeling game for deriving theory-consistent,
  microstructure-based traction--separation laws via deep reinforcement
  learning.
\newblock {\em Computer Methods in Applied Mechanics and Engineering},
  346:216--241, 2019.

\bibitem{wang2019cooperative}
K.~Wang, W.~Sun, and Q.~Du.
\newblock A cooperative game for automated learning of elasto-plasticity
  knowledge graphs and models with ai-guided experimentation.
\newblock {\em Computational Mechanics}, 64:467--499, 2019.

\bibitem{wang2021non}
K.~Wang, W.~Sun, and Q.~Du.
\newblock A non-cooperative meta-modeling game for automated third-party
  calibrating, validating and falsifying constitutive laws with parallelized
  adversarial attacks.
\newblock {\em Computer Methods in Applied Mechanics and Engineering},
  373:113514, 2021.

\bibitem{wang2024multi}
M.~Wang, Y.~Feng, S.~Guan, and T.~Qu.
\newblock Multi-layer perceptron-based data-driven multiscale modelling of
  granular materials with a novel frobenius norm-based internal variable.
\newblock {\em Journal of Rock Mechanics and Geotechnical Engineering}, 2024.

\bibitem{wang2024machine}
M.~Wang, K.~Kumar, Y.~Feng, T.~Qu, and M.~Wang.
\newblock Machine learning aided modeling of granular materials: A review.
\newblock {\em Archives of Computational Methods in Engineering}, pages 1--38,
  2024.

\bibitem{wang2022data}
M.~Wang, T.~Qu, S.~Guan, T.~Zhao, B.~Liu, and Y.~Feng.
\newblock Data-driven strain--stress modelling of granular materials via
  temporal convolution neural network.
\newblock {\em Computers and Geotechnics}, 152:105049, 2022.

\bibitem{Wang2022b}
S.~Wang and S.~Ji.
\newblock {A unified level set method for simulating mixed granular flows
  involving multiple non-spherical {DEM} models in complex structures}.
\newblock {\em Comput. Methods Appl. Mech. Eng.}, 393:114802, 2022.

\bibitem{wang2024cvit}
S.~Wang, J.~H. Seidman, S.~Sankaran, H.~Wang, G.~J. Pappas, and P.~Perdikaris.
\newblock Cvit: Continuous vision transformer for operator learning.
\newblock 2024.

\bibitem{wang2024lno}
T.~Wang and C.~Wang.
\newblock Latent neural operator for solving forward and inverse pde problems,
  2024.

\bibitem{webb2024continuum}
W.~Webb, B.~Turnbull, and C.~Johnson.
\newblock Continuum modelling of a just-saturated inertial column collapse:
  capturing fluid-particle interaction.
\newblock {\em Granular Matter}, 26(1):21, 2024.

\bibitem{Weinhart2020}
T.~Weinhart, L.~Orefice, M.~Post, M.~P.~M. {van Schrojenstein Lantman},
  I.~I.~F. Denissen, D.~D.~R. Tunuguntla, J.~Tsang, H.~Cheng, M.~Y.~M. Shaheen,
  H.~Shi, P.~Rapino, E.~Grannonio, N.~Losacco, J.~Barbosa, L.~L. Jing, J.~J.~E.
  {Alvarez Naranjo}, S.~Roy, W.~K.~W. den Otter, and A.~A.~R. Thornton.
\newblock {Fast, flexible particle simulations — An introduction to
  MercuryDPM}.
\newblock {\em Comput. Phys. Commun.}, 249:107129, dec 2020.

\bibitem{weyn2020improving}
J.~A. Weyn, D.~R. Durran, and R.~Caruana.
\newblock {Improving data-driven global weather prediction using deep
  convolutional neural networks on a cubed sphere}.
\newblock {\em Journal of Advances in Modeling Earth Systems},
  12(9):e2020MS002109, 2020.

\bibitem{Wilke2021kds}
D.~N. Wilke, P.~W. Cleary, and N.~Govender.
\newblock From discrete element simulation data to process insights.
\newblock {\em EPJ Web of Conferences}, 249:15001, 2021.

\bibitem{10.1007/978-3-030-85584-0_5}
D.~N. Wilke, P.~S. Heyns, and S.~Schmidt.
\newblock The role of untangled latent spaces in unsupervised learning applied
  to condition-based maintenance.
\newblock In A.~Hammami, P.~S. Heyns, S.~Schmidt, F.~Chaari, M.~S. Abbes, and
  M.~Haddar, editors, {\em Modelling and Simulation of Complex Systems for
  Sustainable Energy Efficiency}, pages 38--49, Cham, 2022. Springer
  International Publishing.

\bibitem{xiao2015non}
D.~Xiao, F.~Fang, A.~G. Buchan, C.~C. Pain, I.~M. Navon, and A.~Muggeridge.
\newblock Non-intrusive reduced order modelling of the navier--stokes
  equations.
\newblock {\em Computer Methods in Applied Mechanics and Engineering},
  293:522--541, 2015.

\bibitem{Tian2018}
T.~Xie and J.~C. Grossman.
\newblock Hierarchical visualization of materials space with graph
  convolutional neural networks.
\newblock {\em The Journal of Chemical Physics}, 149(17):174111, 11 2018.

\bibitem{xu2022improved}
D.~Xu and Y.~Shen.
\newblock An improved machine learning approach for predicting granular flows.
\newblock {\em Chemical Engineering Journal}, 450:138036, 2022.

\bibitem{GAN_Application}
Z.~Yang, X.~Li, L.~C. Brinson, A.~N. Choudhary, W.~Chen, and A.~Agrawal.
\newblock Microstructural materials design via deep adversarial learning
  methodology.
\newblock {\em Journal of Mechanical Design}, 140(11), 2018.

\bibitem{ye2024gaussian}
D.~Ye and M.~Guo.
\newblock Gaussian process learning of nonlinear dynamics.
\newblock {\em Communications in Nonlinear Science and Numerical Simulation},
  page 108184, 2024.

\bibitem{YE2024112639}
D.~Ye, V.~Krzhizhanovskaya, and A.~G. Hoekstra.
\newblock Data-driven reduced-order modelling for blood flow simulations with
  geometry-informed snapshots.
\newblock {\em Journal of Computational Physics}, 497:112639, 2024.

\bibitem{yu2019non}
J.~Yu, C.~Yan, and M.~Guo.
\newblock Non-intrusive reduced-order modeling for fluid problems: A brief
  review.
\newblock {\em Proceedings of the Institution of Mechanical Engineers, Part G:
  Journal of Aerospace Engineering}, 233(16):5896--5912, 2019.

\bibitem{Zeni:23}
C.~Zeni, R.~Pinsler, D.~Z{\"u}gner, A.~Fowler, M.~Horton, X.~Fu, S.~Shysheya,
  J.~Crabb{\'e}, L.~Sun, J.~Smith, et~al.
\newblock Mattergen: a generative model for inorganic materials design.
\newblock {\em arXiv preprint arXiv:2312.03687}, 2023.

\bibitem{zhang2024multifidelity}
P.~Zhang, Z.-Y. Yin, and B.~Sheil.
\newblock Multifidelity constitutive modeling of stress-induced anisotropic
  behavior of clay.
\newblock {\em Journal of Geotechnical and Geoenvironmental Engineering},
  150(3):04024003, 2024.

\bibitem{zhang2024science}
X.~Zhang, L.~Wang, J.~Helwig, Y.~Luo, C.~Fu, Y.~Xie, M.~Liu, Y.~Lin, Z.~Xu,
  K.~Yan, K.~Adams, M.~Weiler, X.~Li, T.~Fu, Y.~Wang, H.~Yu, Y.~Xie, X.~Fu,
  A.~Strasser, S.~Xu, Y.~Liu, Y.~Du, A.~Saxton, H.~Ling, H.~Lawrence,
  H.~Stärk, S.~Gui, C.~Edwards, N.~Gao, A.~Ladera, T.~Wu, E.~F. Hofgard, A.~M.
  Tehrani, R.~Wang, A.~Daigavane, M.~Bohde, J.~Kurtin, Q.~Huang, T.~Phung,
  M.~Xu, C.~K. Joshi, S.~V. Mathis, K.~Azizzadenesheli, A.~Fang,
  A.~Aspuru-Guzik, E.~Bekkers, M.~Bronstein, M.~Zitnik, A.~Anandkumar,
  S.~Ermon, P.~Liò, R.~Yu, S.~Günnemann, J.~Leskovec, H.~Ji, J.~Sun,
  R.~Barzilay, T.~Jaakkola, C.~W. Coley, X.~Qian, X.~Qian, T.~Smidt, and S.~Ji.
\newblock Artificial intelligence for science in quantum, atomistic, and
  continuum systems, 2024.

\bibitem{Zhao2013}
J.~Zhao and N.~Guo.
\newblock {Unique critical state characteristics in granular media considering
  fabric anisotropy}.
\newblock {\em G{\'{e}}otechnique}, 8(8):695--704, 6 2013.

\bibitem{Zhao2021a}
S.~Zhao and J.~Zhao.
\newblock {Sudo{DEM}: Unleashing the predictive power of the discrete element
  method on simulation for non-spherical granular particles}.
\newblock {\em Comput. Phys. Commun.}, 259:107670, 2021.

\bibitem{zhao2017coupled}
T.~Zhao.
\newblock {\em Coupled {DEM}-{CFD} Analyses of Landslide-Induced Debris Flows}.
\newblock Springer, english edition, May 2017.
\newblock Hardcover.

\bibitem{zhuang2021model}
Q.~Zhuang, J.~M. Lorenzi, H.-J. Bungartz, and D.~Hartmann.
\newblock Model order reduction based on {R}unge-{K}utta neural networks.
\newblock {\em Data-Centric Engineering}, 2, 2021.

\end{thebibliography}

\end{document}